\documentclass[11pt,prd,aps,nofootinbib,floatfix]{revtex4-1} 

\usepackage{epsfig}
\usepackage{graphicx}
\usepackage{multirow} 
\usepackage{color}
\usepackage{hyperref}
\usepackage{amsmath}
\usepackage{amssymb}
\usepackage{macros}
\usepackage{rotating}
\usepackage[normalem]{ulem}
\usepackage{slashed}
\graphicspath{{./figs//}}

\newcommand\higgs{Higgs Centre for Theoretical Physics, School of Physics \& Astronomy, \\
  University of Edinburgh, EH9 3FD, UK}

\newcommand\york{Department of Physics and Astronomy, York University, Toronto, Ontario, M3J 1P3, Canada}

\newcommand\glasgow{SUPA, School of Physics and Astronomy, University of Glasgow, Glasgow, G12 8QQ, UK}
\newcommand\bnl{Physics Department, Brookhaven National Laboratory, Upton, NY 11973, USA}
\newcommand\liverpool{Theoretical Physics Division, Department of Mathematical Sciences,\\ University of Liverpool, Liverpool L69 3BX, UK}

\begin{document}


\title{Neutral kaon mixing beyond the Standard Model with \texorpdfstring{$n_f=2+1$}{nf=2+1} chiral fermions \\
  part II: Non Perturbative Renormalisation of the \texorpdfstring{$\Delta F=2$}{Delta-F=2} four-quark  operators}

\author{P.A.~Boyle~$^1$, N.~Garron~$^{2}$\footnote{nicolas.garron@liverpool.ac.uk}, R.J.~Hudspith~$^3$,
  C.~Lehner~$^4$, A.T.~Lytle~$^5$.}
\affiliation{$^1$\higgs,}
\affiliation{$^2$\liverpool,}
\affiliation{$^3$\york,}
\affiliation{$^4$\bnl,}
\affiliation{$^5$\glasgow.}

\collaboration{The RBC and UKQCD collaborations}

\date{\today}

\begin{abstract}  
  We compute the renormalisation factors ($Z$-matrices) of the $\Delta F=2$ four-quark operators
  needed for Beyond the Standard Model (BSM) kaon mixing.
  We work with $n_f=2+1$ flavours of Domain-Wall fermions whose chiral-flavour properties are essential to maintain
  a continuum-like mixing pattern.
  We introduce new RI-SMOM renormalisation schemes, which we argue are better behaved compared to the commonly-used corresponding
  RI-MOM one.
  We find that, once converted to $\msbar$, the Z-factors computed through these RI-SMOM schemes are in good agreement
  but differ significantly from the ones computed through the RI-MOM scheme. 
  The RI-SMOM $Z$-factors presented here have been used to compute the BSM neutral kaon mixing matrix elements
  in the companion paper~\cite{Garron:2016mva}.
  We argue that the renormalisation procedure is responsible for the discrepancies observed
  by different collaborations, we will investigate and elucidate
  the origin of these differences throughout this work.

\end{abstract}


\maketitle

\section{Introduction}
\label{sec:intro}
Numerical simulations of Quantum ChromoDynamics (QCD) allow for first-principle evaluations
of hadronic matrix elements, which play a crucial r\^ole in theoretical calculations as they
encapsulate the low-energy physics of a process. Computation of such matrix elements is usually
done in two steps: Firstly, the bare quantities of interest are computed at finite lattice spacing $a$,
whose inverse plays the r\^ole an ultra-violet regulator. 
Secondly, these quantities have to be renormalised in order to be divergence-free 
and have a well-defined continuum limit ($a^2\rightarrow 0$).
There are two known non-perturbative methods to perform this renormalisation:
the Schr\"odinger Functional (SF) scheme and the other being the Rome-Southampton method~\cite{Martinelli:1994ty}.
We choose to work with the latter, for practical reasons 
(the interested reader can find a recent study of the same set of operators within the SF
in~\cite{Mainar:2016uwb} and~\cite{Papinutto:2016xpq}).
In phenomenological applications the renormalised quantities are then matched to a scheme in which the corresponding
short distance contributions can be computed, this is commonly performed in the modified minimal subtraction scheme
$\msbar$, see for example \cite{Ciuchini:1997bw,Buras:2000if}.

Let us begin by considering the matrix element of an operator $\la O \ra$ which renormalises multiplicatively,
and with $ \la O \ra^\text{bare}(a)$ being a bare matrix element computed at finite lattice spacing
$a$. We denote $Z^\text{RI}$ the corresponding renormalisation factor computed
on the same lattice (following the Rome-Southampton method) in a regularisation independent (RI) scheme.
The precise definition of the schemes (RI-MOM or a RI-SMOM)
will be given in the next section.
Within our conventions, at some renormalisation scale $\mu$, the renormalised matrix element is
given by 
\be
\label{eq:ZRI}
\la O \ra^\text{RI}(\mu,a) = Z^\text{RI}(\mu,a) \la O \ra^\text{bare}(a) \;,
\ee
which now has a well defined continuum limit
\be
\la O \ra^\text{RI}(\mu) = \lim_{a^2\to 0}\la O \ra^\text{RI}(\mu,a)\;.
\ee
Suppose now that this operator occurs in the determination
of some physical quantity, say an amplitude.
For example in a typical phenomenological application  the hadronic matrix element
has to be combined with a Wilson coefficient $C(\mu)$ computed in continuum perturbation theory 
(the hadronic matrix element describes the long-distance effetcts
and the Wilson coefficient the short-distance ones).
Both of these must be computed in a common scheme, $\msbar$,
to be matched to a physical quantity. Schematically we have
\begin{equation}\label{eq:match}
\begin{aligned}
    \text{Amplitude} &= C^{\msbar} (\mu) \la O \ra^{\msbar} (\mu)\;,\\
    &= C^{\msbar} (\mu) R^{\msbar \leftarrow \text{RI}}(\mu) \la O \ra^\text{RI} (\mu)\;,\\
    &= C^{\msbar} (\mu) R^{\msbar \leftarrow \text{RI}}(\mu)
    \lim_{a^2\to0} \left[Z^\text{RI}(\mu,a) \la O \ra^{bare} (a) \right] \;,
    \end{aligned}
\end{equation}
where
$R$ is the conversion factor from the RI scheme to $\msbar$. 
Eq.~\ref{eq:match} can easily be generalised to the operator mixing case where
$\la O \ra$ and $C$ become vectors, and $R$ and $Z$ become matrices.
We remind the reader that although the renormalisation is performed
non-perturbatively, the matching to $\msbar$ from the RI scheme ($R^{\msbar \leftarrow \text{RI}}(\mu)$)
has to be done using continuum perturbation theory as $\msbar$ is not possible to implement on the lattice.

Accurate matching of lattice operators using the Rome-Southampton technique requires the matching scale $\mu$
(given by the magnitude of a momentum $\mu=\sqrt{p^2}$) to be well-separated from both the scales where 
non-perturbative effects of QCD such as chiral symmetry-breaking become important and the (inverse) lattice
scale where cut-off effects dominate;
ideally one would impose
\be
\Lambda_\text{QCD}^2 \ll \mu^2 \ll \(\frac{\pi}{a}\)^2.
\label{npr_window}
\ee
The first condition ensures that a perturbative treatment of the matching to $\msbar$ is justified, 
while the latter ensures that the lattice artifacts are under control~\footnote{This restriction known at
  the Rome-Southampton window and can be relaxed with step-scaling methods
\cite{Arthur:2010ht,Arthur:2011cn}}.
The $\msbar$ renormalisation factors should be independent of the intermediate (RI) scheme used;
however, in practice there will be some dependence due to systematic uncertainties in the lattice matching
step as well as perturbative truncation errors in the continuum matching.

We compute the $Z$-matrix needed to renormalise the operators required for the determination of neutral
kaon mixing beyond the Standard Model (BSM). 
As is usually done by the RBC-UKQCD collaboration, we implement momentum sources and partially-twisted
boundary conditions. The use of momentum sources (introduced by QCDSF in~\cite{Gockeler:1998ye})
results in very low statistical noise while the use of partially-twisted boundary conditions allows
us to change $\mu=p^2$ smoothly while keeping the orientation of $p$ fixed
\cite{Bedaque:2004kc,deDivitiis:2004kq,Sachrajda:2004mi}. 
In this way we do not discontinuously `jump' into different hypercubic representations as $p^2$ varies,
resulting in $Z$s which are smooth functions of $p^2$.

\begin{table}[t]
  \begin{tabular}{c | c|  c | c }
    \toprule
    $N_f$ &   interm. scheme  & $R_4$ & $R_5$ \\ 
    \hline
    $2$   &  RI-MOM            &  28.5(9)  &  7.3(4) \\
    $2+1$ &  RI-MOM            &  34.6(22) &  8.5(9) \\
    $2+1$ &  $\gmugmu$         &  43.1(25) & 11.0(9) \\
    $2+1$ &  $\qq$             &  44.3(25) & 10.7(9) \\
    \botrule
  \end{tabular}
  \caption{Example of results for the ratio of the BSM matrix elements over the SM one
    $R_i=\la \bar K | O_i^{BSM} | K \ra / \la \bar K | O^{SM} K\ra$, in $\msbar$ at 3 GeV
    in the SUSY basis. The statistical and systematic errors have been combined in quadrature.
    Although in principle these quantities should agree up to $\alpha_s^2$ errors,
    the RI-MOM results differ significantly from the $\gmugmu$ and $\qq$ ones,
    which are consistent with each other.
    The latter are RI-SMOM schemes whose precise definitions are given in this work.
    The $N_f=2+1$ results quoted here are obtained with exactly the same framework apart from the intermediate
    renormalisation scheme, see~\cite{Garron:2016mva}. We argue that the difference comes the
    renormalisation and we suggest to discard the results obtained with the RI-MOM scheme. 
    Not included are results obtained with $N_f=2+1+1$ flavours by the ETM collaboration~\cite{Carrasco:2015pra},
    which are roughly consistent with the $N_f=2$ RI-MOM results,
    and by SWME with $N_f=2+1$ \cite{Bae:2013tca,Jang:2015sla}, which are in a good agreement
    with our RI-SMOM results, see text for discussion.
  }
  \label{table:Rresults}
\end{table}

In principle, after extrapolation to the continuum and conversion to $\msbar$
(or any common scheme) at a given scale, the results should be universal - up to truncation
error of the perturbative series - and in particular should not depend on the details of the discretisation.
The physical results could still depend on the number of dynamical flavours
but, by experience, we do not expect this dependence to be important for the weak matrix elements
discussed in this work. In the past few years, these matrix elements have been
computed by three different collaborations and some discrepancy has been observed
for two of the four relevant four-quark operators.
The first results with dynamical quarks was reported by our collaboration
in \cite{Boyle:2012qb}, it was done with $N_f=2+1$ flavours of dynamical
quarks at a single value of the lattice spacing.
Shortly after our work was published, the ETM collaboration published
their study with $N_f=2$ flavours and several lattice spacings~\cite{Bertone:2012cu},
they found compatible results (within $2\sigma$ for $O_5$).
Then the SWME collaboration~\cite{Bae:2013tca} reported on their computation, using $N_f=2+1$ flavours 
of improved staggered and again several lattice spacings. They find an important
disagreement for two of these matrix elements.
The ETM collaboration has then repeated their computation with $N_f=2+1+1$ flavours~\cite{Carrasco:2015pra}
and found roughly the same results as in their previous study
(again only within $\sim 2 \sigma$ for $O_5$ and the new result is now in perfect agreement with our old result).

In~\cite{Hudspith:2015wev,Garron:2016mva}, we added another lattice spacing
and investigated the origin of the discrepancy.
In particular for the non-perturbative renormalisation procedure,
in addition to the traditional RI-MOM scheme, we have
implemented new intermediate renormalisation schemes,
called $\gmugmu$ and $\qq$
which satisfy the RI-SMOM condition, and therefore exhibit non-exceptional kinematics.
As summarised in Table~\ref{table:Rresults},
we find that the results depend significantly on the intermediate renormalisation
scheme:
\begin{itemize}
\item If we use the traditional RI-MOM scheme with exceptional kinematics,
  we reproduce our old result and are compatible with ETMc,
  who used the same RI-MOM scheme .
\item With the RI-SMOM schemes, our results for $O_4$ and $O_5$ are significantly different from our old
  RI-MOM results, but are consistent with each other.
\item 
  Our new RI-SMOM results are also in good agreement with SWME,
  who perform the renormalisation at one-loop in perturbation theory.
  This has been confirmed by the update of SWME~\cite{Jang:2015sla}.
  Therefore, one of our main conclusions in~\cite{Hudspith:2015wev,Garron:2016mva}
  is that the renormalisation  procedure is the source of the discrepancy
  and we suggest to discard the results obtained with exceptional kinematics
  due to the systematic uncertainty in the pion pole subtraction.
\end{itemize}

In Table~\ref{table:Rresults}, we choose to compare the results
for $R_i$~\cite{Babich:2006bh}
as they give directly the deviation of new physics with respect to the SM contribution.
Since we could not find these quantities in~\cite{Carrasco:2015pra,Bae:2013tca,Jang:2015sla},
we do not show the results from ETMc ($N_f=2+1+1$) and SWME.
However such a comparison for the bag parameters can be found in~\cite{Garron:2016mva}.

The main purpose of this work is the definition of RI-SMOM schemes for the BSM
operators, generalising what has been done for the Standard Model $B_K$ and for $K\to\pi\pi$
matrix elements~\cite{Arthur:2011cn,Boyle:2011cc,Boyle:2011kn,Arthur:2012opa,Garron:2012ex,Lytle:2013oqa,Blum:2012uk,Blum:2011ng,Blum:2015ywa}.
These RI-SMOM schemes use non-exceptional kinematics with a symmetric point
and have much better infrared behaviour, resulting in the suppression of pion pole contribution
and wrong-chirality operator mixing~\cite{Aoki:2007xm,Sturm:2009kb}.
We argue in this work that at this point, results obtained using
the RI-MOM scheme should be approached with skepticism or, if possible even discarded,
at least for these quantities (the renormalisation of BSM kaon mixing operators).
In addition, we define two new NPR schemes which both have different perturbative truncation systematics,
upon comparing the two we can cleanly estimate the systematic from the renormalisation procedure

The paper is organised as follows: in the next section we
explain our procedure to obtain the $Z$ factors.
In Section~\ref{s3:scheme} we give the explicit definitions of the projectors, which
complete the definition of the schemes.
The numerical results can be found in Section~\ref{s4:results}.
In Section~\ref{s5} we discuss the pole subtraction and the advantages of using
the RI-SMOM schemes.
Section~\ref{s6:conclusions} contains our conclusions.
Further details can be found in the appendices,
where we give the relevant Z-factors for the bag parameters,
the non-perturbative scale evolution of our renormalisation matrices,
its comparison with perturbation theory, and finally the Fierz relations
for the operators considered here.

\section{Methodology}\label{sec:methodology}

The Non-Perturbative-Renormalization (NPR) procedure works as follows:
we compute numerically the Landau-gauge-fixed Green's functions of the operators of interest 
between incoming and outgoing quarks in a given kinematic configuration. 
After amputation of the external legs, projection onto the Dirac-colour structure
and extrapolation to the chiral limit, 
we require that the renormalised Green's functions are equal to their tree-level values.
Since we renormalise a set of four-quark operators which can mix, 
this renormalisation condition defines a matrix of renormalisation factors.
We will discuss importance of the choice of kinematics; in particular the renormalisation 
condition is imposed for a certain momentum transfer $p$ which defines the 
renormalisation scale $\mu =\sqrt{p^2}$. 
For comparison we will also implement the original RI-MOM scheme~\cite{Martinelli:1994ty},
for which
results at a single lattice spacing were presented in \cite{Boyle:2012qb}, 
but we chose to discard them for our final result in \cite{Garron:2016mva} as we will argue herein they appear to 
suffer from large systematic errors.

In the Standard Model only one operator contributes to
neutral kaon mixing ($a$ and $b$ are colour indices)
\be
\label{eq:Q1}
Q_1 = \; (\overline s_a \gamma_\mu (1- \gamma_5) d_a)\, (\overline s_b\gamma_\mu (1- \gamma_5) d_b).\\
\ee
Beyond the Standard Model, under reasonable assumptions, four other four-quark operators are required
(seven if parity is not conserved).
Different choices of basis are possible but 
since we are concerned here with renormalisation,
we find it convenient to only consider color-unmixed operators,
i.e. those with the same colour structure as $Q_1$.
In Appendix~\ref{appendix:fierz} we give the relation between the colour-mixed
and colour-unmixed operators. In order to simplify the equations,
we do not explicitly write the colour indices, 
the contraction over spin and colour indices is simply indicated by the parentheses.
We define the BSM operators
(see for example~\cite{Ciuchini:1997bw}):
\begin{equation}
\label{eq:renbasis}
\begin{aligned}
Q_2 &= (\overline s \gamma_\mu(1-\gamma_5)  d)\, (\overline s \gamma_\mu(1+ \gamma_5) d),\\
Q_3 &= (\overline s(1-\gamma_5) d) \, (\overline s(1+\gamma_5) d),\\
Q_4 &= (\overline s(1-\gamma_5) d)\,  (\overline s(1-\gamma_5) d),\\
Q_5 &= \frac{1}{4} (\overline s (\sigma_{\mu \nu}(1-\gamma_5)) d)\, (\overline s (\sigma_{\mu \nu}(1-\gamma_5)) d)\,.
\end{aligned}
\end{equation}
where $\sigma_{\mu \nu} =\frac{1}{2} [\gamma_\mu,\gamma_\nu]$.
In practice we only consider the parity-even  part of these operators,
\begin{equation}\label{eq:renbasis_even}
\begin{aligned}
   Q_1^+ &= (\overline s\gamma_\mu d )\, (\overline s\gamma_\mu d)
   + (\overline s\gamma_\mu \gamma_5 d )\, (\overline s\gamma_\mu \gamma_5 d) \,,\\
   Q_2^+ &=
   (\overline s\gamma_\mu d )\, (\overline s\gamma_\mu d)
   - (\overline s\gamma_\mu \gamma_5 d )\, (\overline s\gamma_\mu \gamma_5 d) \,,\\
   Q_3^+ &= (\overline sd) \, (\overline sd) -  (\overline s\gamma_5 d)\,  (\overline s\gamma_5 d) \,,\\
 Q_4^+ &= (\overline sd) \, (\overline sd) +  (\overline s\gamma_5 d)\,  (\overline s\gamma_5 d),\\
 Q_5^+ &= \sum_{\nu>\mu} (\overline s \gamma_\mu \gamma_\nu  d)\, (\overline s \gamma_\mu \gamma_\nu d).
\end{aligned}
\ee
We will refer to Eq.~\ref{eq:renbasis_even} as the NPR basis (the relation between the SUSY and the NPR basis
can be found in Appendix~\ref{appendix:fierz}).
The factor $1/4$ in $Q_5$ of Eq.~(\ref{eq:renbasis}) ensures that our definition matches the ususal lattice convention:
\begin{equation}
\begin{aligned}
\sigma_{\mu\nu}(1-\gamma_5) \times \sigma_{\mu\nu}(1-\gamma_5)
&= 2 \sigma_{\mu\nu}\times \sigma_{\mu\nu} + \mbox{ parity odd terms} \\
&= 4 \sum_{\nu > \mu} \gamma_\mu \gamma_\nu  \times  \gamma_\mu \gamma_\nu  + \mbox{ parity odd terms}
\end{aligned}
\end{equation}
These four-quark operators
mix under renormalisation and - in a massless scheme - the mixing pattern is given by the chiral properties
of these operators. They belong to three different representations
of $SU_L(3) \times SU_R(3)$:
It is well-known that 
the Standard Model operator $Q_1$ transforms as $(27,1)$ and renormalises multiplicatively. 
Similarly, one can easily see that 
$Q_{2,3}$ transform like $(8,8)$ while $Q_{4,5}$ transform like $(6, \bar{6})$.

If chiral symmetry were perfectly maintained in the lattice theory, 
the mixing pattern would consist solely of a single $Z$-factor
for the $(27,1)$ 
while the $(8,8)$s and $(6, \bar{6})$s will
each mix among themselves producing a block-diagonal structure with a single element
in $Z_{11}$ and two blocks of $2 \! \times \! 2$ mixing matrices in $Z_{2/3,2/3}$ and $Z_{4/5,4/5}$. 
Since the Domain-Wall action exhibits a  continuum-like chiral-flavor symmetry (to a very good approximation), 
we expect to find a mixing pattern very close to the continuum one.
However the effects of spontaneous chiral symmetry breaking 
will be present at some level and could introduce some forbidden mixing 
(mixing between operators which belong to different
representations of $SU_L(3) \times SU_R(3)$).
These unwanted infrared contaminations
decrease as the renormalisation scale is increased beyond the typical interaction scale of QCD ($\Lambda_\text{QCD}$). 

Such unphysical mixings are strongly suppressed in SMOM schemes (compared to the RI-MOM scheme),
where the choice of kinematic prevents the contribution
of exceptional momentum configurations~\cite{Aoki:2007xm}.
In practice, we will take the degree to which the expected continuum mixing pattern is satisfied as a quantitative 
indicator of the degree to which the NPR condition Eq~(\ref{npr_window}) is satisfied.

\subsection{Choice of kinematic and vertex functions}

The choice of kinematic for the RI-SMOM schemes is illustrated in \fig{fig:SMOMkin}. There are two different momenta $p_1$ and $p_2$ such 
that the momentum transfer is $p^2 = (p_2 - p_1)^2$.
In this way a single renormalisation scale $\mu=\sqrt{p^2}$ is maintained and momentum flows 
through the vertex, which suppresses unwanted non-perturbative behaviour compared to the 
original RI-MOM scheme~\footnote{In the orginial RI-MOM scheme there is no momentum transfer $p_1=p_2$.}.
In practice we need two (momentum source) propagators, we associate a momentum to a given flavour,
here $p_1$ for the d-quark and $p_2$ for the s-quark.
The momenta used are of the form
(the Euclidian-time component is the last coordinate)
\be
p_1 = \frac{2\pi}{L}  (-m, 0, m, 0)\; \mbox{ and } \;
p_2 = \frac{2\pi}{L}  ( 0, m, m, 0) \;,
\ee
so that $p=p_2-p_1 =  \frac{2\pi}{L} (m,m,0,0)$ .
Since we use twisted boundary conditions in the valence sectors
the momenta are not restricted to the Fourier modes. 
Our conventions are such that $m=n+\pi/2$, with $n\in \mathbb Z$ and
$\theta \in \mathbb R$.
\be
  \label{eq:mom}
p = \frac{2\pi}{L}(-n, 0, n, 0) + \frac{\pi}{L} (-\theta, 0, \theta, 0) \,,
\ee
where $\theta$ is the twist angle of the boundary condition and n is an integer Fourier mode.

Our choice of convention is the following:
with respect to the position of the vertex $x$,
\begin{enumerate}
\item
  An incoming $s$ quark with momentum $p_2$ is denoted by 
  \begin{equation}
    G_x(p_2) = \sum_y G^{(s)} (x,y) {\rm e}^{ i p_2.(y-x)}
  \end{equation}
\item An outgoing $d$ quark with momentum $-p_1$ is denoted by
  \begin{equation}
    \bar G_x(p_1)  = \gamma_5 G(x,p_1)^\dagger \gamma_5  = \sum_y  {\rm e}^{ -i p_1.(y-x)} G^{(d)} (y,x)
  \end{equation}
\end{enumerate}	
For each operator $Q_i$ of Eq.\ref{eq:renbasis} we compute the following Green's function
(where we define $\tilde{x}_i = x_i - x$)
\begin{equation}
\label{eq:Contraction}
\begin{aligned}
M_i^{\delta \gamma ;\beta \alpha} (q^2)
=&
\sum_{x,x_1,\ldots,x_4}
\la 0 | s^\delta (x_4) \bar d^\gamma (x_3) 
\left[  Q_i(x)  \right]
s^\beta (x_2) \bar d^\alpha (x_1)| 0 \ra
{\rm e}^{ -i p_1.\tilde x_1 +i p_2.\tilde x_2 -ip_1.\tilde x_3 +ip_2.\tilde x_4},
\\
=&
2\sum_x\left(\la
\left[ \bar G_x(p_2) \Gamma^1_{(i)} G_x(p_1) \right]^{\delta \gamma}
\left[ \bar G_x(p_2) \Gamma^2_{(i)} G_x(p_1) \right]^{\beta  \alpha}
\ra \right. \\
&\qquad - \left.
 \la 
 \left[ \bar G_x(p_2)  \Gamma^1_{(i)} G_x(p_1) \right]^{\delta \alpha}
  \left[ \bar G_x(p_2) \Gamma^2_{(i)} G_x(p_1) \right]^{\beta  \gamma}
  \ra
  \right),
\end{aligned}
\end{equation}
where the Greek letters denote combined spin-color indices.
The color-Dirac structure of the four-quark operator $Q_i$ is encoded
in $\Gamma^{1,2}_{(i)}$, (there is no summation over $i$ in Eq.~\ref{eq:Contraction}).
For the numerical implementation, we have only considered four-quark operators that are color unmixed
(the color partners can be obtained by Fierz transformation, see Appendix~\ref{appendix:fierz}).
For example, 
if $i,j,k,l$ are Dirac indices
and $a,b,c,d$ are color indices, then
for the operator $Q_2$, we have
\be
(\Gamma^1)^{ab}_{ij} \times (\Gamma^2)^{cd}_{kl} =
\left[ \gamma_\mu(1-\gamma_5)\right]_{ij} \delta^{ab}
\times
\left[ \gamma_\mu(1+\gamma_5) \right]_{kl} \delta^{cd} \;.
\ee
The vertex functions are then amputated
\be
\Pi_i^{\bar \delta \bar \gamma;  \bar \beta \bar \alpha}
=
\la  \bar G(p_2)^{-1} \ra^{\bar \delta \delta } \;
\la G(p_1)^{-1} \ra ^{\gamma \bar \gamma} \;
\la  \bar G(p_2)^{-1} \ra^{\bar \beta \beta } \;
\la G(p_1)^{-1} \ra ^{\alpha \bar \alpha} \;
M_i^{\delta \gamma ;\beta \alpha} (q^2),
\ee
where we have introduced the inverse of the ``full momentum propagators''
\be
G(p) = \sum_x G_x(p) \quad \text{and} \quad \bar{G}(p) = \sum_x \bar{G}_x(p) \;.
\ee
We still have to project these amputated vertex functions in order to obtain
the renormalisation matrix. This is described in the next section.

\begin{figure}[!t]
\begin{center}
\includegraphics[width=5cm]{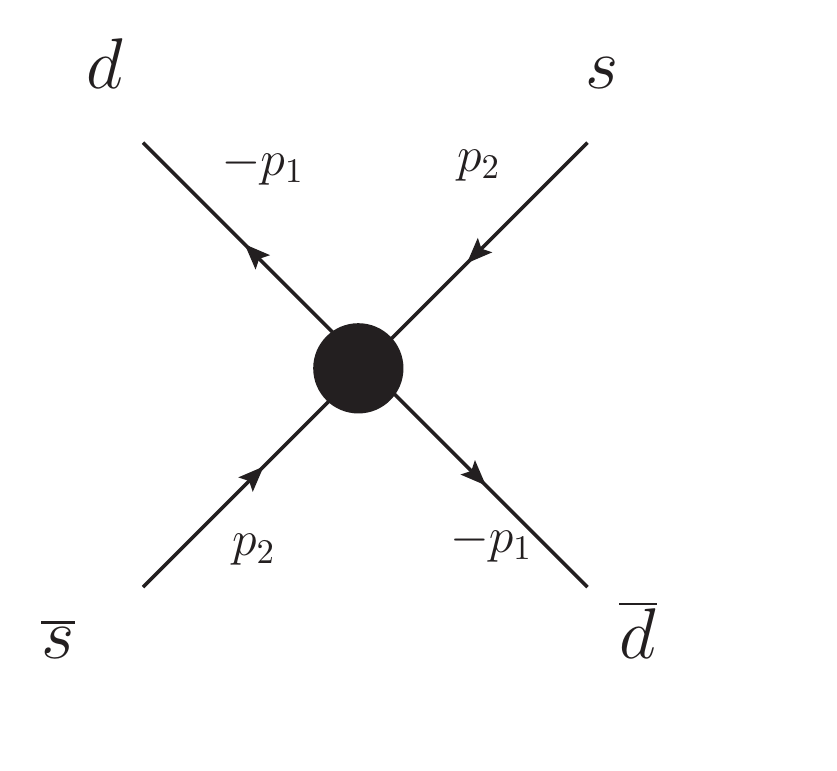}
\caption[]{Illustration of the choice of kinematics in an RI-SMOM scheme
  for a four quark operator contributing to the process $(\bar s d) \to (\bar d s)$.
  We choose the momenta $p_1 \ne p_2$ such that $p_1^2=p_2^2=(p_1-p_2)^2$.
  This configuration prevents the existence of a channel with 
  zero-momentum transfer.
}
\label{fig:SMOMkin}
\end{center}
\end{figure}

\subsection{Projection}

Following Eq.~\ref{eq:ZRI}, we introduce the renormalisation matrix $Z$ which relates the renormalised 
four-quark operators to the bare ones
(we drop the superscript RI for the $Z$ factors)
\begin{equation}
  \langle Q_i\rangle^\text{RI} (\mu,a)  = {Z_{ij}(\mu,a )} \langle Q_j\rangle^\text{bare} (a),
  \label{eq:renormalisation}
\end{equation}
Denoting by $\Pi^\text{bare}_i$ the bare amputated Green's function of the four quark operator $Q_i$,
the matrix $Z_{ij}$ is defined by imposing
the renormalisation condition~\footnote{In order to define a massless renormalisation scheme, 
the renormalisation condition is actually imposed in the chiral limit. 
However the quark mass dependence of the vertex functions $\Pi$s is very mild
and the chiral extrapolations are perfectly under control. In order to simplify 
the discussion, we omit any reference to the finite quark mass effects in this section}:
\begin{equation}
\label{eq:Z_cond}
  P_k\left[ \frac{Z_{ij}(\mu,a )}{Z_q^2(\mu,a)} \Pi_j^\text{bare}(a,p^2)\right]_{p^2=\mu^2} = F_{ik} \;,
\end{equation}
where $\sqrt{Z_q}$ is the quark wave function renormalisation. 
In the previous equation, $P_i$ projects onto the tree-level spin-colour structure of $Q_i$:
\be
P_k\left[\Pi_i^{(0)}\right] = F_{ik} \;,
\ee
where the superscript $(0)$ denotes the tree-level value.
The fact that there is a non-vanishing momentum transfer in the vertex gives
us more freedom for the choice of projectors.
In this work, we introduce two different sets of projectors:
$P^{(\gamma^\mu)}$ and $P^{(\s{q})}$, they are defined explicitly below.
We also need a prescription for the quark wave function $Z_q$. 
This is done in two steps: first we cancel the factors of $Z_q$ in \eqref{eq:Z_cond} 
using the vertex function of the local vector current. The value of $Z_V$ is then determined from 
some Ward identity in~\cite{Blum:2014tka}.
We implement two projectors $P_V^{(\gamma_\mu)}$ and $P_V^{(\s{q})}$ to obtain $Z_V/Z_q$.
The choices of projectors for the four-quark operators and for the vector 
current define the non-perturbative scheme. 
Denoting by ${\cal A}$ and ${\cal B}$ the choices of projectors, ie  ${(\gamma_\mu)}$ or $(\s{q})$,
for both the four-quark operators and the vector current, 
the NPR condition for the scheme $({\cal A},{\cal B})$ reads
\begin{equation}
\label{eq:Z_condscheme}
\frac{ Z_{ij}^{({\cal A},{\cal B})}(\mu,a ) } { Z_V(a)^2}
\times
\left.
\frac{ P_k^{\cal A}\left[ \Pi_j^{bare}(a,p^2)\right] }
{( P_V^{\cal B}\left[ \Pi_V^{bare}(a,p^2)\right] )^2}
\right|_{p^2=\mu^2} = \frac {F_{ik}^{\cal A}}{F_V^{\cal B}} \;.
\end{equation}
The matrix $Z^{({\cal A},{\cal B})}$ converts the bare four-quark operators 
onto the renormalised four-quark operators in the RI-SMOM scheme ${(\cal A,\cal B})$. \\

In \cite{Garron:2016mva} the primary quantities we presented were the ratios
of particular BSM matrix elements over the SM one~\footnote{
  In~\cite{Garron:2016mva} the results are given in the SUSY basis, here we worked
  in the NPR basis of Eq.~\ref{eq:renbasis_even}.
  In particular $O_4$ and $O_5$ in the SUSY basis are now related
  to $Q_2$ and $Q_3$ in the NPR basis. The change of basis is given explicitly
  in Appendix~\ref{appendix:fierz}
  }
\be
R_i = \frac{\la Q_i \ra}{\la Q_1 \ra }.
\ee
So we now consider the $Z$ factors needed for these ratios.
Introducing some notation for the projected vertex functions
\begin{equation}\label{eq:def_Lambda}
\Lambda^{\cal A}_{ij} = P_j^{\cal A}\left[ \Pi_i^{bare}\right]\;,\quad
{\cal Z}_{ij}^{({\cal A},{\cal B})} = \frac{Z_{ij}^{({\cal A},{\cal B})}}{Z_{11}^{({\cal A},{\cal B})}} \;.
\end{equation}
From Eq.~(\ref{eq:Z_condscheme}), neglecting the mixing of the $(27,1)$ with the
other operators, 
one finds that the quantity
\be
{\cal Z}^{({\cal A},{\cal B})}(\mu,a ) 
=
\frac{ \Lambda_{11}^{\cal A} (\mu,a ) } {F_{11}^{\cal A}} \,
\times
F^{\cal A} \times
{\left(\Lambda^{\cal A}(\mu,a ) \right)^{-1}},
\ee
is independent of ${\cal B}$, which is the choice of the projector for the
denominator of Eq.(\ref{eq:Z_condscheme}).
Therefore, although in principle we have defined
four RI-SMOM schemes $(\gamma_\mu,\gamma_\mu),(\gamma_\mu,\s{q}),(\s{q},\gamma_\mu),(\s{q},\s{q})$,
in this work we mainly consider the ``diagonal'' schemes, for which ${\cal A} = {\cal B}$,
namely $(\gamma_\mu,\gamma_\mu)$ and $(\s{q},\s{q})$.

\section{Non-exceptional schemes}
\label{s3:scheme}
\subsection{Choice of projectors}

For the quark wave function renormalisation, we make use of two different definitions of $Z_q$.
The factors $Z_q/Z_V$ are determined by imposing the condition
\be
\frac{Z_V}{Z_q} P_V\left[\Pi_V\right]  = F_V \;.
\ee
The two projectors we use are $ P_V^{(\gamma^\mu)}$ and  $P_V^{(\s{q})}$,
they are defined explicitly by:
\begin{equation}
\label{eq:Zq_gamma}
\begin{aligned}
\frac{Z_q^{(\gamma^\mu)}}{Z_V} &=  \frac{1}{F_V^{(\gamma^\mu)}} P_V^{(\gamma^\mu)}\left[\Pi_V\right] 
=  \frac{1}{48} \Tr \[\g^{\mu} \Pi_V^\mu\]\,, \\
\frac{Z_q^{(\s{q})}}{Z_V} &= \frac{1}{F_V^{(\s{q})}} P_V^{(\s{q})} \left[\Pi_V \right] 
= \frac{q^\mu}{12q^2} \Tr \[\s{q} \Pi_V^\mu \] \,,
\end{aligned}
\end{equation}
where $\Pi_V$ is the amputated Green's functions of the vector and axial-vector current.  \\

The basis of the four-quark operators is given in~Eq.~(\ref{eq:renbasis}), our convention is such that
that all the operators are ``colour-unmixed''.
The definition of the $\gamma_\mu$-projectors is straighforward:
they are defined with the same spin-colour structure as their respective operators.
Explicitly, for the SM operator we have 
\be
  \left[ \P_1^{(\gamma^\mu)}\right]_{\beta\alpha ; \delta \gamma}^{ba;dc}
=
 \left[ (\gamma^\mu)_{\beta\alpha} (\gamma^\mu)_{\delta\gamma} 
        + (\gamma^\mu \gamma^5)_{\beta\alpha}(\gamma^\mu\gamma^5)_{\delta\gamma} \right] \delta^{ba}\delta^{dc}
\;.
\label{P_gamma1}
\ee
For the $\s{q}$ schemes, following~\cite{Aoki:2010pe}, we replace the $\gamma_\mu$ 
matrices by $\s{q}/\sqrt{q^2}$, for example 
\be
\left[ \P_1^{(\s{q})}\right]_{\beta\alpha ; \delta \gamma}^{ba;dc}
    = \frac{1}{q^2}
      \left[ (\s{q})_{\beta\alpha} (\s{q})_{\delta\gamma} 
        + (\s{q} \gamma^5)_{\beta\alpha}(\s{q}\gamma^5)_{\delta\gamma} \right] \delta^{ba}\delta^{dc}
\;.
\ee
Similarly for the $(8,8)$ doublet we have
\begin{equation}
\begin{aligned}
\left[ \P_2^{(\gamma^\mu)}\right]_{\beta\alpha ; \delta \gamma}^{ba;dc} 
&=
\left[ (\gamma^\mu)_{\beta\alpha} (\gamma^\mu)_{\delta\gamma} 
  - (\gamma^\mu \gamma^5)_{\beta\alpha}(\gamma^\mu\gamma^5)_{\delta\gamma} \right]
\delta^{ba}\delta^{dc}\,,
\\
\left[ \P_3^{(\gamma^\mu)}\right]_{\beta\alpha ; \delta \gamma}^{ba;dc}
&=
\left[  
\delta_{\beta\alpha} \delta_{\delta\gamma} - (\gamma^5)_{\beta\alpha}(\gamma^5)_{\delta \gamma}  \right]
\delta^{ba}\delta^{dc}\,.
\end{aligned}
\end{equation}
For the $\s{q}$ projectors, in the case of $P_2$, we apply the same recipe as
the previous operator. For $P_3$, we take advantage of the  Fierz arrangements
to ``trade'' the $S$ and $P$ Dirac matrices for the vector and axial ones.
Explicitly we define
\begin{equation}
\begin{aligned}
\left[ \P_2^{(\s{q})}\right]_{\beta\alpha ; \delta \gamma}^{ba;dc} 
&=  \frac{1}{q^2}
\left[ (\s{q})_{\beta\alpha} (\s{q})_{\delta\gamma} - 
  (\s{q} \gamma^5)_{\beta\delta}(\s{q}\gamma^5)_{\delta \gamma} \right] \delta^{ba}\delta^{dc}\,,
\\
\left[ \P_3^{(\s{q})}\right]_{\beta\alpha ; \delta \gamma}^{ba;dc} 
&= \frac{1}{q^2}
\left[ (\s{q})_{\beta\alpha} (\s{q})_{\delta\gamma} - 
  (\s{q} \gamma^5)_{\beta\delta}(\s{q}\gamma^5)_{\delta \gamma} \right]\delta^{bc}\delta^{da}\;.
\end{aligned}	
\end{equation}
Where the latter is now ``colour-mixed''
(this set of projector has already been introduced in ~\cite{Blum:2011ng,Blum:2012uk} 
in the context of $K\to\pi\pi$ decays).

Finally for the $(6,\bar 6)$ operators we define
\begin{equation}
\begin{aligned}
\left[ \P_4^{(\gamma^\mu)}\right]_{\beta\alpha ; \delta \gamma}^{ba;dc} 
&=
\left[ \delta_{\beta\alpha} \delta_{\delta\gamma} + 
  (\gamma^5)_{\beta\alpha}(\gamma^5)_{\delta \gamma} \right] \delta^{ba}\delta^{dc}\,,
\\
\left[ \P_5^{(\gamma^\mu)}\right]_{\beta\alpha ; \delta \gamma}^{ba;dc} 
&=
\left[ \frac{1}{2} (\sigma^{\mu \nu})_{\beta\alpha} (\sigma^{\mu \nu})_{\delta \gamma} \right] \delta^{ba}\delta^{dc} \;,
\end{aligned}
\end{equation}
and
\begin{equation}
\begin{aligned}
\left[ \P_4^{(\s{q})}\right]_{\beta\alpha ; \delta \gamma}^{ba;dc} 
&=  \frac{1}{p_1^2 p_2^2 - (p_1.p_2)^2}
\left[ 
  \left(p_1^\mu (\sigma^{\mu\nu} P_L )  p_2^{\nu} \right)_{\beta \alpha}
  \left(p_1^\rho (\sigma^{\rho \sigma} P_L) p_2^\sigma \right)_{\delta \gamma}
  \right] \delta^{bc}\delta^{da} \;,
\\
\left[ \P_5^{(\s{q})}\right]_{\beta\alpha ; \delta \gamma}^{ba;dc} \,
&=  \frac{1}{p_1^2 p_2^2 - (p_1.p_2)^2}
\left[ 
  \left(p_1^\mu (\sigma^{\mu\nu} P_L )  p_2^{\nu} \right)_{\beta \alpha}
  \left(p_1^\rho (\sigma^{\rho \sigma} P_L) p_2^\sigma \right)_{\delta \gamma}
\right] \delta^{ba}\delta^{dc}\;,
\end{aligned}
\end{equation}
where $P_{R,L} = \frac{1}{2}(1\pm\gamma_5)$.
Imposing Eq.~\ref{eq:Z_cond} with the projectors given above 
defines the various schemes $(\cal A, \cal B)$ where ${\cal A}$ and ${\cal B}$ 
are either $\gamma_\mu$ or $\s{q}$

\subsection{Tree-level values}

For SM operator the tree-level vertex function reads:
\begin{equation}
\begin{aligned}
\left[\Pi_1^{(0)}\right]_{\alpha \beta ; \gamma \delta}^{ab;cd} \;
= & \;
2\left[(\gamma^\mu)_{\alpha\beta} (\gamma^\mu)_{\gamma\delta} + (\gamma^\mu\gamma_5)_{\alpha\beta} (\gamma^\mu\gamma_5)_{\gamma\delta} \right]
\delta^{ab}\delta^{cd}\\
&\;-
2\left[(\gamma^\mu)_{\alpha\delta} (\gamma^\mu)_{\gamma\beta} + (\gamma^\mu\gamma_5)_{\alpha\delta} (\gamma^\mu\gamma_5)_{\gamma\beta} \right]
\delta^{ad}\delta^{cb} \;, 
\end{aligned}
\end{equation}
and equivalently for the other Dirac structures.
The projectors act on the vertex functions by simply tracing over the Dirac and colour indices,
explicitly the tree-level version of Eq.~\ref{eq:Z_cond} is
\begin{equation}
P_j\left[\Pi_i^{(0)}\right] =
\left[ P_j \right]_{\beta \alpha ; \delta \gamma}^{ba;dc} 
\left[\Pi_i^{(0)}\right]_{\alpha \beta ; \gamma \delta}^{ab;cd} = F_{ij} \,.
\end{equation} 
The corresponding tree-level matrices ($N=3$ is the number of colours) are
\begin{equation}
F^{(\gamma^\mu)}=\left(
\begin{array}{ccccc}
256N(N+1)   &  0        &    0    &  0   &   0    \\
0           &  256N^2   &   -128N  &  0   &   0    \\
0           &  -128N     &   64N^2 &  0   &   0    \\
0           &  0   &  0  & 32N(2N-1)   &   96N     \\
0           &  0   &  0  &  96N       & 96N(2N+1)
\end{array}
\right)\;,
\end{equation}
and
\begin{equation}
  F^{(\s{q})}=\left(
  \begin{array}{ccccc}
    64N(N+1) &  0   &    0   &0 &0 \\
    0           &  64N^2  &   64N & 0 & 0  \\
    0           &  -32N   &   -32N^2 & 0 & 0 \\
    0 & 0 & 0 & 8N^2     & 8N       \\
    0 & 0 & 0 & 8N(N+2)  & 8N(2N+1)
  \end{array}
  \right)\;.
\end{equation}

\section{Numerical results}
\label{s4:results}
\subsection{Non-perturbative Z factors}
\begin{table}[t]
  \centering
      {
\begin{tabular}{  c | c | c | c| c | c  }
\toprule
    Volume & $a^{-1}$ [GeV] & $am^{\text{sea}}_{ud} \, (= am^{\text{val}}_{ud})$ &
    $m_\pi$ [MeV] & $am^{\text{sea}}_{s}$ &
    $am^{\text{phys}}_{s}$\\
    \hline
    $24^3 \times 64 \times 16$ & 1.785(5)  
    & 0.005, 0.01, 0.02 &340, 430, (560)& 0.04
    & 0.03224(18) \\ 
    $32^3 \times 64 \times 16$ & 2.383(9) 
    & 0.004, 0.006, 0.008 &300, 360, 410 &0.03
    & 0.02477(18) \\ 
    \botrule
\end{tabular}
}
\caption{
  Summary of the lattice ensemble used in this work.
  Since the renormalisation is performed with momentum sources, only a few configurations
  are needed (between ten and  twenty for each ensemble).
  \label{tab:lattparam}}
\end{table}

\begin{table}
  \begin{tabular}{c| ccccccccccc}
    \toprule
    $\theta$  & -0.1875 & 0      &   0.1875  &  0.3750 &   0.5625 &   0.7500 &   0.9375 &   1.1250 & 1.3125 &   1.5000  \\
    $(ap)^2$  &  1.1578 & 1.2337 &   1.3120  &  1.3927 &   1.4759 &   1.5614 &   1.6494 &   1.7397 & 1.8325 &   1.9277  \\
    $p [GeV]$ &  1.92   & 1.98   &   2.04   &   2.11   &   2.17   &   2.23   &   2.29   &   2.35   & 2.42   &   2.48    \\
    \hline
    $\theta$  &  1.6875 & 1.8750 &   2.0625 &   2.2500  &  2.4375 &  2.6250  &  2.8125  &  3.0000  &  3.1875  &  3.3750 &  \\
    $(ap)^2$  &  2.0252 & 2.1252 &   2.2276 &   2.3325  &  2.4397 &  2.5493  &  2.6614  &  2.7758  &  2.8927  &  3.0120 &   \\
    $p [GeV]$ &  2.54   & 2.60   &   2.66   &   2.73    &  2.79   &  2.85    &  2.91    &  2.97    &  3.04    &  3.10   &   \\
    \botrule
  \end{tabular}
  \caption{List of momenta for the $24^3$ lattices. Here we fix the Fourier mode to $n=3$ and only change
  the twist angle $\theta$, see Eq.\ref{eq:mom}.  }
\label{Table:listmom24}
\end{table}

\begin{table}
  \begin{tabular}{c| cccccccccc}
    \toprule
    $n$       &  3      & 3      & 3      & 3      & 3      & 3      &  3     & 3      & 3      & 3      \\
    $\theta$  & -0.1    & 0      & 0.2    & 0.4    & 0.6    & 0.8    & 1.   0 & 1.2    & 1.4    & 1.6    \\
    $(ap)^2$  &  0.6710 & 0.6940 & 0.7410 & 0.7896 & 0.8397 & 0.8913 & 0.9446 & 0.9993 & 1.0556 & 1.1134 \\
    $p [GeV]$ &  1.95   & 1.99   & 2.05   & 2.12   & 2.18   & 2.25 8 & 2.32   & 2.38   & 2.45   & 2.51   \\
    \hline
    $n$       &  3      & 4      & 4      & 4      &  4     & 4      & 4      & 4      & 4      & 4      \\
    $\theta$  &  1.8    & 0      & 0.2    & 0.4    & 0.6    & 0.8    & 0.9    & 1.   0 & 1.1    & 1.2    \\
    $(ap)^2$  &  1.1728 & 1.2337 & 1.2962 & 1.3602 & 1.4257 & 1.4928 & 1.5269 & 1.5614 & 1.5963 & 1.6316 \\
    $p [GeV]$ &  2.58   & 2.65   & 2.71   & 2.78   & 2.84   & 2.92   & 2.94   & 2.98   & 3.01   & 3.04   \\
    \botrule
  \end{tabular}
  \caption{List of momenta for the $32^3$ lattices.}
\label{Table:listmom32}
\end{table}

The renormalisation is performed on the same ensembles as in~\cite{Garron:2016mva},
the parameters are summarised in Table~\ref{tab:lattparam}.
We implement numerically Eq.~\ref{eq:Z_condscheme} and obtain the $\Lambda$ matrices
(as defined in Eq~\ref{eq:def_Lambda})
at finite quark mass for the the list of momenta listed in Tables~\ref{Table:listmom24} and~\ref{Table:listmom32}.
The parameters for these ensembles are summarised in Table~\ref{tab:lattparam}.
We perform a chiral extrapolation, invert the result and then interpolate to the desired scale of $3$ GeV. 
Strictly speaking, there is mismatch from $m^{\text{sea}}_{s}\ne m_{ud}^{\text{sea}}$,
however the quark mass dependence is dominated by the valence sector, the sea contribution
plays very little r\^ole here. 
Furthermore, for the $\rismom$ schemes the light quark mass-dependence is very mild,
practically invisible at our renormalisation scale even within our high statistical resolution,
and so we consider any associated systematic to be negligible.

Due to the use of partially twisted boundary conditions, we can simulate momenta arbitrarily close
to the targeted point, hence only a very small, well controlled, interpolation (performed with a quadratic Ansatz) is 
required. We illustrate these points in Fig.~\ref{fig:Zexample}. 
The numerical results for the $Z$ factors at $3\,\GeV$ are given in 
tables~\ref{tab:ZBK_24},\ref{tab:Z88_24}, \ref{tab:Z66_24}, \ref{tab:ZBK_32},\ref{tab:Z88_32} and \ref{tab:Z66_32}.

\begin{figure}[!t]
\begin{tabular}{cc}
\includegraphics[width=8cm]{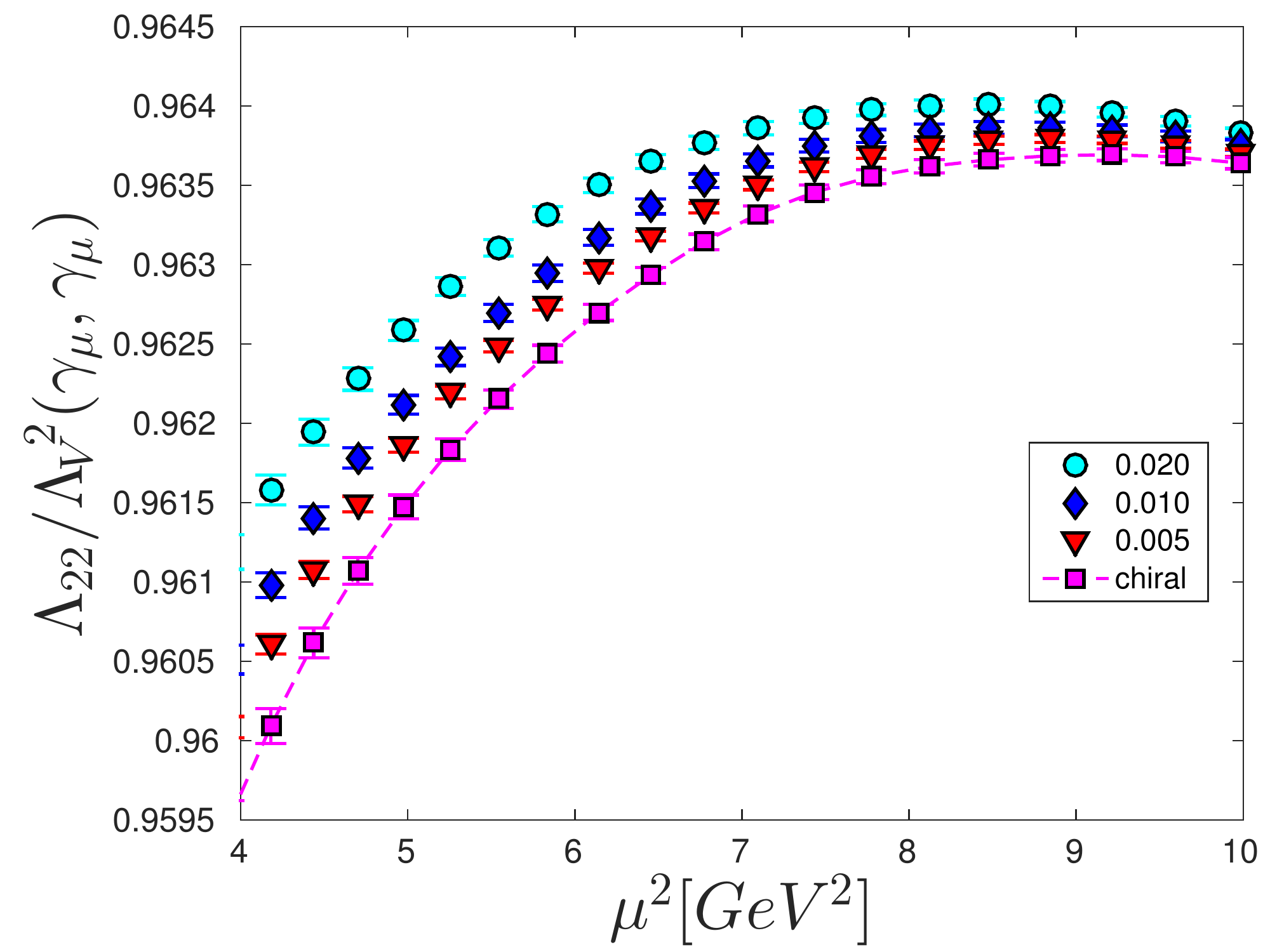} &
\includegraphics[width=8cm]{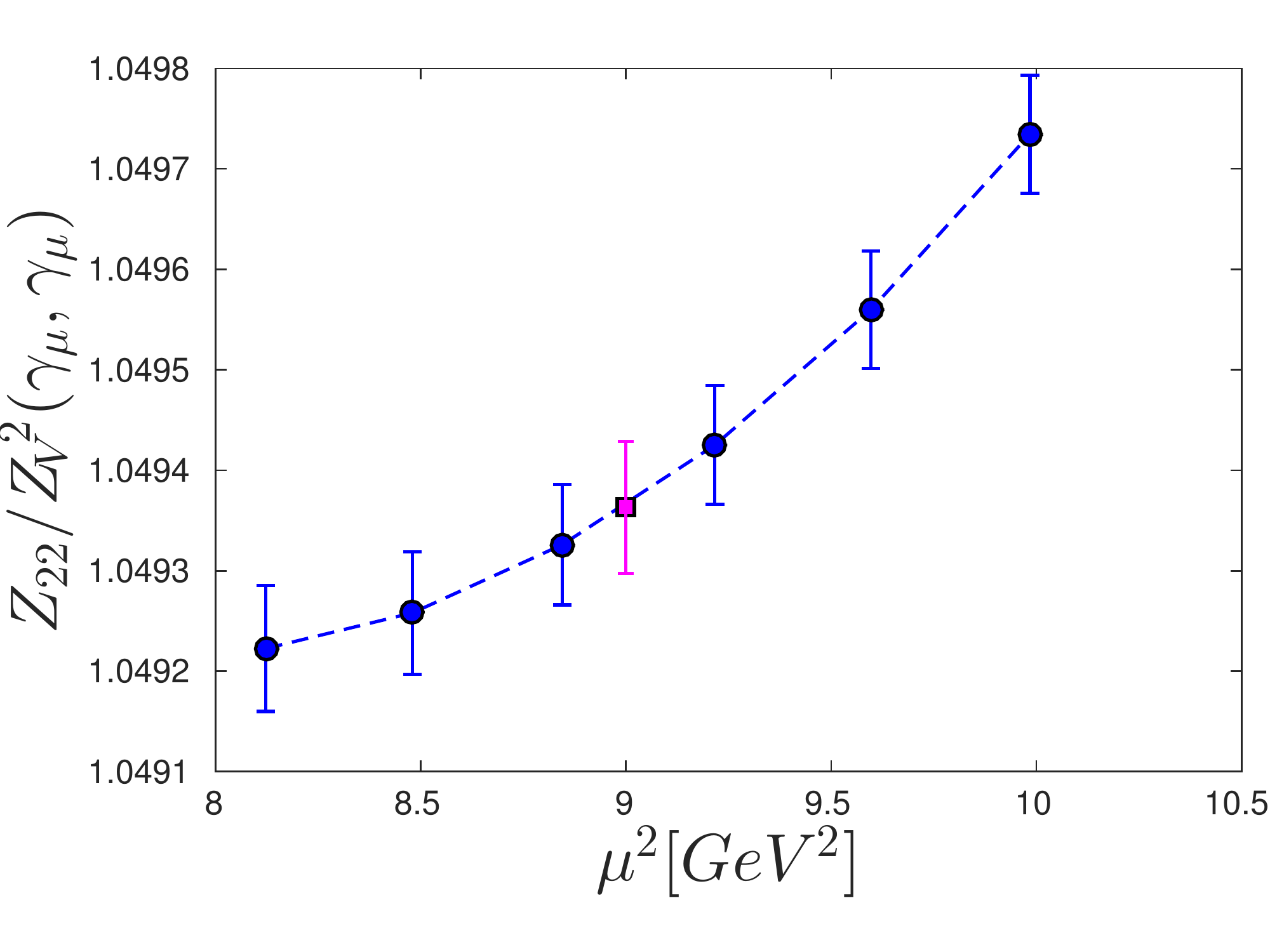}
\end{tabular}
\caption[]{Example of amputated and projected vertex functions
  at the simulated momenta and quark masses (left)
and interpolation of a $Z$ matrix element to the $3\,\GeV$-scale after chiral extrapolation (right).
Results are shown for the $\rismomgg$ scheme on the $24^3$ lattice.}
\label{fig:Zexample}
\end{figure}

In principle we only need momenta close to the scale we wish to present our final results at
(here $\mu=3\, \GeV$), 
however it is useful to compute the $Z$ factors for a larger range,
say between $2$ and $3$ GeV. We can then compare the non-perturbative 
scale evolution to its perturbative approximation and estimate 
the effects of truncating the perturbative series for the various schemes. 
Furthermore, since the running has a continuum limit, we also obtain a nice handle 
on the discretisation effects. 

\subsection{Conversion to \texorpdfstring{$\msbar$}{ms-bar} }

It is commonplace to convert the renormalised matrix elements computed 
on the lattice to the $\msbar$ scheme. 
In that way, the Wilson coefficients
can be combined with the matrix elements  to produce phenomenological predictions. 
The conversion from the $\rimom$ or $\rismom$ to $\msbar$ is done in continuum 
perturbation theory. The matching coefficients are known at the one-loop level
for $\rimom$ from~\cite{Ciuchini:1993vr} and \cite{Buras:2000if}.
The situation is different for the $\rismom$ schemes:
the relevant conversion factors of the $(27,1)$ operator have been computed in~\cite{Aoki:2010pe}. 
The conversion matrix for the $(8,8)$ operators can be extracted from~\cite{Lehner:2011fz} 
where the conversion was computed for the $\Delta S=1$ $K\to \pi\pi$ four-quark operators.
For the $(6,\bar 6)$ operators, the coefficients were unknown
and have been computed for this work. 
The full expression can be found in Appendix~\ref{appendix:msbarmatch}.

To obtain $\alpha_s$ at $\mu=3$ GeV in the three-flavour theory, we
start from $\alpha_s(M_Z) = 0.1185(6)$, we use the four-loop
running given in~\cite{vanRitbergen:1997va,Chetyrkin:1997sg} to compute the scale evolution down to
the appropriate charm scale, while changing the number of flavours
when crossing a threshold, and then run back up to 3 GeV in the three-flavour theory.

The values of the one-loop conversion matrices and the $Z$ factors in $\msbar$ 
(ie the $Z$ factors which convert our bare matrix elements to $\msbar$)
are given in
tables~\ref{tab:ZBK_24},\ref{tab:Z88_24}, \ref{tab:Z66_24}, \ref{tab:ZBK_32},\ref{tab:Z88_32} and \ref{tab:Z66_32}.
For completeness, we also give the conversion factor for the original $\rimom$ scheme
(the equivalent of the second columns of the above-listed tables)  
\begin{equation}
R^{\msbar \leftarrow \text{RI-MOM} }(3\;\text{GeV}) = 
\begin{pmatrix}
1.01711 & 0       & 0         & 0       & 0        \\
0       & 0.97795 & -0.13228  & 0       & 0        \\
0       & 0.00599 & \m1.21233 & 0       & 0        \\
0       & 0       & 0         & 1.11023 & 0.016719 \\
0       & 0       & 0         & 0.06318 & 1.052524
\end{pmatrix} \,.
\end{equation}

The conversion to $\msbar$ is then given by 
$
Z^{\msbar} = R^{\msbar \leftarrow ({\rm scheme})} 
\times Z^{(\rm scheme)}
$
where $({\rm scheme})$ can be $\rimom$, 
$(\gamma_\mu, \gamma_\mu)$, $(\s{q},\gamma_\mu)$,$(\gamma_\mu,\s{q})$ or $(\s{q},\s{q})$.

We observe that in general, the ``diagonal'' schemes $(\gamma_\mu, \gamma_\mu)$ and
$(\s{q},\s{q})$ have a better perturbative convergence than the off-diagonal ones.
At $3$ GeV, the conversion matrices are rather close to the identity
(which probably explains why our results agree so well with SWME).
For our two favorite schemes, we find that after conversion to $\msbar$,
the numbers agree rather well.
The convergence of the perturbative series
and the effects of the lattice artefacts could also be estimated
by looking at the step-scaling matrices, which we do in the
next section (see also Appendix~\ref{app:ssfplot}).

\subsection{Non-perturbative scale evolution and comparison with perturbation theory}

The scale evolution matrix, $\sigma(\mu_1,\mu_2)$ is a rich source of information,
in particular it helps us to estimate the systematic errors affecting the renormalisation procedure.
We define
\be
\label{eq:ssf}
\sigma(\mu_1,\mu_2,a) = Z(\mu_1,a) Z^{-1}(\mu_2,a)\;,
\ee
where $Z$ is the $5\times 5$ matrix defined in Eq.~\ref{eq:Z_condscheme}.
(Although in practice we take the chiral limit of the right hand side of Eq.~\ref{eq:ssf},
once again in order to simplify the notation, we discard any reference 
to the quark masses.)

The scale evolution matrix has a universal continuum limit and may
be directly compared to continuum perturbation theory.
The continuum extrapolation
\be
\sigma(\mu_1,\mu_2) = \lim_{a^2\to 0} \sigma(\mu_1,\mu_2,a) \;.
\ee
is performed assuming a linear behaviour in $a^2$.
For this step the use of twisted boundary conditions is essential,
since it allows us to vary $\mu$ continuously holding the momentum
orientation (and $\mathcal{O}(a^2)$ coefficients) fixed.

The continuum extrapolation of $\sigma_{ii}(2\,\GeV,\mu)$, where $2\, \GeV \le \mu \le 3\,\GeV$, is shown in
Figs.~\ref{fig:ssf_11_2to3}-\ref{fig:ssf_44_2to3}, compared with continuum perturbation theory.
We find in general good agreement with the perturbative series,
indicating that the $a^2$ extrapolation is valid and discretisation
effects are under control.
An example of off-diagonal matrix elements can be found in Fig~\ref{fig:ssf_32_2to3}.

By comparing the non-perturbative running to its perturbative approximation,
we can estimate the quality of the perturbative series for the various schemes.
This is important in view of the perturbative macthing of the NPR factors to $\msbar$.
In order to compare the scale evolution matrix to the perturbative estimates,
it is useful to construct the quantity
$\sigma(\mu_1, \mu_2) \sigma_{\text{PT}}^{-1}(\mu_1, \mu_2)$,
which is equal to $1_{5 \times 5}$ up to higher-order terms not included
in the perturbative expansions, residual discretisation effects,
and non-perturbative contributions.
These quantities are shown in 
Figs.~\ref{fig:ssfoverpt_11}-\ref{fig:ssfoverpt_44}.
When running from 3 to 2 GeV, we find that these effects are typically of order
a few percents, and in many instances much less.

\begin{table}[!t]
\begin{tabular}{c c c c}
\toprule
$Z^{\msbar}$     & $R^{\msbar \leftarrow {\rm RI-SMOM}}$ &   $Z^{\rm RI-SMOM}$  &   scheme  \\
\hline
$0.92022(26)$  & $1.00414$  & $0.91642(26)$        &   $(\gamma_\mu,\gamma_\mu)$\\
$0.97675(48)$  & $0.95205$  & $1.02593(51)$        &   $(\gamma_\mu,\s{q})$     \\
$0.89123(23)$  & $1.04320$  & $0.85432(22)$        &   $(\s{q},\gamma_\mu)$     \\
$0.94796(34)$  & $0.99112$  & $0.95645(34)$        &   $(\s{q},\s{q})$          \\
\botrule
\end{tabular}
\caption[]{$Z/Z_V^2$ factors for the $(27,1)$ operators at $3\, \GeV$ for $a=a_{\bf 24}$.}
\label{tab:ZBK_24}
\end{table}

\begin{table}[!t]
\ars{1.1}
\begin{tabular}{c c c c}
\toprule
$Z^{\msbar}$     & $R^{\msbar \leftarrow {\rm RI-SMOM}}$ &   $Z^{\rm RI-SMOM}$  &   scheme  \\
\hline
$
\begin{pmatrix}
1.05043(7)  &  0.28197(37) \\
0.05654(23) &  0.95348(189)\\
\end{pmatrix}$ \;
& 
$\begin{pmatrix}
1.00084 &   0.00506 \\
0.01576 &   1.08781 \\ 
\end{pmatrix}$ \;
&
$\begin{pmatrix}
1.04936(7)  &  0.27732(38) \\
0.03677(21) &  0.87249(174) \\
\end{pmatrix}$ \;
&   $(\gamma_\mu,\gamma_\mu)$ 
\vspace{0.1cm} \\
$\begin{pmatrix}
1.11482(33) &  0.29951(46)  \\
0.06115(24) &  1.01655(183)  \\
\end{pmatrix}$ \;
&
$\begin{pmatrix}
0.94876 &  0.00506 \\
0.01576 &  1.03572 \\
\end{pmatrix}$ \;
&
$\begin{pmatrix}
1.17481(35) &  0.31048(50) \\ 
0.04117(23) &  0.97677(178) \\
\end{pmatrix}$ \;
&
$(\gamma_\mu,\s{q})$ 
\vspace{0.1cm} \\
$\begin{pmatrix}
0.98777(27) &  0.26988(33)  \\
0.06483(20) &  0.95664(185) \\
\end{pmatrix}$ \;
&
$\begin{pmatrix}
1.05293  &  0.00506 \\
0.00599  &  1.08130 \\
\end{pmatrix}$ \;
&
$\begin{pmatrix}
0.93785(26) &  0.25207(32)  \\
0.05475(18) &  0.88332(171) \\
\end{pmatrix}$ \;
&
$(\s{q},\gamma_\mu)$ 
\vspace{0.1cm} \\
$\begin{pmatrix}
1.05116(7)  & 0.28745(40)  \\
0.06938(21) & 1.01946(179) \\
\end{pmatrix}$ \;
&
$\begin{pmatrix}
1.00084 & 0.00506 \\
0.00599 & 1.02921 \\ 
\end{pmatrix}$ \;
&
$\begin{pmatrix}
1.04996(7)  &  0.28221(41)  \\ 
0.06129(20) &  0.98888(174) \\
\end{pmatrix}$ \;
&
$(\s{q},\s{q})$ \\
\botrule
\end{tabular}
\caption[]{$Z/Z_V^2$ matrices for the $(8,8)$ operators at $\mu=3\, \GeV$ for $a=a_{\bf 24}$.}
\label{tab:Z88_24}
\end{table}


\begin{table}[!t]
\ars{1.1}
\begin{tabular}{c c c c}
\toprule
$Z^{\msbar}$     & $R^{\msbar \leftarrow {\rm RI-SMOM}}$ &   $Z^{\rm RI-SMOM}$  &   scheme  \\
\hline
$
\begin{pmatrix}
 0.93350(166) &  -0.02688(26) \\
-0.33065(42)  &   1.16123(94) \\
\end{pmatrix}$ \;
& 
$\begin{pmatrix}
 1.02004 & 0.00968 \\
-0.05621 & 1.11206 \\
\end{pmatrix}$ \;
&
$\begin{pmatrix}
 0.91754(162)  & -0.03625(25) \\
-0.25096(45)   &  1.04239(86) \\
\end{pmatrix}$ \;
&   $(\gamma_\mu,\gamma_\mu)$ 
\vspace{0.1cm} \\
$\begin{pmatrix}
  0.99155(159) &  -0.02798(27) \\
 -0.35554(54)  &  1.23927(126) \\
\end{pmatrix}$ \;
&
$\begin{pmatrix}
 0.96796 & 0.00968  \\
-0.05621 & 1.05997  \\
\end{pmatrix}$ \;
&
$\begin{pmatrix}
 1.02719(164) &  -0.04058(27)  \\
-0.28096(58) &   1.16700(121) \\
\end{pmatrix}$ \;
&
$(\gamma_\mu,\s{q})$ 
\vspace{0.1cm} \\

$\begin{pmatrix}
 0.92457(145) & -0.02343(19) \\
-0.41226(142) &  1.19248(49) \\
\end{pmatrix}$ \;
&
$\begin{pmatrix}
 1.01556 &  0.01118 \\
-0.07860 &  1.11952 \\
\end{pmatrix}$ \;
&
$\begin{pmatrix}
0.91375(143)  & -0.03477(18) \\
-0.30410(119) &  1.06273(45) \\
\end{pmatrix}$ \;
&
$(\s{q},\gamma_\mu)$
\vspace{0.1cm} \\

$\begin{pmatrix}
 0.98180(136) & -0.02422(20) \\
-0.44382(148) &  1.27309(83) \\
\end{pmatrix}$ \;
&
$\begin{pmatrix}
 0.96348 &  0.01118 \\
-0.07860 &  1.06744 \\
\end{pmatrix}$ \;
&
$\begin{pmatrix}
1.02296(142)   & -0.03894(20) \\
-0.34046(131)  &  1.18980(79) \\
\end{pmatrix}$ \;
&
$(\s{q},\s{q})$ \\
\botrule
\end{tabular}
\caption[]{$Z/Z_V^2$ matrices for the $(6,\bar 6)$ operators at $\mu=3\, \GeV$ for $a=a_{\bf 24}$.}
\label{tab:Z66_24}
\end{table}



\begin{table}[!t]
\begin{tabular}{c c c c}
\toprule
$Z^{\msbar}$     & $R^{\msbar \leftarrow {\rm RI-SMOM}}$ &   $Z^{\rm RI-SMOM}$  &   scheme  \\
\hline
$0.94526(26)$  & $1.00414$  & $0.94137(26)$        &   $(\gamma_\mu,\gamma_\mu)$\\
$0.99554(67)$  & $0.95205$  & $1.04568(70)$        &   $(\gamma_\mu,\s{q})$     \\
$0.91915(53)$  & $1.04320$  & $0.88109(51)$        &   $(\s{q},\gamma_\mu)$     \\
$0.96999(32)$  & $0.99112$  & $0.97868(32)$        &   $(\s{q},\s{q})$          \\
\botrule
\end{tabular}
\caption[]{$Z/Z_V^2$ factors for the $(27,1)$ operators at $3\, \GeV$ for $a=a_{\bf 32}$.}
\label{tab:ZBK_32}
\end{table}
\begin{table}[!t]
\ars{1.1}
\begin{tabular}{c c c c}
\toprule
$Z^{\msbar}$     & $R^{\msbar \leftarrow {\rm RI-SMOM}}$ &   $Z^{\rm RI-SMOM}$  &   scheme  \\
\hline
$
\begin{pmatrix}
1.04740(14)  &  0.27818(76) \\
0.04391(15) &  0.87386(157)\\
\end{pmatrix}$ \;
& 
$\begin{pmatrix}
1.00084 &   0.00506 \\
0.01576 &   1.08781 \\ 
\end{pmatrix}$ \;
&
$\begin{pmatrix}
1.04639(14)  &  0.27391(76) \\
0.02521(13) &  0.79935(145) \\
\end{pmatrix}$ \;
&   $(\gamma_\mu,\gamma_\mu)$ 
\vspace{0.1cm} \\
$\begin{pmatrix}
1.10288(81) &  0.29313(92)  \\
0.04731(16) &  0.92432(154)  \\
\end{pmatrix}$ \;
&
$\begin{pmatrix}
0.94876 &  0.00506 \\
0.01576 &  1.03572 \\
\end{pmatrix}$ \;
&
$\begin{pmatrix}
1.16230(85) &  0.30423(97) \\ 
0.02799(15) &  0.88781(149) \\
\end{pmatrix}$ \;
&
$(\gamma_\mu,\s{q})$ 
\vspace{0.1cm} \\
$\begin{pmatrix}
0.99359(64) &  0.26772(73)  \\
0.05169(22) &  0.87793(156) \\
\end{pmatrix}$ \;
&
$\begin{pmatrix}
1.05293  &  0.00506 \\
0.00599  &  1.08130 \\
\end{pmatrix}$ \;
&
$\begin{pmatrix}
0.94344(61) &  0.25037(70)  \\
0.04258(20) &  0.81054(144) \\
\end{pmatrix}$ \;
&
$(\s{q},\gamma_\mu)$ 
\vspace{0.1cm} \\
$\begin{pmatrix}
1.04908(16)  & 0.28286(83)  \\
0.05496(25) & 0.92821(156) \\
\end{pmatrix}$ \;
&
$\begin{pmatrix}
1.00084 & 0.00506 \\
0.00599 & 1.02921 \\ 
\end{pmatrix}$ \;
&
$\begin{pmatrix}
1.04795(16)  &  0.27806(84)  \\ 
0.04730(24) &  0.90024(152) \\
\end{pmatrix}$ \;
&
$(\s{q},\s{q})$ \\
\botrule
\end{tabular}
\caption[]{$Z/Z_V^2$ matrices for the $(8,8)$ operators at $\mu=3\, \GeV$ for $a=a_{\bf 32}$.}
\label{tab:Z88_32}
\end{table}


\begin{table}[!t]
\ars{1.1}
\begin{tabular}{c c c c}
\toprule
$Z^{\msbar}$     & $R^{\msbar \leftarrow {\rm RI-SMOM}}$ &   $Z^{\rm RI-SMOM}$  &   scheme  \\
\hline
$
\begin{pmatrix}
 0.86595(130) &  -0.01245(18) \\
-0.32627(79)  &   1.21084(92) \\
\end{pmatrix}$ \;
& 
$\begin{pmatrix}
 1.02004 & 0.00968 \\
-0.05621 & 1.11206 \\
\end{pmatrix}$ \;
&
$\begin{pmatrix}
 0.85132(127)  & -0.02253(17) \\
-0.25036(76)   &  1.08769(83) \\
\end{pmatrix}$ \;
&   $(\gamma_\mu,\gamma_\mu)$ 
\vspace{0.1cm} \\
$\begin{pmatrix}
  0.91256(128) &  -0.01252(19) \\
 -0.34790(98)  &  1.28201(145) \\
\end{pmatrix}$ \;
&
$\begin{pmatrix}
 0.96796 & 0.00968  \\
-0.05621 & 1.05997  \\
\end{pmatrix}$ \;
&
$\begin{pmatrix}
 0.94555(132) &  -0.02502(19)  \\
-0.27808(96) &   1.20814(137) \\
\end{pmatrix}$ \;
&
$(\gamma_\mu,\s{q})$ 
\vspace{0.1cm} \\
$\begin{pmatrix}
 0.86318(125) & -0.01153(18) \\ 
-0.34762(135) &  1.21953(90) \\
\end{pmatrix}$ \;
&
$\begin{pmatrix}
1.05293  &  0.00506 \\
0.00599  &  1.08130 \\
\end{pmatrix}$ \;
&
$\begin{pmatrix}
 0.85271(123) & -0.02332(17) \\  
-0.25065(122) & +1.08770(81) \\
\end{pmatrix}$ \;
&
$(\s{q},\gamma_\mu)$
\vspace{0.1cm} \\
$\begin{pmatrix}
 0.90945(122) & -0.01146(19)  \\ 
-0.37164(161) &  1.29174(159) \\
\end{pmatrix}$ \;
&
$\begin{pmatrix}
 0.96348 &  0.01118 \\
-0.07860 &  1.06744 \\
\end{pmatrix}$ \;
&
$\begin{pmatrix}
 0.94715(127) & -0.02590(19)   \\
-0.27842(149) &  1.20822(149)  \\
\end{pmatrix}$ \;
&
$(\s{q},\s{q})$
\vspace{0.1cm} \\
\botrule
\end{tabular}
\caption[]{$Z/Z_V^2$ matrices for the $(6,\bar 6)$ operators at $\mu=3\, \GeV$ for $a=a_{\bf 32}$.}
\label{tab:Z66_32}
\end{table}

\begin{table}[!t]
\begin{tabular}{c c c}
\toprule
$\sigma(2\GeV,3\GeV)^{\msbar}$ \;    & \; $\sigma(2\GeV,3\GeV)^{\rm RI-SMOM}$ \; &   scheme  \\
\hline
$1.0194(9)$  & $1.0186(9)$  &   $(\gamma_\mu,\gamma_\mu)$\\
$1.0649(32)$ & $1.0761(32)$ &   $(\gamma_\mu,\s{q})$     \\
$0.9963(26)$ & $0.9879(25)$ &   $(\s{q},\gamma_\mu)$     \\
$1.0428(12)$ & $1.0448(12)$ &   $(\s{q},\s{q})$          \\
\botrule
\end{tabular}
\caption[]{Continuum running factor  between 3 and 2 $\GeV$ for the $(27,1)$ operator and the
  various intermediate schemes. In the first column we give the results converted to $\msbar$,
  whereas in the middle column the results are in the RI-SMOM scheme and are purely non-perturbative. }
\label{tab:sigma_BK}
\end{table}

\begin{table}[!t]
\ars{1.1}
\begin{tabular}{c c c}
  \toprule
  $\sigma(2\GeV,3\GeV)^{\msbar}$     & $\sigma(2\GeV,3\GeV)^{\rm RI-SMOM}$  &   scheme  \\
\hline
$
\begin{pmatrix}
1.0177(11)  &  0.1453(22)  \\
0.0095(7)   &  0.7873(44)  \\
\end{pmatrix}$ \;
& 
$\begin{pmatrix}
1.0198(11) &   0.1583(24)  \\
0.0023(6)  &   0.7718(44)  \\
\end{pmatrix}$ \;
&   $(\gamma_\mu,\gamma_\mu)$ 
\vspace{0.1cm} \\
$\begin{pmatrix}
1.0634(37) &   0.1506(22)  \\ 
0.0105(7)  &   0.8256(65)  \\
\end{pmatrix}$ \;
&
$\begin{pmatrix}
1.0778(37) &  0.1667(22) \\
0.0024(7) &   0.8168(65) \\
\end{pmatrix}$ \;
&
$(\gamma_\mu,\s{q})$ 
\vspace{0.1cm} \\
$\begin{pmatrix}
0.9743(35) &   0.1347(29)  \\
0.0134(18) &   0.7863(49)  \\
\end{pmatrix}$ \;
&
$\begin{pmatrix}
0.9650(34)  &  0.1371(29)  \\
0.0108(17)  &  0.7735(49)  \\
\end{pmatrix}$ \;
&
$(\s{q},\gamma_\mu)$ 
\vspace{0.1cm} \\
$\begin{pmatrix}
1.0199(10) &    0.1397(28) \\
0.0142(18) &    0.8241(74) \\
\end{pmatrix}$ \;
&
$\begin{pmatrix}
1.0205(10) &   0.1438(30) \\
0.0114(18) &   0.8184(73) \\
\end{pmatrix}$ \;
&
$(\s{q},\s{q})$ \\
\botrule
\end{tabular}
\caption[]{Same for the running matrix of the $(8,8)$ operators. }
\label{tab:sigma88}
\end{table}


\begin{table}[!t]
\ars{1.1}
\begin{tabular}{c c c}
  \toprule
    $\sigma(2\GeV,3\GeV)^{\msbar}$     & $\sigma(2\GeV,3\GeV)^{\rm RI-SMOM}$  &   scheme  \\
\hline
$
\begin{pmatrix}
   0.8402(36) &   0.0068(5)  \\
  -0.1356(18) &   1.1446(16) \\
\end{pmatrix}$ \;
& 
$\begin{pmatrix}
  0.8379(35)  &  0.0025(6) \\
 -0.1285(18)  &  1.1203(16) \\
\end{pmatrix}$ \;
&   $(\gamma_\mu,\gamma_\mu)$ 
\vspace{0.1cm} \\
$\begin{pmatrix}
  0.8795(58)  &   0.0075(5)  \\
 -0.1426(21)  &   1.1968(38) \\
\end{pmatrix}$ \;
&
$\begin{pmatrix}
   0.8868(59) &   0.0028(6)   \\
  -0.1354(20) &   1.1819(37)  \\
\end{pmatrix}$ \;
&
$(\gamma_\mu,\s{q})$ 
\vspace{0.1cm} \\

$\begin{pmatrix}
  0.8288(62) &   0.0084(19) \\
 -0.1795(76) &   1.1602(26) \\
\end{pmatrix}$ \;
&
$\begin{pmatrix}
  0.8278(63)  &  0.0033(20) \\
 -0.1705(72)  &  1.1337(25) \\
\end{pmatrix}$ \;
&
$(\s{q},\gamma_\mu)$
\vspace{0.1cm} \\

$\begin{pmatrix}
  0.8612(60) &   0.0092(19) \\
 -0.1853(84) &   1.2155(60) \\
\end{pmatrix}$ \;
&
$\begin{pmatrix}
  0.8697(61) &   0.0034(21) \\
 -0.1772(80) &   1.1982(59) \\
\end{pmatrix}$ \;
&
$(\s{q},\s{q})$ \\
\botrule
\end{tabular}
\caption[]{Same for the running matrix of the $(6,\bar 6)$ operators. }
\label{tab:sigma66}
\end{table}

\section{RI-MOM renormalisation scheme}
\label{s5}
In addition to the RI-SMOM renormalisation schemes used
to obtain our main results~\cite{Garron:2016mva}, we also implemented
RI-MOM renormalisation conditions for the intermediate scheme.
The RI-MOM scheme differs in the kinematic configuration of the
vertex functions, which depend on a single momentum vector (obtained
by setting $p_1 = p_2$ in Eq.~(\ref{eq:Contraction})).
Vertex functions in this ``exceptional" configuration
can have large contributions from infrared poles which go as inverse
powers of the quark mass ($m_\pi^2$) and momenta; 
as our renormalisation matrices are defined in the chiral limit $m \to 0$
(here and throughout this section $m=m_{\rm bare}+m_{\rm res}$)
we have an unphysical divergence due to this scheme, which must be subtracted.
These pole contributions
are suppressed by powers of $p^2$ but in practice turn out to be large for
momenta accessible in our Rome Southampton window.

As the 
$m \to 0$ limit is approached the raw RI-MOM data
clearly suffers from pole contamination, 
the effect of these pion poles is clearly visible in our data, in particular
in the $\Lambda_{i3}$ and $\Lambda_{i4}$ elements
(Fig.~\ref{fig:Lambdas_sub});
in contrast the RI-SMOM data have only a weak
mass dependence and tend to $Z^{-1}$ in the $m \to 0$ limit (Fig~\ref{fig:Lambda_ne}).

At large $\mu = \sqrt {p^2}$ the matrix of vertex functions $\Lambda$ 
will become block diagonal in the chiral limit if
the effects of spontaneous chiral symmetry breaking are suppressed.
In the RI-MOM scheme chiral symmetry breaking effects can be extremely enhanced
in the $m \to 0$ limit; as a result the chiral structure
is strongly broken.
This can be seen for example in Fig.~\ref{fig:chi_e} (right).

We focus first on the chiral extrapolation and work at fixed momentum.
In order to extract $Z_{ij}$ from the RI-MOM data, we fit the mass dependence
of the vertex functions $\Lambda_{ij}$.
In principle we expect the vertex function to exhibit poles which go like
$1/m$ and $1/m^2$~(see for example \cite{Bertone:2012cu}), 
and so will be described by the general form
\begin{equation} \label{zpoles}
\Lambda_{ij}(a,\mu,m) = 
Z_{ij}^{-1}(a,\mu) + \frac{B_{ij}(a,\mu)}{(am)} + 
\frac{C_{ij}(a,\mu)}{(am)^2} + D_{ij}(a,\mu)(am) + \O((am)^2) \,.
\end{equation}

First, we observe that not all the matrix elements require a pole subtraction.
In that case, we just perform a linear fit in the quark mass (ie $B=C=0$) with the three (lightest) unitary
\footnote{
  Strictly speaking the setup is unitary in the light quark sector $m_{\rm light}^{\rm val}=m_{\rm light}^{\rm sea}$ ,
  but partially quenched for the strange   as $m_{\rm s}^{val} = m_{\rm light}^{\rm val} \ne m_{\rm s}^{\rm sea}$,
  however we have checked this effect is negligible within our systematic errors
}.
quark masses:
$am_{\rm bare } = 0.005, 0.01, 0.02$ on the $24^3$ and $0.004, 0.006, 0.008$ on the $32^3$.
The chirally-allowed elements which suffer from those pole contaminations are
$\Lambda_{23}, \Lambda_{33}, \Lambda_{44}$ and $\Lambda_{55}$
and the chirally-forbidden are  $\Lambda_{24}, \Lambda_{34}, \Lambda_{44}$ and $\Lambda_{53}$.
In this case, our main results are obtained on the same data with a single pole fit Ansatz $C=D=0$.

More explicitly, we used a `linear fit' method~\cite{Boyle:2012qb,Lytle:2013oqa} 
to remove the $1/m$ contributions to the RI-MOM
vertex functions. 
Here we multiply the data by $am$ and fit
$am\Lambda_{ij} = (am)Z^{-1}_{ij} + B_{ij} + \O((am)^2)$ to a straight line
to determine $Z^{-1}_{ij}$. This gives results equivalent
to fitting $\Lambda$
to the form $A+B/m$. We observed that $am\Lambda$ is consistent
with linear $am$ behavior to justify neglecting the $(am)^2$ term,
and also found the data after subtracting the pole contribution is linear.
After subtracting the pole we find good restoration of the chiral block structure
for the $32^3$ ensemble (Table~\ref{tab:bayes32}),
The chiral restoration is not as good on the $24^3$ ensemble,
the residual matrix elements are of the order of a few $\%$.
However we observed that they do affect the physical matrix elements,
and that different fit procedure give the same residual, see below
and Table~\ref{tab:bayes24}.

Since this infrared contamination completely dominates some of the raw data in the RI-MOM scheme,
we investigated the effect of this pole subtraction,
in particular we want to have a reasonable estimate of the systematic
error associated with the procedure. 
On the $24^3$, we used another ensemble $m_{\rm light}^{\rm val}=m_{\rm light}^{\rm sea}=0.03$
and have implemented different fit forms.
We fit each of the $\Lambda_{ij}$ with the forms $A +D m$ (\fit{0}), $A + B/m + D m$ (\fit{1}), 
and $A+ C/m^2 + Dm$ (\fit{2}).
We find that in cases with significant singular behavior, the
\fit{1} has $\chi^2 < 1$ and \fit{2} has $\chi^2 \gg 1$.
For $j=1,2,5$ there is no evidence of $1/m$ behavior and the results
are compatible with \fit{0}.
The fits are shown in Fig.~\ref{fig:Lambdas_sub} for the chirally allowed
elements $\Lambda_{23}$, $\Lambda_{33}$, $\Lambda_{44}$, and $\Lambda_{54}$.
From this we conclude that any $1/m^2$ dependence is to a large degree 
suppressed in the range of $\mval$ for which
we have data, and we determine $Z_{ij}^{-1}$ assuming the form of
{\bf fit-1}.

As a check on the procedure we compare the \fit{1} results on the
$24^3$ ensemble to the linear fit procedure. 
For the linear fit method
we threw out the heaviest ($am=0.03$) mass point 
because we found a degradation in the $\chi^2$ (though central values
remain consistent),
and which we attribute
to neglecting the quadratic term. 
The results from the two subtraction methods
are shown in Fig.~\ref{fig:Lambdas_sub} and Table~\ref{tab:bayes24}.
There is a slight tension
in the extrapolated results which highlights that some
uncontrolled systematic due to specifics of the subtraction procedure
may remain.

We also implemented Bayesian fits using the {\tt lsqfit} 
package~\footnote{\href{https://github.com/gplepage/lsqfit}{https://github.com/gplepage/lsqfit}}
to include additional terms from
Eq.~\eqref{zpoles} without requiring the number of data
points to exceed the number of fit parameters.
Table~\ref{tab:bayes24} compares results of frequentist and Bayesian fitting
on the $24^3$ ensemble,
for both chirally allowed and forbidden elements. 
The Bayesian fit
of the full form~\eqref{zpoles} is
consistent with the results of the other methods
but with larger
uncertainties. For the chirally-forbidden elements, the single pole fits (\fit{1} and Lin. fit) 
find values which differ
significantly from zero, whereas the Bayesian method finds best-fit values
very close to zero, but with errors comparable to the size of the central
values in the single pole case.

\begin{figure}[th]
  \centering
  \begin{tabular}{cc}
    \includegraphics[height=0.4\textwidth,width=0.49\textwidth]{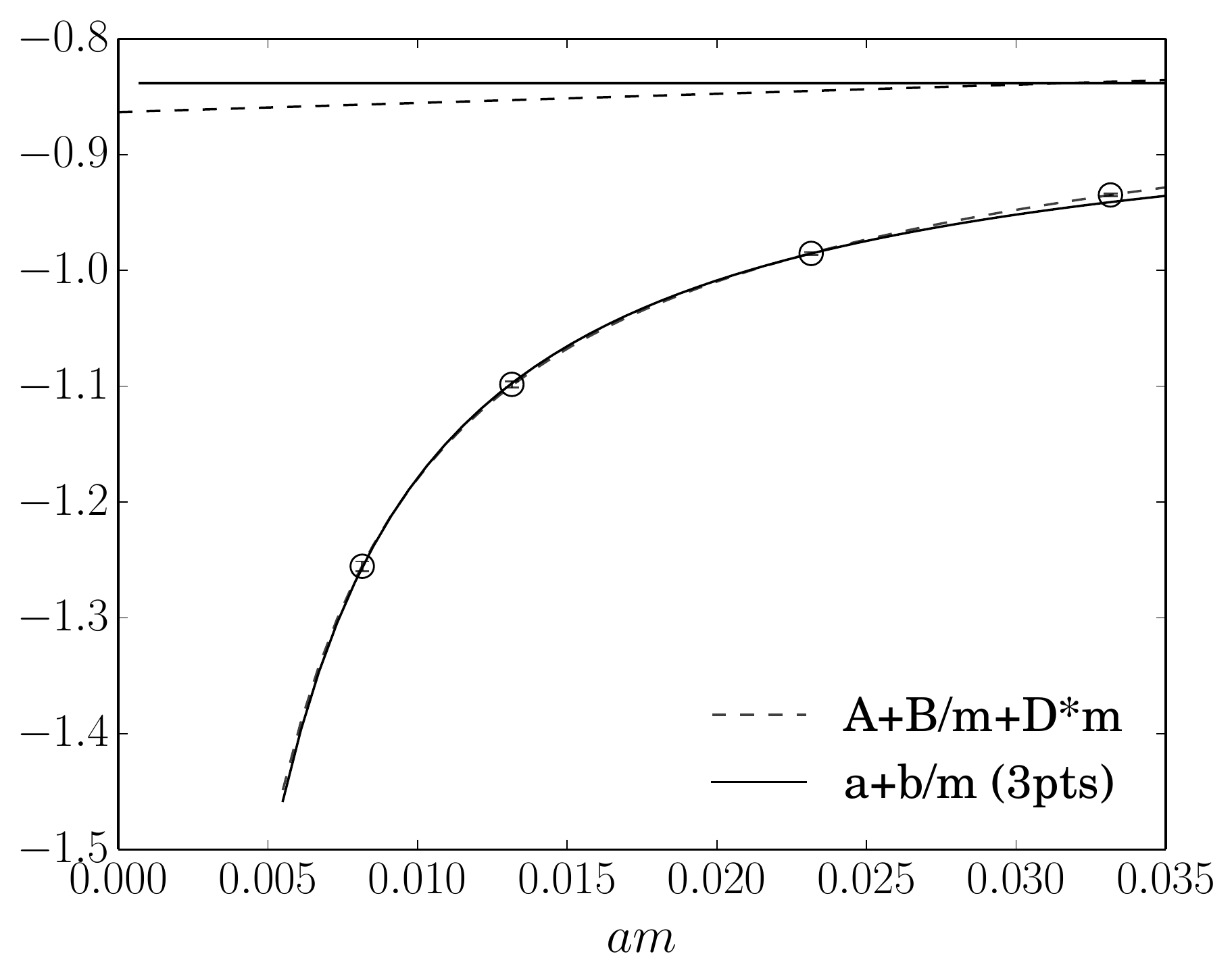}    &
    \includegraphics[height=0.4\textwidth,width=0.49\textwidth]{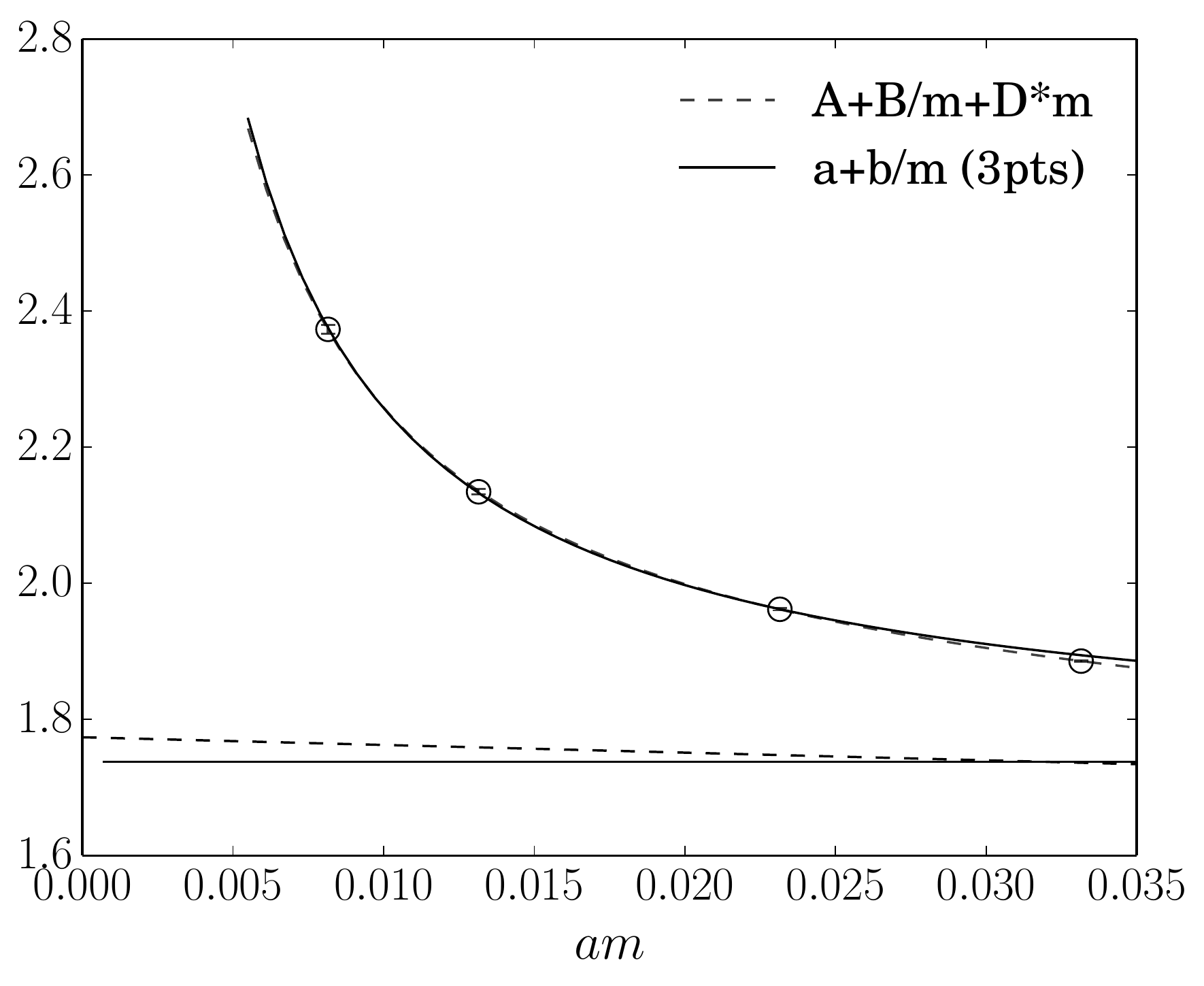}    \\
    \includegraphics[height=0.4\textwidth,width=0.49\textwidth]{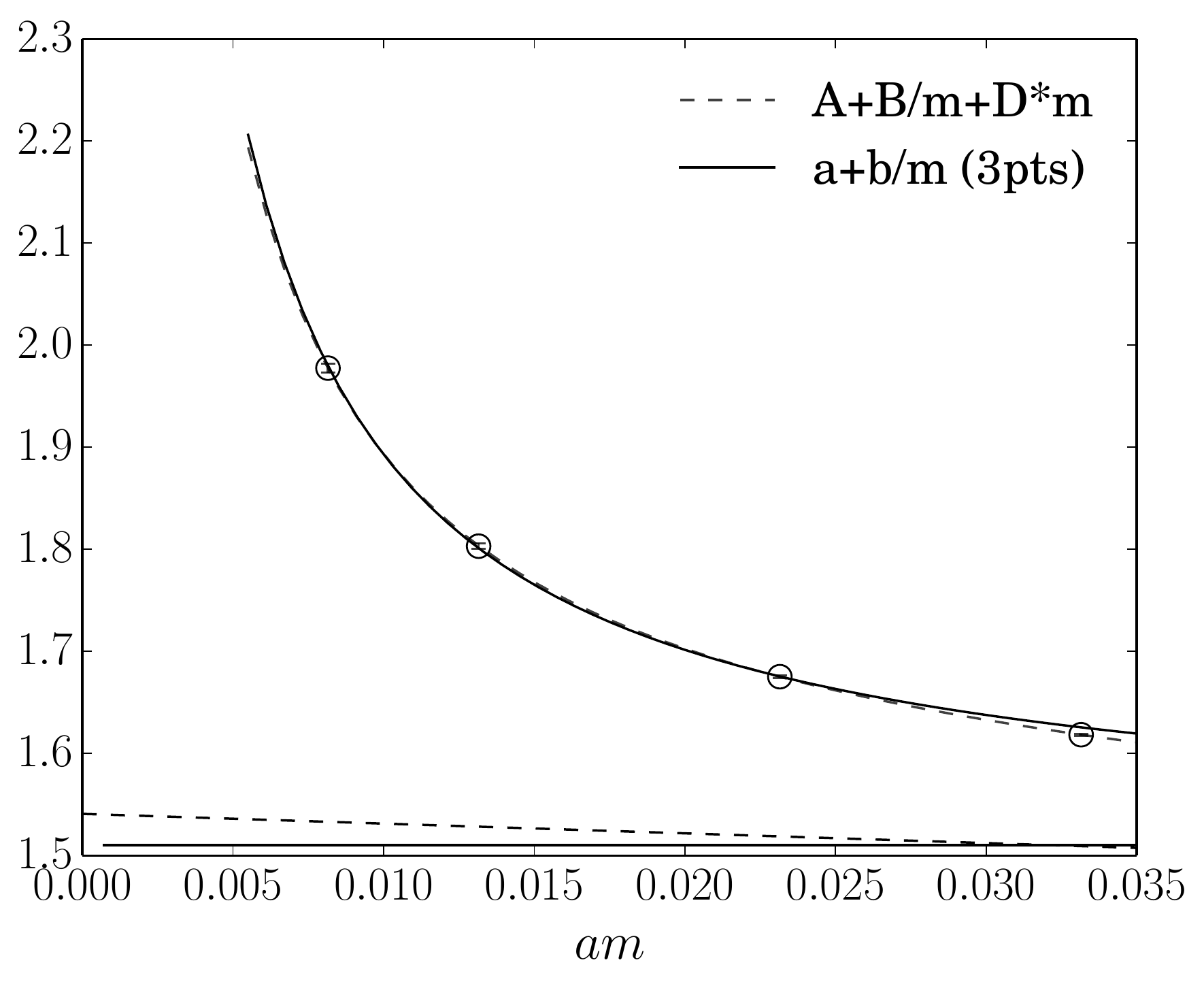}    &
    \includegraphics[height=0.4\textwidth,width=0.49\textwidth]{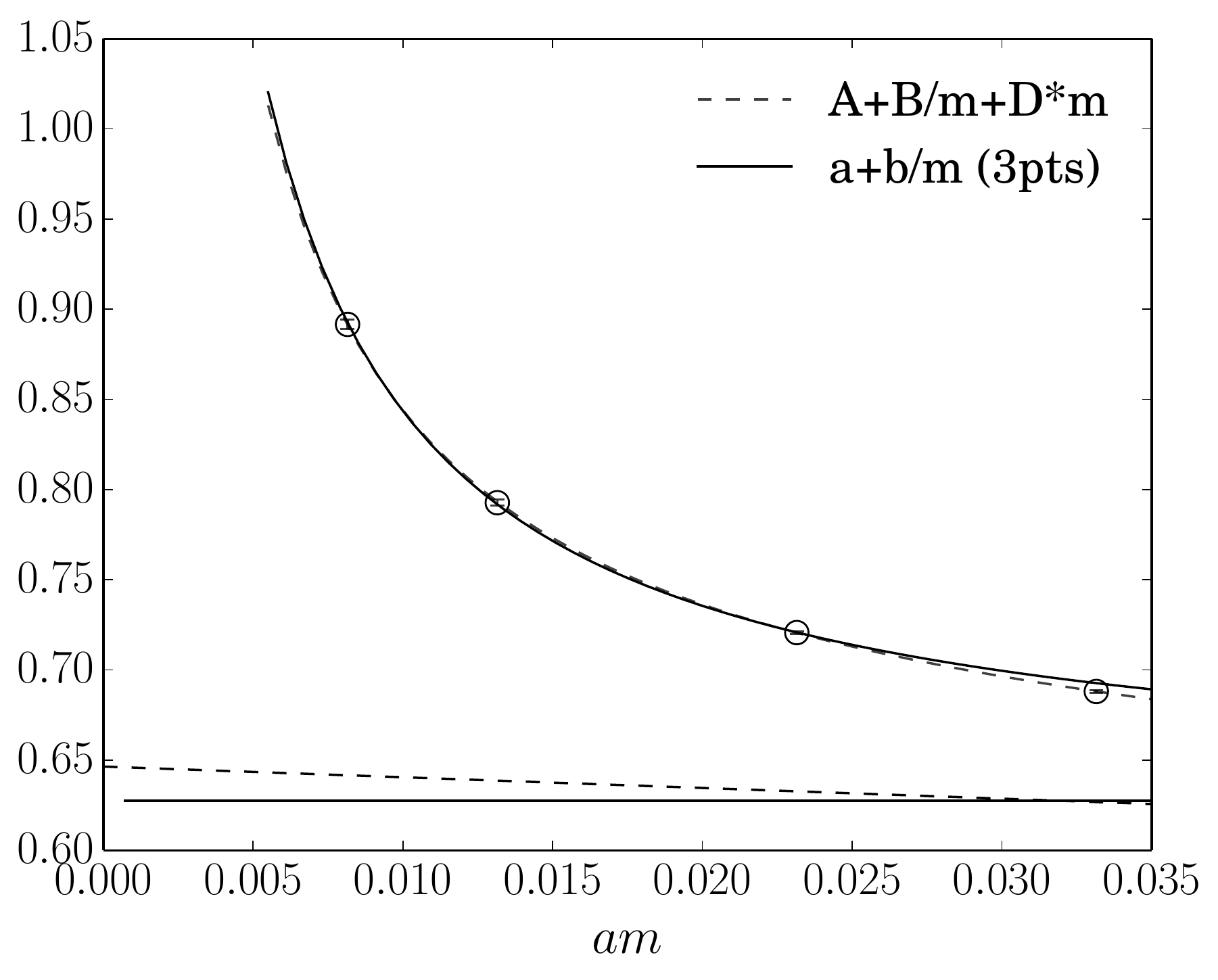}
  \end{tabular}
  \caption{Chirally-allowed RI-MOM vertex functions with singular behaviour
    from the $24^3$ ensemble. 
    The result of fitting the raw data (circles) to
    \fit{1} (dotted line) and a fit to the lightest three
    points with the form $a+b/m$ (solid line), 
    along with the result of subtracting
    the single pole contribution from each of the fits 
    (same line type as respective fits through data).
    Quantities shown from left to right are $\Lambda_{23}, \Lambda_{33}$
      (first row) and $\Lambda_{44} ,\Lambda_{54}$  (second row) at fixed momentum close
      to 3 GeV.}
      \label{fig:Lambdas_sub}
\end{figure}

\begin{figure}[th]
  \centering
  \begin{tabular}{cc}
    \includegraphics[width=.49\textwidth]{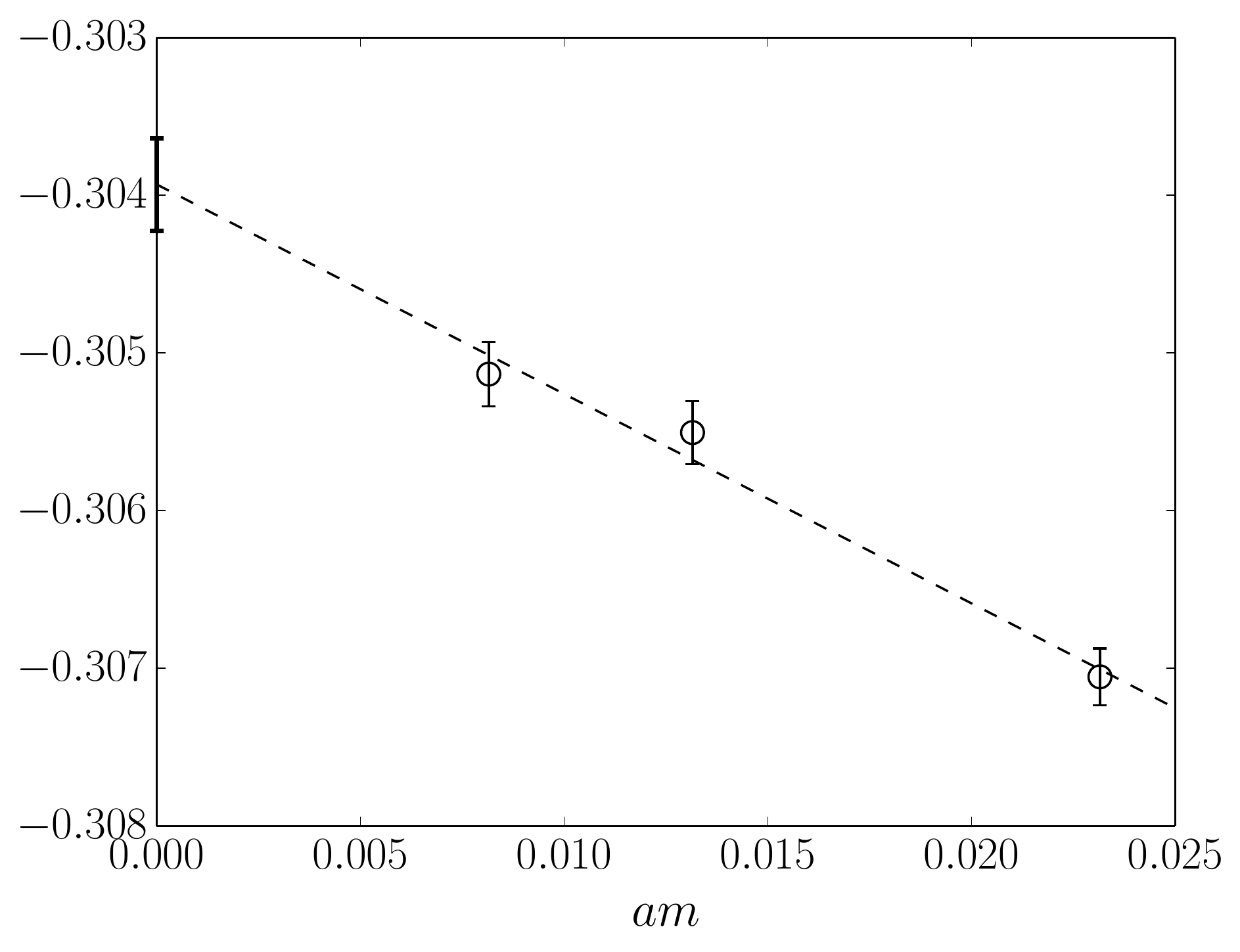} &
    \includegraphics[width=.49\textwidth]{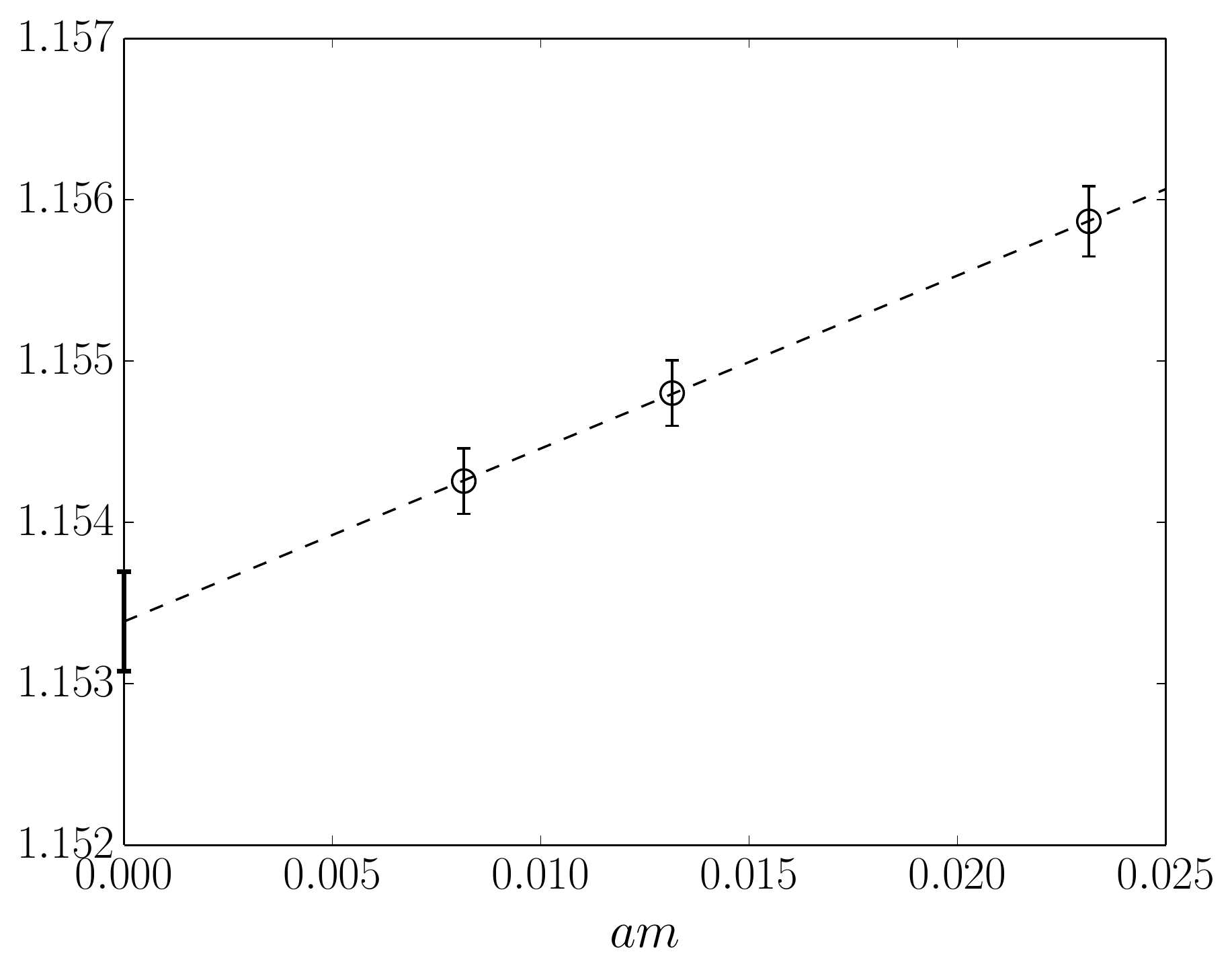} \\
    \includegraphics[width=.49\textwidth]{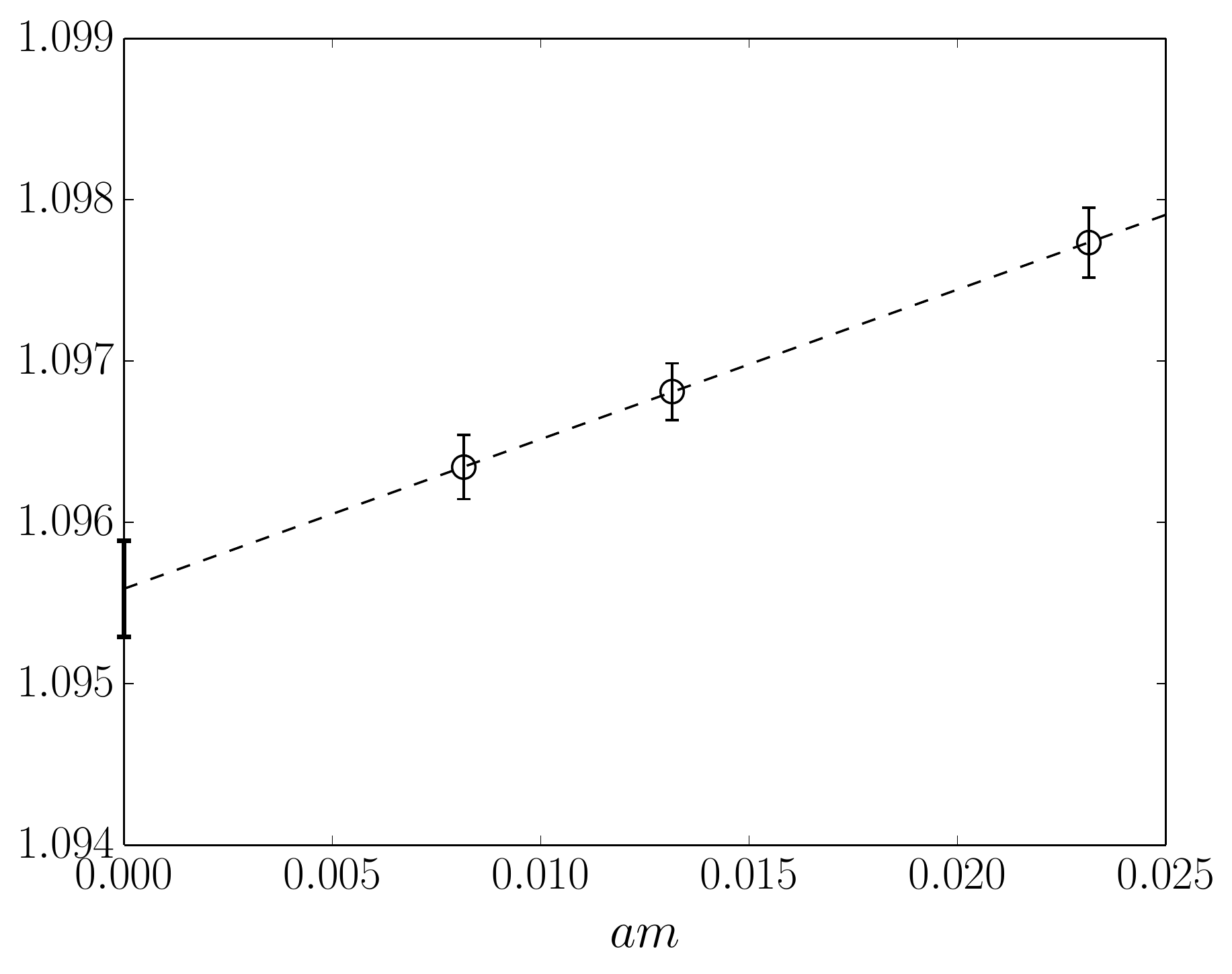} &
    \includegraphics[width=.49\textwidth]{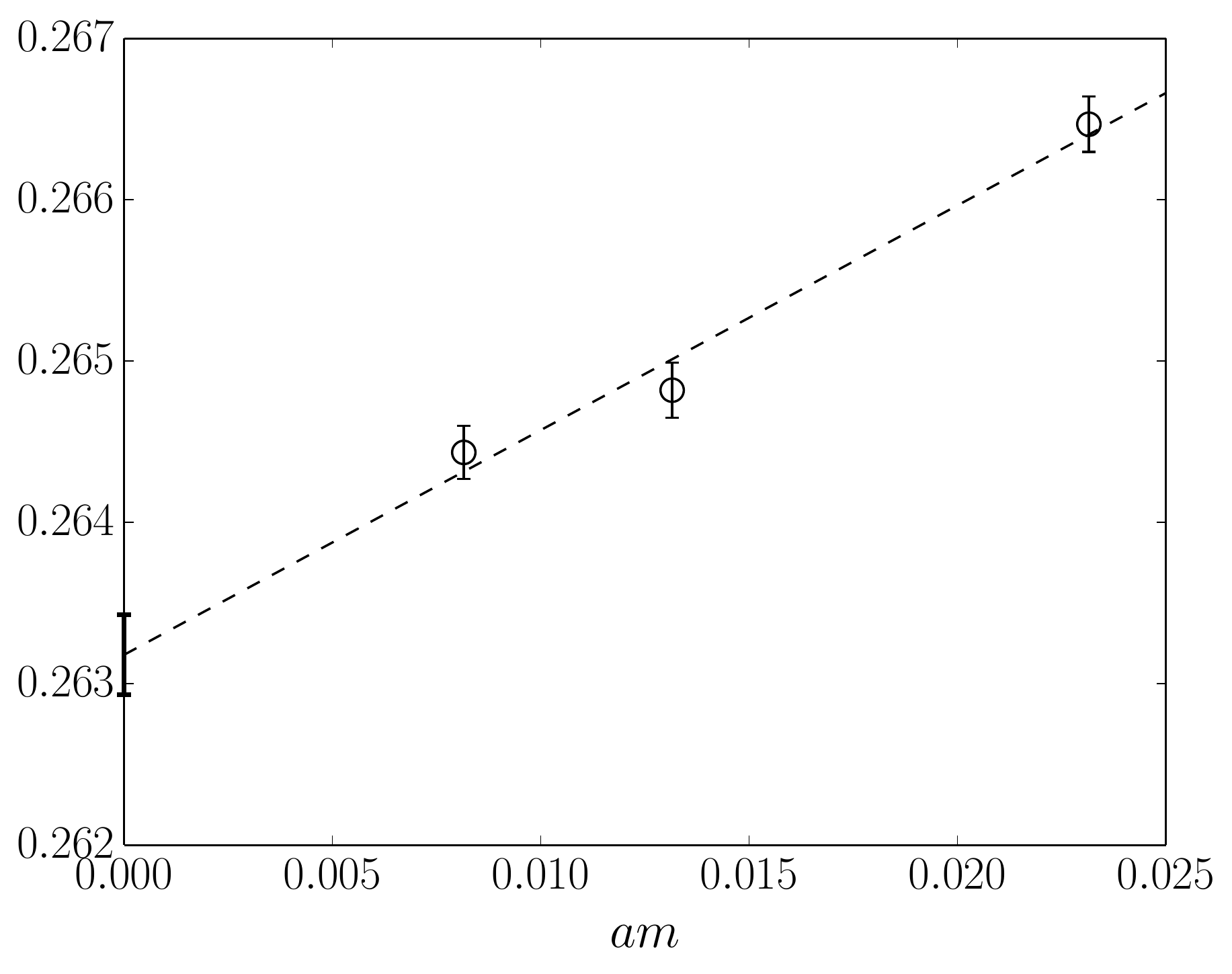}
  \end{tabular}
  \caption{
    Same as Fig.~\ref{fig:Lambdas_sub},
    from left to right: $\Lambda_{23}, \Lambda_{33}$ 
    (first row) and $\Lambda_{44} ,\Lambda_{54}$  (second row),      
    for the non-exceptional $(\gamma_\mu,\gamma_\mu)$ scheme.
    Here we fix the momentum $\mu$ close to 3 GeV..
    In that case we observed a very mild, linear, quark mass dependence.
    In contrast to the RI-MOM case, no pole subtraction is required
    (we show the vertex function without applying any pole strubaction procedure).
  }
\label{fig:Lambda_ne}
\end{figure}

\begin{table}[th]
\begin{tabular}{l | c c cc}
$Z_V^2Z^{-1}_{ij}$ &Lin.\ Fit meth.\ &Freq.\ $\frac{1}{m}$ &
Bayes* $\frac{1}{m}$ &Bayes $\frac{1}{m}$,$\frac{1}{m^2}$ \\
\hline
23 & -0.835(3) & -0.863(10) & -0.849(28) & -0.880(41) \\
33 & 1.733(4) & 1.774(14) & 1.758(40) & 1.791(58)  \\
44 & 1.506(3) & 1.541(11) & 1.535(29) & 1.548(42) \\
54 & 0.625(2) & 0.646(7) & 0.639(18) & 0.657(26) \\
\hline
24 & 0.052(5) & 0.063(16) & 0.100(45) & 0.011(65) \\
34 & -0.077(7) & -0.091(24) & -0.143(68) & 0.006(98) \\
43 & -0.065(5) & -0.080(19) &-0.127(51) & 0.003(74) \\
53 & -0.038(3) & -0.047(11) &-0.076(29) & -0.001(43)\\
\hline
\end{tabular}

\caption{Comparison of fit results on $24^3$ using
``linear fit method'', frequentist fit with $1/m$ term
($1/m^2$ term set to zero), Bayesian fit with only $1/m$ term
(* result uses only lightest three masses),
and Bayesian fit with both $1/m$ and $1/m^2$ terms.
The lower set of values corresponds to chirally-forbidden elements.
\label{tab:bayes24}}
\end{table}

As another consistency check of the method, we should also find an approximate
recovery of the block diagonal structure expected from chiral symmetry 
after removing the singular
parts of the data. 
Although to a decent approximation the terms that are chirally-forbidden
are suppressed after the pole subtraction, we find that the values are
statistically non-zero and the magnitude of chirally-forbidden elements
tend to be larger for the pole-subtracted ($\Lambda_{i,3/4}$) compared to
elements that do not require pole subtraction ($\Lambda_{i,1/2/4}$).
Fig.~\ref{fig:chi_e} shows the mass
and $\mu$ dependence of chirally-forbidden RI-MOM vertex functions for 
a case without discernible singular structure
($\Lambda_{12}$, left), and where the pole behavior is clearly visible
($\Lambda_{24}$, right). 
These results should be contrasted with the RI-SMOM results
shown in Fig.~\ref{fig:chi_ne},
where in all cases the chirally-forbidden elements extrapolate very nearly
to zero.

\begin{figure}[th]
\centering
\includegraphics[height=0.4\textwidth,width=.49\textwidth]{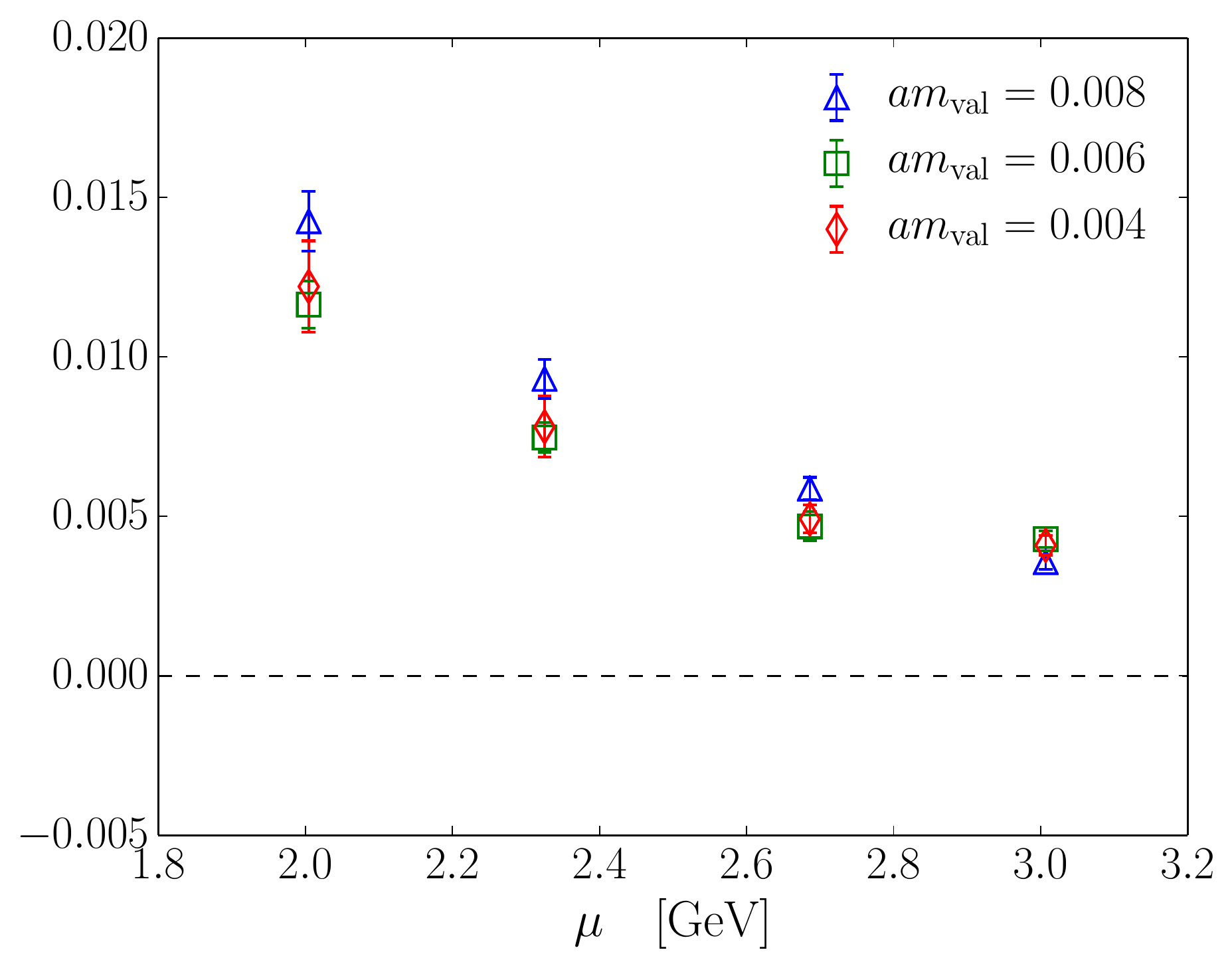}
\includegraphics[height=0.4\textwidth,width=.49\textwidth]{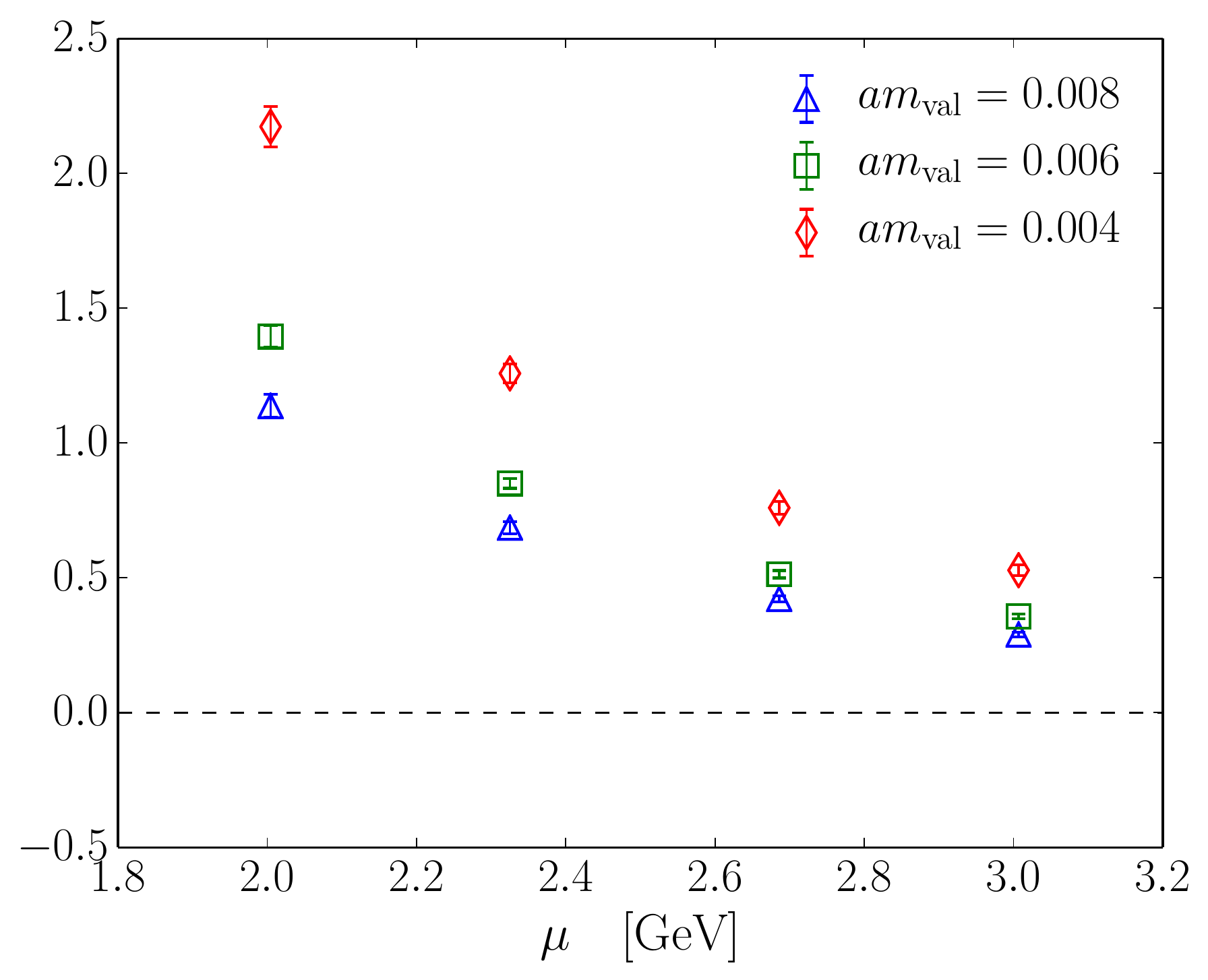}
\caption{
Left:
Example of an amputated and projected Green function in the exceptional
RI-MOM scheme at finite quark mass (on $32^3$ ensemble)
for different momenta.
This specific quantity should vanish
if chiral symmetry is exact.
Right: Example of a RI-MOM vertex function with strong singular behavior.
This specific quantity should also  vanish if chiral symmetry is exact
but is affected by large infrared contaminations. 
\label{fig:chi_e}
}
\end{figure}

\begin{figure}[th]
\centering
\includegraphics[height=0.4\textwidth,width=.49\textwidth]{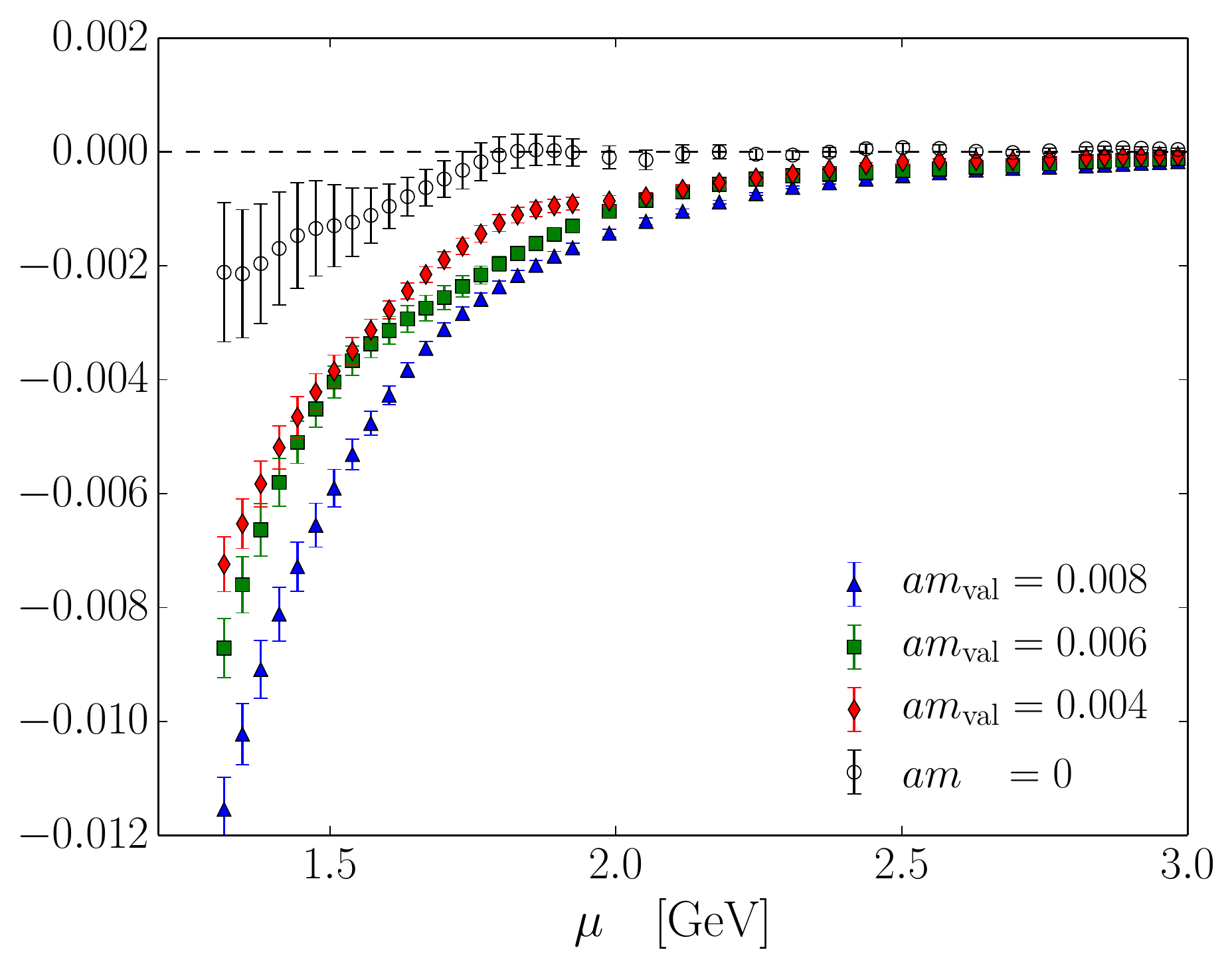}
\includegraphics[height=0.4\textwidth,width=.49\textwidth]{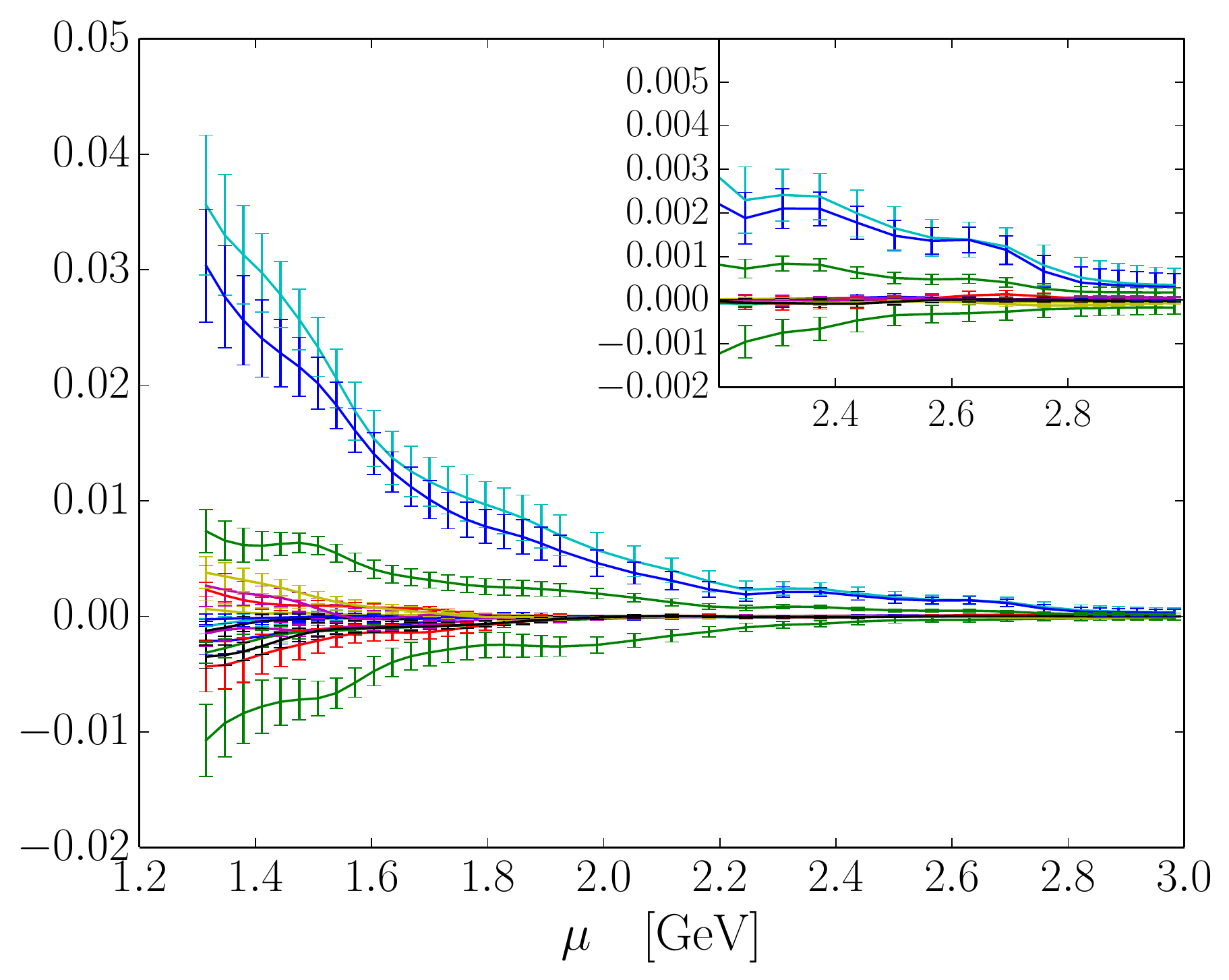}
\caption{
Left: Example of RI-SMOM $(\gamma_\mu, \gamma_\mu)$
$Z$-factor at finite quark mass and in the chiral limit (on $32^3$ ensemble)
for various momenta.
Right: Chirally-forbidden renormalisation factors in the 
$(\gamma_\mu, \gamma_\mu)$ scheme after chiral extrapolation
for various momenta.
\label{fig:chi_ne}
}
\end{figure}

\begin{table}[th]
\begin{tabular}{l | c c}
$Z_V^2Z^{-1}_{ij}$ &Lin.\ Fit meth.\ &Bayes $\frac{1}{m}$ \\
\hline
23 & -0.778(21) & -0.691(107)  \\
33 & 1.769(31) & 1.701(121)  \\
44 & 1.538(21) & 1.461(107)  \\
54 & 0.570(12) & 0.511(81) \\
\hline
24 & 0.006(34)  & -0.036(124) \\
34 & -0.007(53) & 0.046(135) \\
43 & -0.018(38) &  0.048(127) \\
53 & -0.011(21) &  0.072(107) \\
\hline
\end{tabular}

\caption{Comparison of fit results on $32^3$ using ``linear fit method'',
and Bayesian fit including $1/m$ term. 
The lower set of values corresponds to
chirally-forbidden elements at $\mu=3.01$ GeV.
\label{tab:bayes32}}
\end{table}

On the $32^3$ ensemble we also compare results of including the single pole
or both poles using a Bayesian fit, and results from the linear fit method, 
shown in Table~\ref{tab:bayes32}. 
The results again agree with the linear fit results
but have larger associated uncertainties. Note here the chirally-forbidden
elements obtained from the linear fit method 
are much smaller than in the $24^3$ case and are in fact zero within errors.
We also tried including the $1/m$ terms in `global' fits by constraining
the $1/m$ coefficient
coefficient in $\Lambda_{i3}$ to be the
negative of the coefficient in $\Lambda_{i4}$, which we observed
to be the case. Althought this strategy seems to improve somewhat the fit quality,
the numerical resuls were essentially unchanged.

We choose two options when we compute our renormalisation matrices: 
firstly we invert the whole matrix of fit parameters for all $\Lambda_{ij}$ 
and secondly we invert only the block diagonal elements of the matrix, 
zeroing by hand the chirally-forbidden elements. 
We will label these as the Not Block-Diagonal (NBD) 
and the Block Diagonal (BD).

Here we list the results for $Z$-matrices obtained in RI-MOM from the linear fit method.
On the $24^3$ ensemble at a fixed value of $\mu$ close to 3 GeV ,
\begin{equation}
\begin{aligned}
\frac{Z^{BD}}{Z_V^2} &= 
\begin{pmatrix}
0.88768(13) & 0 & 0 & 0 & 0 \\ 
0 & 1.10238(24) & 0.5318(22) & 0 & 0 \\ 
0 & 0.04812(16) & 0.5988(14) & 0 & 0 \\ 
0 & 0 & 0 & 0.6807(13) & -0.04444(14) \\ 
0 & 0 & 0 & -0.4940(18) & 1.18877(50) \\ 
\end{pmatrix}\;,
\\
\frac{Z^{NBD}}{Z_V^2} &= 
\begin{pmatrix}
0.88771(13) & -0.00732(27) & -0.00305(14) & 0.000479(29) & 0.000017(7) \\ 
-0.00524(20) & 1.10240(24) & 0.5313(22) & -0.0112(44) & 0.00068(31) \\ 
0.000335(24) & 0.04817(16) & 0.5999(14) & 0.0297(29) & -0.00125(20) \\ 
0.000055(7) & 0.00135(18) & 0.0254(22) & 0.6820(13) & -0.04449(14) \\ 
0.000077(9) & 0.00047(23) & 0.0080(26) & -0.4936(18) & 1.18875(50) \\
\end{pmatrix}
\;.
\end{aligned}
\end{equation}

And the $32^3$ at $\mu=3.01$ GeV, 
\begin{equation}
\begin{aligned}
\frac{Z^{BD}}{Z_V^2} &= 
\begin{pmatrix}
0.92128(28) & 0 & 0 & 0 & 0 \\ 
0 & 1.0887(10) & 0.480(14) & 0 & 0 \\ 
0 & 0.03506(56) & 0.5787(85) & 0 & 0 \\ 
0 & 0 & 0 & 0.6592(76) & -0.02996(47) \\ 
0 & 0 & 0 & -0.448(10) & 1.2072(12) \\
\end{pmatrix}\;,
\\
\frac{Z^{NBD}}{Z_V^2} &= 
\begin{pmatrix}
0.92129(28) & -0.00310(61) & -0.00101(29) & 0.000219(75) & 0.000000(27) \\ 
-0.00251(47) & 1.0886(10) & 0.480(14) & -0.004(27) & 0.0001(13) \\ 
0.000249(72) & 0.03507(57) & 0.5789(85) & 0.008(17) & 0.00032(82) \\ 
0.000021(47) & 0.00038(82) & 0.010(12) & 0.6593(76) & -0.02996(47) \\ 
0.000033(53) & 0.00006(96) & 0.004(15) & -0.448(10) & 1.2072(12) & \\
\end{pmatrix}
\;.
\end{aligned}
\end{equation}

It is evident that the BD and NBD Z-matrices are not too dissimilar, 
we take the difference in results of the operators renormalised using 
either of these as a systematic for our final RI-MOM results. 
Our results in the RI-MOM scheme after chiral extrapolation and interpolation to $\mu=3$ GeV
read
\be
\frac{ Z^{\mbox{RI-MOM}}}{Z_V^2} (a_{\bf 24}) =
\begin{pmatrix}
  0.88989(134)  & 0 & 0 & 0 & 0 \\
  0             & 1.1015(6) & 0.5299(14) &  0 & 0 \\
  0             & 0.0470(8) & 0.5931(49) &  0 & 0 \\
  0             & 0           & 0        &  0.6744(50) &  -0.0429(10) \\ 
  0             & 0           & 0        & -0.4929(9)  &   1.1918(24) \\
\end{pmatrix}\;,
\ee

\be
\frac{Z^{\mbox{RI-MOM}}}{Z_V^2} (a_{\bf 32}) =
\begin{pmatrix}
  0.9213(11) & 0 & 0 & 0 & 0 \\
  0  & 1.08879(64) & 0.4792(47) & 0 & 0 & 0 \\ 
  0  & 0.03512(70) & 0.580(11)  & 0 & 0 & 0 \\
  0  & 0           & 0          &  0.6602(101) &  -0.0301(8) \\
      0  & 0           & 0          & -0.4476(43)  &   1.2080(51) \\
\end{pmatrix}\,.
\ee

In conclusion, the infrared 
contamination in some of the RI-MOM vertex functions makes
it difficult to extract the $Z$-factors precisely in the $m \to 0$ limit,
where these contributions diverge. These effects also strongly break
the chiral structure one expects to recover for 
$\mu \gg \Lambda_{\text{QCD}}$, though this
structure is restored (albeit imperfectly)
after subtraction of the pole contributions.
For these reasons, we find that the RI-MOM scheme (with exceptional kinematics) suffer from
systematic errors which are difficult estimate. 
Applying different strategies to subtract the poles, we find that final results
vary by $5\%$ in the worse case. 

In contrast the SMOM procedure
strongly suppresses these infrared effects --
evidence of chiral symmetry breaking
disappears in the $am \to 0$ limit at sufficiently large $\mu$ 
(Fig.~\ref{fig:chi_ne}), and the chirally-allowed $Z$-factors have
very mild linear mass dependence.
We also note that the SMOM to $\overline{\text{MS}}$ matching factors
are much closer to unity, suggesting a better behaved perturbative series
and a reduced perturbative matching uncertainty.
Therefore we strongly advocate using SMOM renormalisation conditions,
which are theoretically much cleaner.

We have argued that the discrepancies from results of~\cite{Boyle:2012qb,Bertone:2012cu,Carrasco:2015pra}
come the renormalisation procedure. 
Because these discrepancies appear in those matrix elements affected by these issues, 
we suggest avoiding the RI-MOM renormalisation conditions, at least for this set of operators.
\footnote{
A similar discrepancy was also 
recently observed
in $D$-mixing calculations using 
RI-MOM~\cite{Carrasco:2015pra,Carrasco:2014uya}
vs.\ `mostly nonperturbative' (mNPR)~\cite{Bazavov:2017weg} matching.}

Even we assess a rather conservative $5\%$ systematic error from the renormalisation procedure
in RI-MOM, our results are still not compatible with the RI-SMOM ones.
It remains the possibility of a conspiracy between these infrared artefacts and
omitted term in the perturbative matching.
Even if the latter should be of order $\alpha_s^2$, the anomalous dimensions of those operators
are rather large. Since a computation at the next order is technically very challenging,
this systematic error is difficult to control without using multiple schemes.

\section{Conclusions}
\label{s6:conclusions}

In this work we have defined and investigated new RI-SMOM intermediate schemes for the renormalisation
of $\Delta F=2$ four-quark operators needed for neutral kaon mixing beyond the standard model studies.
These schemes can easily be generalised to other processes.
We have implemented these different schemes and shown that they lead to consistent results
after continuum extrapolation and conversion to $\msbar$. These results are, however,
inconsistent with those obtained using the intermediate RI-MOM scheme.

Although the theoretical advantages of the RI-SMOM schemes - as compared to RI-MOM - have been
known for a long time, we have provided further numerical evidences in the context BSM kaon mixing: 
\bi
\item No pole subtraction is required.
\item The chirally-forbidden matrix elements are largely suppressed.
\item The Z and conversion matrices are closer to the identity matrix; 
  the scale-evolution between 2 and 3 GeV is relatively close to the perturbative prediction
  (known at next-to-leading order). 
\ei

On the other hand, in the RI-MOM scheme the effects of chiral symmetry breaking can be large even at large momentum,
and a procedure must be used to remove infrared contributions that dominate some vertex functions in the chiral limit.
We investigated the effect of different subtraction procedures in our
RI-MOM data and found some dependence on the procedure, which may be at
least partly responsible for the discrepancies in $O_4$ and $O_5$.
These effects are particularly important in the $(S+P)$ and $(S-P)$ channels. 
We have shown that the RI-SMOM procedure is superior because the unwanted infrared behaviour is nearly completely
suppressed (and has better pertubative behaviour).

Our study indicates these discrepancies in $O_4$ and $O_5$ could be due to a conspiracy of systematic errors in
the RI-MOM scheme, the dominant ones being the infrared contamination and the truncation error of the perturbative
series in the matching to $\msbar$ (as these operators have rather large anomalous dimension).

In other to have a better control on the physical point extrapolation, we are currently
investigating the effects of including physical pion-mass ensembles and a finer lattice
spacing. Our preliminary analysis~\cite{Boyle:2017wkz,JuliaLatt17}
shows that our results are stable and we hope to decrease
the uncertainties on the BSM matrix elements by at least a factor of two.
We are also investigating a strategy to run through the charm threshold with $n_f=2+1+1$
flavours~\cite{Frison:2014esa,Frison:2015grl}.

\section*{Acknowledgements}
N.G. is supported by the Leverhulme Research grant RPG-2014-118, 
RJH by by the Natural Sciences and Engineering Research Council of Canada.
C.L. is supported in part by US DOE Contract \#AC-02-98CH10886(BNL) and in part through a DOE Office of Science Early Career Award.
Simulations were performed on the on the STFC funded “DiRAC” BG/Q system in the Advanced Computing Facility at the University of Edinburgh.
We thank our colleagues of the RBC and UKQCD collaborations.


\section{Appendices}
\subsection{Renormalisation of the Bag parameters}

By solving numerically Eq.\ref{eq:Z_condscheme}, we obtained the
NPR matrix $Z_{ij}/Z_V^2$, such that the first element $i=j=1$
corresponds to $Z_{B_K}$.
However, the vacuum saturation approximation of the BSM matrix elements
involve the pseudo-scalar density rather than the axial current.
Therefore we compute the BSM renormalisation matrix
in the following way
\be
\label{eq:ZB}
{\mathfrak Z}_{ij}^{(\cal A)} 
= \frac{Z_V^2}{Z_S^{({\cal B})}}
 \frac{ Z_{ij} ^{({\cal A},{\cal B})} }{Z_V^2} \;,
\ee 
where we used the fact that $Z_S=Z_P$ in the chiral limit.
  The BSM bag parameters are then renormalised by
\be
B_i^{({\cal A})} = {\mathfrak Z}^{({\cal A})}_{ij} B_j^{{bare }} \;,  \qquad i,j\ge2\;.
\ee
For clarity, we note that Eq.(\ref{eq:ZB}) is equivalent to imposing the renormalisation condition
\be
{{\mathfrak Z}_{ij}}^{ ( {\cal A} )}  (\mu,a )
\left[ 
\frac{ P_k^{\cal A}\left[ \Pi_j^{bare}(a,p^2)\right] }
{( P_S\left[ \Pi_S^{bare}(a,p^2)\right] )^2}
\right]_{p^2=\mu^2} = 
\frac {F_{ik}^{\cal A}}{F_S^2} \;.
\ee

The factor $Z_S$ in Eq.(\ref{eq:ZB}) is computed through
\be
\label{eq:ZS}
Z_S^{({\cal A})} = Z_V 
\times \frac {P_V^{({\cal A})}\[\Pi_V\]}{F_V^{{\cal A}}} 
\times \frac{F_S}{P_S\[\Pi_S\]} \;, \qquad {\rm where } \;{\cal A}\in \[ \gamma_\mu, \s{q} \]\;.
\ee
Note that the choice ${\cal A}=\s{q}$ (resp. $\gamma_{\mu}$)
corresponds to the scheme called ${\rm RI-SMOM}$  (resp. ${\rm RI-SMOM}_{\gamma_\mu}$) in~\cite{Sturm:2009kb}. 
We find
\begin{equation}
 \begin{aligned}
  Z_S^{(\gamma_\mu)}(3\;\text{GeV},a_{\mathbf{24}}) = 0.6563(6), \quad
  Z_S^{(\s{q})}(3\;\text{GeV},a_{\mathbf{24}}     ) = 0.6945(5)\;,\\
  Z_S^{(\gamma_\mu)}(3\;\text{GeV},a_{\mathbf{32}}) = 0.6585(6),\quad
  Z_S^{(\s{q})}(3\;\text{GeV},a_{\mathbf{32}}     )= 0.6940(6)\;.
 \end{aligned}
\end{equation}

In order to match our results to $\msbar$,
we also need the conversion factor for $Z_S $, 
as can seen from Eq.~\ref{eq:ZB}. The one-loop coefficient can extracted from~\cite{Sturm:2009kb} 
whereas the next-to-next-to-leading-order corrections are known from~{\cite{Gorbahn:2010bf,Almeida:2010ns}.
  Here we follow~\cite{Arthur:2012opa} and with
  $\alpha_s(3{\rm GeV})=0.24544$, we find 
  \be
  \begin{aligned}
    \label{eq:convertZsgamma}
    R_S^{\msbar \leftarrow (\gamma_\mu) }(3\;\text{GeV}) &=& 1.05259 \\
    R_S^{\msbar \leftarrow  (\s{q}) }(3\;\text{GeV}) &=& 1.01664
    \end{aligned}
  \ee
up to $\alpha_s^3$ terms.
Putting everything together, we find
\begin{equation}
 \begin{aligned}
  Z_S^{\msbar \leftarrow (\gamma_\mu)}(3\;\text{GeV},a_{\mathbf{24}}) = 0.6908(6), \quad
  Z_S^{\msbar \leftarrow (\s{q})}(3\;\text{GeV},a_{\mathbf{24}}     ) = 0.7060(6)\;,\\
  Z_S^{\msbar \leftarrow (\gamma_\mu)}(3\;\text{GeV},a_{\mathbf{32}}) = 0.6931(6),\quad
  Z_S^{\msbar \leftarrow (\s{q})}(3\;\text{GeV},a_{\mathbf{32}}     )= 0.7056(6)\;.
\end{aligned}
\end{equation}

For the reader's convenience, we report the values used in this analysis in table~\ref{tab:Lambdabil},
\begin{table}[t]
  \centering
  \begin{tabular}{ | c | c  c | }
    \toprule
    & $a_{\mathbf{24}}$ & $a_{\mathbf{32}}$ \\ 
     \hline 
     ${F_V^{{(\gamma_\mu)}}} / \Lambda_V^{({\gamma_\mu})} (3\;\text{GeV},a) $ & $0.94952(17)$   & $0.96339(16)$ \\
     ${F_V^{(\s{q})}} / \Lambda_V^{(\s{q})}           (3\;\text{GeV},a)$  & $0.89737(19)$   & $0.91410(37)$ \\
     ${F_S}/{\Lambda_S}                           (3\;\text{GeV},a) $ & $0.8743(7)$  &  $0.8526(8)$ \\
     $Z_V(a)$                  & 0.71273(26) & 0.74404(181) \\
    \botrule
      \end{tabular}
  \caption{Values used for the renormalisation factors of the bilinear needed for the BSM bag parameters.
    }
  \label{tab:Lambdabil}
\end{table}

For completeness, we also give $Z_S$ at $2 \rm GeV$. 
In the RI-SMOM schemes, we find
\begin{equation}
 \begin{aligned}
  Z_S^{(\gamma_\mu)}(2\;\text{GeV},a_{\mathbf{24}}) = 0.5974(9) , \quad
  Z_S^{(\s{q})}(2\;\text{GeV},a_{\mathbf{24}}     ) = 0.6423(8)\;,\\
  Z_S^{(\gamma_\mu)}(2\;\text{GeV},a_{\mathbf{32}}) = 0.6585(6),\quad
  Z_S^{(\s{q})}(2\;\text{GeV},a_{\mathbf{32}}     )= 0.6940(6)\;.
 \end{aligned}
\end{equation}
With $\alpha_s(2 {\rm Gev})=0.2961$, the conversion factors read
\be
\begin{aligned}
R_S^{\msbar \leftarrow (\gamma_\mu) }(2\;\text{GeV}) &= 1.06689 \;,\\
R_S^{\msbar \leftarrow  (\s{q}) }(2\;\text{GeV})   &=  1.02107 \;,
\end{aligned}
\ee
therefore 
\begin{equation}
 \begin{aligned}
  Z_S^{\msbar \leftarrow (\gamma_\mu)}(2\;\text{GeV},a_{\mathbf{24}}) = 0.5924(14) , \quad
  Z_S^{\msbar \leftarrow (\s{q})}(2\;\text{GeV},a_{\mathbf{24}}     ) = 0.6372(17)\;,\\
  Z_S^{\msbar \leftarrow (\gamma_\mu)}(2\;\text{GeV},a_{\mathbf{32}}) = 0.6320(15),\quad
  Z_S^{\msbar \leftarrow (\s{q})}(2\;\text{GeV},a_{\mathbf{32}}     )= 0.6506(17)\;.
\end{aligned}
\end{equation}
 
\subsection{Matching factors between the RI-SMOM schemes and \texorpdfstring{$\overline{\text{MS}}$}{MS-bar}}
\label{appendix:msbarmatch}

The conversion between the RI-SMOM schemes and $\msbar$ (of \cite{Buras:2000if})
is given at one-loop order. 
We define
(we chose a negative sign for historical reasons)
\be
R^{\msbar \leftarrow \rm scheme} = 1 - \frac{\alpha_s}{4\pi} \Delta r^{\msbar \leftarrow \rm scheme} \;.
\ee
In the following expressions, the constant $C_0$ is $C_0=\frac{2 \psi ^{(1)}\left(\frac{1}{3}\right)}{3}-\left(\frac{2 \pi }{3}\right)^2$,
where $\psi$ is the PolyGamma function, $N$ is the number of colors and $\xi$ the usual gauge parameter
(the non-perturbative Z-factors have been computed in the Landau gauge, $\xi=0$).
Note that the  coefficients for the $(27,1)$ and the $(8,8)$  operators were already known
or could be derived from~\cite{Aoki:2010pe,Lehner:2011fz}.
The others are new, they have been computed for this work.
First we have the matching factors for the $(\gamma_\mu,\gamma_\mu)$ scheme,
for the $(27,1)$ operator we have 
\be
\begin{aligned}
\Delta r_{11}^{\overline{\text{MS}}\leftarrow (\gamma_\mu, \gamma_\mu)} 
&= -\frac{8}{N}+\frac{12 \log (2)}{N}+8-12 \log (2)
+ \xi\left (-\frac{C_0}{2N} + \frac{C_0}{2} - \frac{1}{2N} + \frac{4 \log (2)}{N} + \frac{1}{2} - 4 \log (2)\right)\;,
\end{aligned}
\ee
For the $(8,8)$ doublet:
\be
\begin{aligned}
\Delta r_{22}^{\overline{\text{MS}}\leftarrow (\gamma_\mu, \gamma_\mu)} 
&=-\frac{3 C_0}{2 N}+\frac{2}{N}+\frac{2 \log (2)}{N}
+\xi\left( -\frac{C_0}{2N} + \frac{1}{2N}  + \frac{2 \log (2)}{N} \right)\;,
 \\
\Delta r_{23}^{\overline{\text{MS}}\leftarrow (\gamma_\mu, \gamma_\mu)} 
&=-3 C_0+4+4 \log (2)
+\xi\left(-C_0+1+4\log (2)\right)\;,
\\
\Delta r_{32}^{\overline{\text{MS}}\leftarrow (\gamma_\mu, \gamma_\mu)} 
&=\log (2)-\frac{3}{2}
+\xi\left( \log (2) -\frac{C_0}{4} \right)\;,
\\
\Delta r_{33}^{\overline{\text{MS}}\leftarrow (\gamma_\mu, \gamma_\mu)} 
&=\frac{3 C_0 N}{2}-\frac{3 C_0}{2 N}-5 N+\frac{2}{N}+\frac{2 \log(2)}{N}
+ \xi\left (- \frac{C_0}{2N} - \frac{N}{2}  + \frac{1}{2N} + \frac{2 \log (2)}{N} \right)\;,
\end{aligned}
\ee
and for the $(6,\bar 6)$ doublet:
\be
\begin{aligned}
\Delta r_{44}^{\overline{\text{MS}}\leftarrow (\gamma_\mu, \gamma_\mu)} 
&=\frac{3 C_0 N}{2}-\frac{3 C_0}{2 N}-\frac{3 C_0}{4}-5 N+\frac{5}{N}+\frac{2 \log (2)}{N}+7-4 \log (2) \\
& + \xi\left (-\frac{C_0}{2N} - \frac{C_0}{4} - \frac{N}{2} + \frac{1}{2N} + \frac{2 \log (2)}{N}  +\frac{1}{2}  \right)\;,
\\
\Delta r_{45}^{\overline{\text{MS}}\leftarrow (\gamma_\mu, \gamma_\mu)} 
&=4 \left(\frac{C_0}{8 \text{N}}-\frac{C_0}{16}-\frac{7}{6 N}+\frac{5 \log (2)}{6 N}+\frac{7}{12}-\frac{2 \log (2)}{3}\right)
\\&+4\xi\left (\frac{C_0}{16}  - \frac{1}{12N} + \frac{\log (2)}{6N}  +\frac{1}{24} - \frac{\log (2)}{3} \right)\;,
\\
\Delta r_{54}^{\overline{\text{MS}}\leftarrow (\gamma_\mu, \gamma_\mu)} 
&=\frac{1}{4} \left(\frac{6 C_0}{N}+9 C_0-\frac{16}{N}+\frac{40 \log (2)}{N}+4-32 \log (2)\right)
\\+&\frac{1}{4}\xi\left (3C_0 - \frac{4}{N} + \frac{8\log (2)}{N} -2 -16\log(2) \right)\;,
\\
\Delta r_{55}^{\overline{\text{MS}}\leftarrow (\gamma_\mu, \gamma_\mu)} 
&=-\frac{C_0 N}{2}-\frac{C_0}{2 N}-\frac{C_0}{4}+\frac{N}{3}-\frac{7}{3 N}+\frac{26 \log (2)}{3 N}+3-\frac{28 \log (2)}{3}
\\+&\xi\left(-\frac{C_0}{2N} + \frac{C_0}{4} +\frac{N}{6} -\frac{1}{6N} + \frac{10\log(2)}{3N} + \frac{1}{2} - \frac{8\log(2)}{3} \right)
\;.
\end{aligned}
\ee
Secondly are the matching factors to $\overline{\text{MS}}$ of the ($\slashed{q},\slashed{q}$) scheme,
the $(27,1)$ is 
\be
\begin{aligned}
\Delta r_{11}^{\overline{\text{MS}}\leftarrow (\slashed{q},\slashed{q})} 
&= -\frac{9}{N}+\frac{12 \log (2)}{N}+9-12 \log (2)
+ \xi\left (-\frac{C_0}{N} + C_0 + \frac{4 \log (2)}{N}  - 4 \log (2)\right)\;.
\end{aligned}
\ee
For the $(8,8)$ doublet we have 
\be
\begin{aligned}
\Delta r_{22}^{\overline{\text{MS}}\leftarrow (\slashed{q},\slashed{q})} 
&= -\frac{3 C_0}{2 N}+\frac{2}{N}+\frac{2 \log (2)}{N}
+ \xi\left (-\frac{C_0}{2N} +\frac{1}{2N} + \frac{2 \log (2)}{N} \right)\;,
\\
\Delta r_{23}^{\overline{\text{MS}}\leftarrow (\slashed{q},\slashed{q})} 
&= -3 C_0+4+4 \log (2)
+ \xi\left (-C_0 + 1 +4 \log (2)  \right)\;,
\\
\Delta r_{32}^{\overline{\text{MS}}\leftarrow (\slashed{q},\slashed{q})} 
&= \log (2)-1
+ \xi\left(\log(2) - \frac{1}{4} \right)\;,
\\
\Delta r_{33}^{\overline{\text{MS}}\leftarrow (\slashed{q},\slashed{q})} 
&= \frac{3 C_0 N}{2}-\frac{3 C_0}{2 N}-4 N+\frac{2}{N}+\frac{2 \log (2)}{N}
+\xi\left (\frac{C_0 N}{2} - \frac{C_0}{2N} -N +\frac{1}{2N} + \frac{2 \log (2)}{N} \right)\;,
\end{aligned}
\ee
and for the $(6,\bar 6)$ doublet:
\be
\begin{aligned}
\Delta r_{44}^{\overline{\text{MS}}\leftarrow (\slashed{q},\slashed{q})} 
&= \frac{3 C_0 N}{2}-\frac{C_0}{N}-\frac{C_0}{4}- 4 N +\frac{3}{N}+\frac{2 \log (2)}{N}+6- 4\log (2)
\\&+  \xi\left (\frac{C_0N}{2} -\frac{3C_0}{2N} -\frac{3C_0}{4} - N +\frac{2}{N} + \frac{2 \log (2)}{N} +\frac{3}{2}\right)\;,
\\
\Delta r_{45}^{\overline{\text{MS}}\leftarrow (\slashed{q},\slashed{q})} 
&= 4 \left(\frac{C_0}{12 N}-\frac{5C_0}{48}-\frac{13}{12 N}+\frac{5 \log (2)}{6 N}+\frac{2}{3}-\frac{2 \log (2)}{3}\right)
\\&+4 \xi\left (\frac{C_0}{24N}+\frac{5C_0}{48} -\frac{1}{6N} + \frac{\log (2)}{6N} - \frac{1}{24} -\frac{\log (2)}{3} \right)\;,
\\
\Delta r_{54}^{\overline{\text{MS}}\leftarrow (\slashed{q},\slashed{q})} 
&= \frac{1}{4} \left(4C_0 N + \frac{4 C_0}{N} +  11C_0 - 8N -\frac{12}{N} +\frac{40 \log (2)}{N} -32 \log (2)\right)
\\&+\frac{1}{4} \xi\left ( -4 C_0N + \frac{2C_0}{N}+C_0 + 8N -\frac{8}{N} + \frac{8\log (2)}{N} + 2 -16\log (2) \right)\;,
\\
\Delta r_{55}^{\overline{\text{MS}}\leftarrow (\slashed{q},\slashed{q})} 
&= -\frac{5C_0N}{6}  -\frac{C_0}{3 N}-\frac{5C_0}{12} +2N - \frac{11}{3N} 
+\frac{26 \log (2)}{3 N}+\frac{10}{3}-\frac{28 \log (2)}{3}
\\&+ \xi\left ( \frac{5C_0N}{6}-\frac{7C_0}{6N} +\frac{5C_0}{12} -N + \frac{2}{3N} + \frac{10\log (2)}{3N} +\frac{1}{6} -\frac{8\log (2)}{3} \right)
\;.
\end{aligned}
\ee

The factors for other schemes can be obtained trivially if ones knows for example the matching coefficients for the SM operator
\be
\begin{aligned}
  \Delta r_{11}^{\overline{\text{MS}}\leftarrow (\gamma_\mu, \s{q})}
  &= N-\frac{9}{N}+\frac{12 \log (2)}{N}+8-12 \log (2)
  \\ &+ \xi\left ( \frac{C_0N}{2} -\frac{C_0}{N} + \frac{C_0}{2} - \frac{N}{2} + \frac{4 \log (2)}{N} + \frac{1}{2} - 4 \log (2)\right)\;,
\end{aligned}
\ee
from which we can derive
\be
\begin{aligned}
  \Delta r_{11}^{\overline{\text{MS}}\leftarrow (\s{q}, \gamma_\mu)} 
  &= -N-\frac{8}{N}+\frac{12 \log (2)}{N}+9-12 \log (2)
  \\ &+ \xi\left ( - \frac{C_0N}{2} -\frac{C_0}{2N} + C_0 + \frac{N}{2} - \frac{1}{2N} + \frac{4 \log (2)}{N}  - 4 \log (2)\right)\;.
\end{aligned}
\ee
Although they can be obtained from the the previous equations, for completeness
we also list the other matching factors for the $(\slashed{q},\gamma_\mu)$ sheme.
For the $(8,8)$ doublet:
\begin{equation}
\begin{aligned}
\Delta r^{\overline{\text{MS}}\leftarrow(\slashed{q},\gamma_\mu)}_{22} &= -\frac{3C_0}{2N}-N+\frac{3}{N}+\frac{2\log(2)}{N}+\xi\left( -\frac{C_0 N}{2} + \frac{N}{2}+\frac{2\log(2)}{N}\right),\\
\Delta r^{\overline{\text{MS}}\leftarrow(\slashed{q},\gamma_\mu)}_{23} &= -3C_0 + 4 + 4\log(2)+\xi\left( -C_0 + 1 + 4\log(2) \right),\\
\Delta r^{\overline{\text{MS}}\leftarrow(\slashed{q},\gamma_\mu)}_{32} &= -1 + \log(2) + \xi\left(\log(2)-\frac{1}{4}\right),\\
\Delta r^{\overline{\text{MS}}\leftarrow(\slashed{q},\gamma_\mu)}_{33} &= \frac{3C_0 N}{2}-\frac{3C_0}{2N}-5N+\frac{3}{N}+\frac{2\log(2)}{N}+\xi\left( \frac{2\log(2)}{N}- \frac{N}{2}\right),\\
\end{aligned}
\end{equation}
and finally for the $(6,\bar{6})$ doublet:
\begin{equation}
\begin{aligned}
\Delta r^{\overline{\text{MS}}\leftarrow(\slashed{q},\gamma_\mu)}_{44} &= 
\frac{3C_0 N}{2}-\frac{C_0}{N}-\frac{C_0}{4}-5N+\frac{4}{N}+\frac{2\log(2)}{N}+6-4\log(2)\\
&+\xi\left( -\frac{C_0}{N} - \frac{3C_0}{4} - \frac{N}{2}+\frac{3}{2N}+\frac{2\log(2)}{N}+\frac{3}{2}\right),\\
\Delta r^{\overline{\text{MS}}\leftarrow(\slashed{q},\gamma_\mu)}_{45} &= 
4\left( \frac{C_0}{12N} - \frac{5C_0}{48}-\frac{13}{12N}+\frac{5\log(2)}{6N}+\frac{2}{3}-\frac{2\log(2)}{3} \right) \\
&+4\xi\left( \frac{C_0}{24N} + \frac{5C_0}{48} - \frac{1}{6N} + \frac{\log(2)}{6N} - \frac{1}{24} - \frac{\log(2)}{3} \right),\\
\Delta r^{\overline{\text{MS}}\leftarrow(\slashed{q},\gamma_\mu)}_{54} &= 
\frac{1}{4}\left( 4C_0 N + \frac{4C_0}{N} +11C_0 - 8N - \frac{12}{N} + \frac{40\log(2)}{N} - 32\log(2)\right) \\
&+\frac{1}{4}\xi\left( 
-4C_0 N + \frac{2C_0}{N}+ C_0 + 8N - \frac{8}{N} + \frac{8\log(2)}{N} + 2 - 16\log(2) \right),\\
\Delta r^{\overline{\text{MS}}\leftarrow(\slashed{q},\gamma_\mu)}_{55} &= 
-\frac{5C_0 N}{6} - \frac{C_0}{3N} - \frac{5C_0}{12} + N - \frac{8}{3N} + \frac{26\log(2)}{3N} + \frac{10}{3} - \frac{28\log(2)}{3}\\
&+\xi\left( \frac{C_0 N}{3} - \frac{2 C_0}{3N} + \frac{5C_0}{12} - \frac{N}{2} + \frac{1}{6N} + \frac{10\log(2)}{3N} + \frac{1}{6} - \frac{8\log(2)}{3} \right).
\end{aligned}
\end{equation}

\subsection{Figures for the non-perturbative running}
\label{app:ssfplot}

In Fig~\ref{fig:ssf_11_2to3} we show the running between $\mu_1=2\,\GeV$ and $\mu$ 
where $\mu$ varies between $2\,\GeV$ and $3\GeV$.

\begin{figure}[htb]
\begin{center}
\begin{tabular}{cc}
\includegraphics[type=pdf,ext=.pdf,read=.pdf,width=8cm]{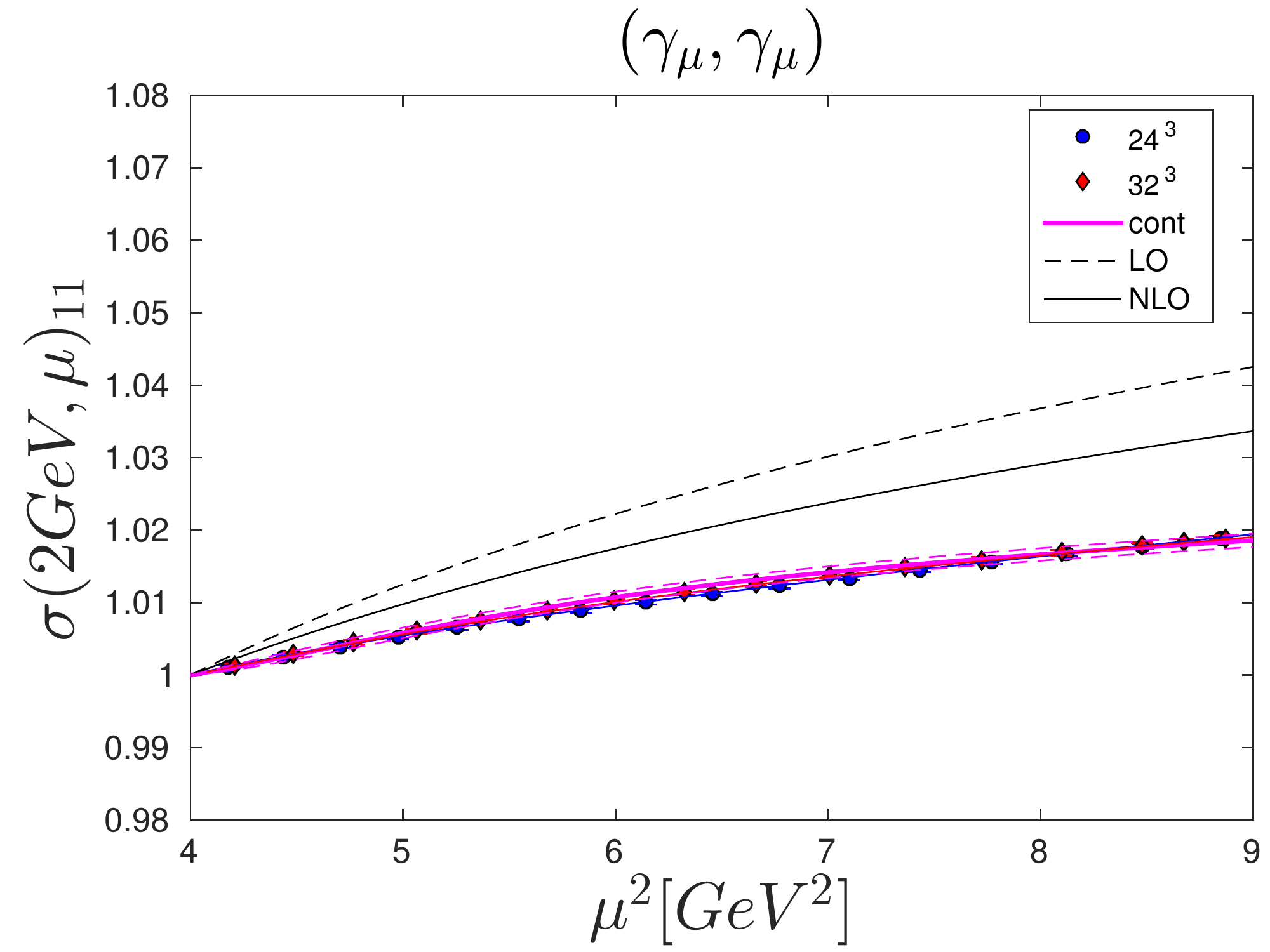} &
\includegraphics[type=pdf,ext=.pdf,read=.pdf,width=8cm]{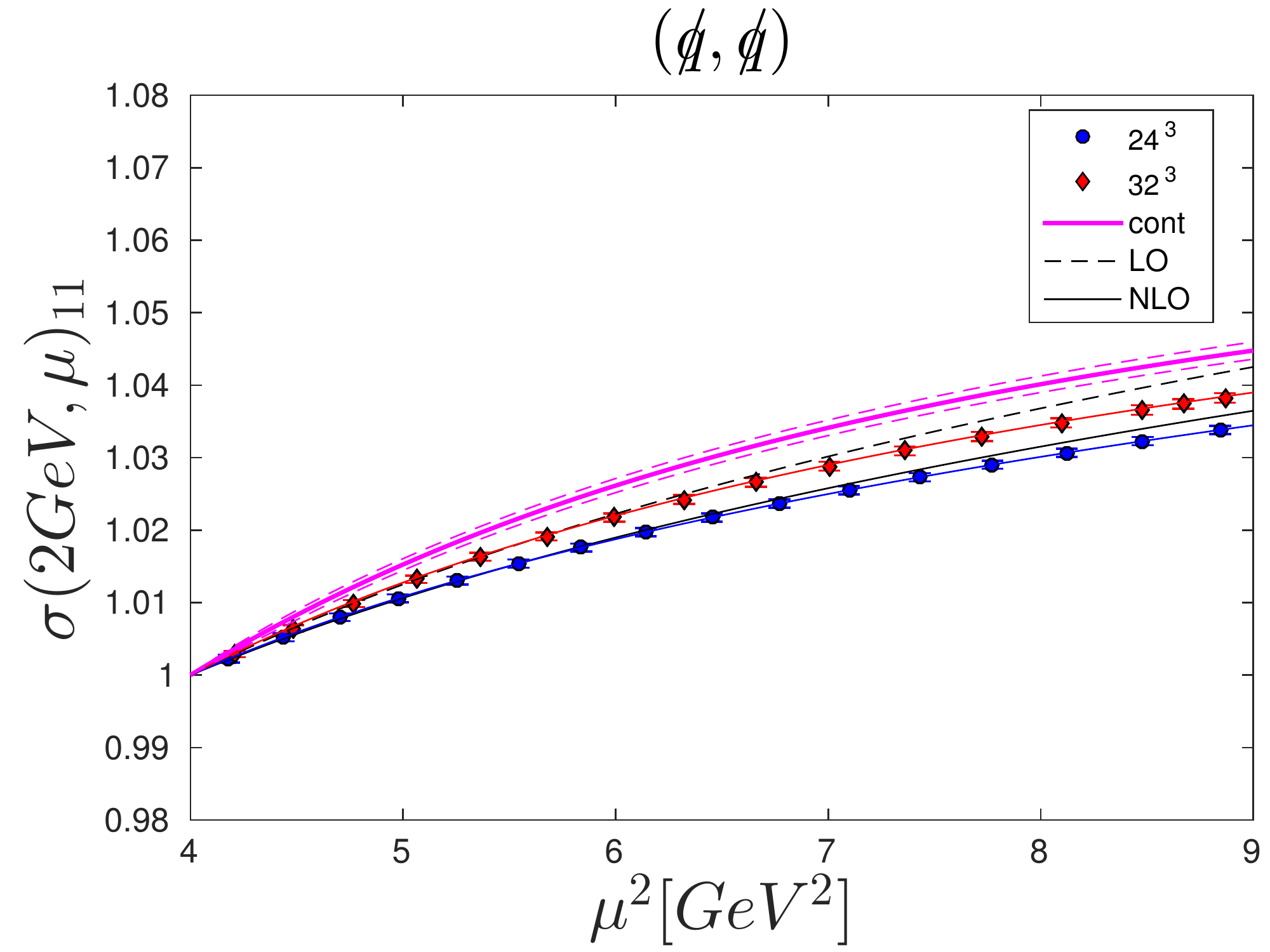}  
\end{tabular} 
\caption{Scale evolution of the $(27,1)$ operator for the various schemes;
left: $(\gamma_\mu,\gamma_\mu)$, right:  $(\s{q}, \s{q})$.
We show the non-perturbative running computed on the coarse lattice, on the fine lattice
and extrapolated to the continuum. We also compare with the perturbative prediction
at leading-order (LO) and next-to-leading-order (NLO). }
\label{fig:ssf_11_2to3}
\end{center}
\end{figure}

\begin{figure}[htb]
\begin{center}
\begin{tabular}{cc}
\includegraphics[type=pdf,ext=.pdf,read=.pdf,width=8cm]{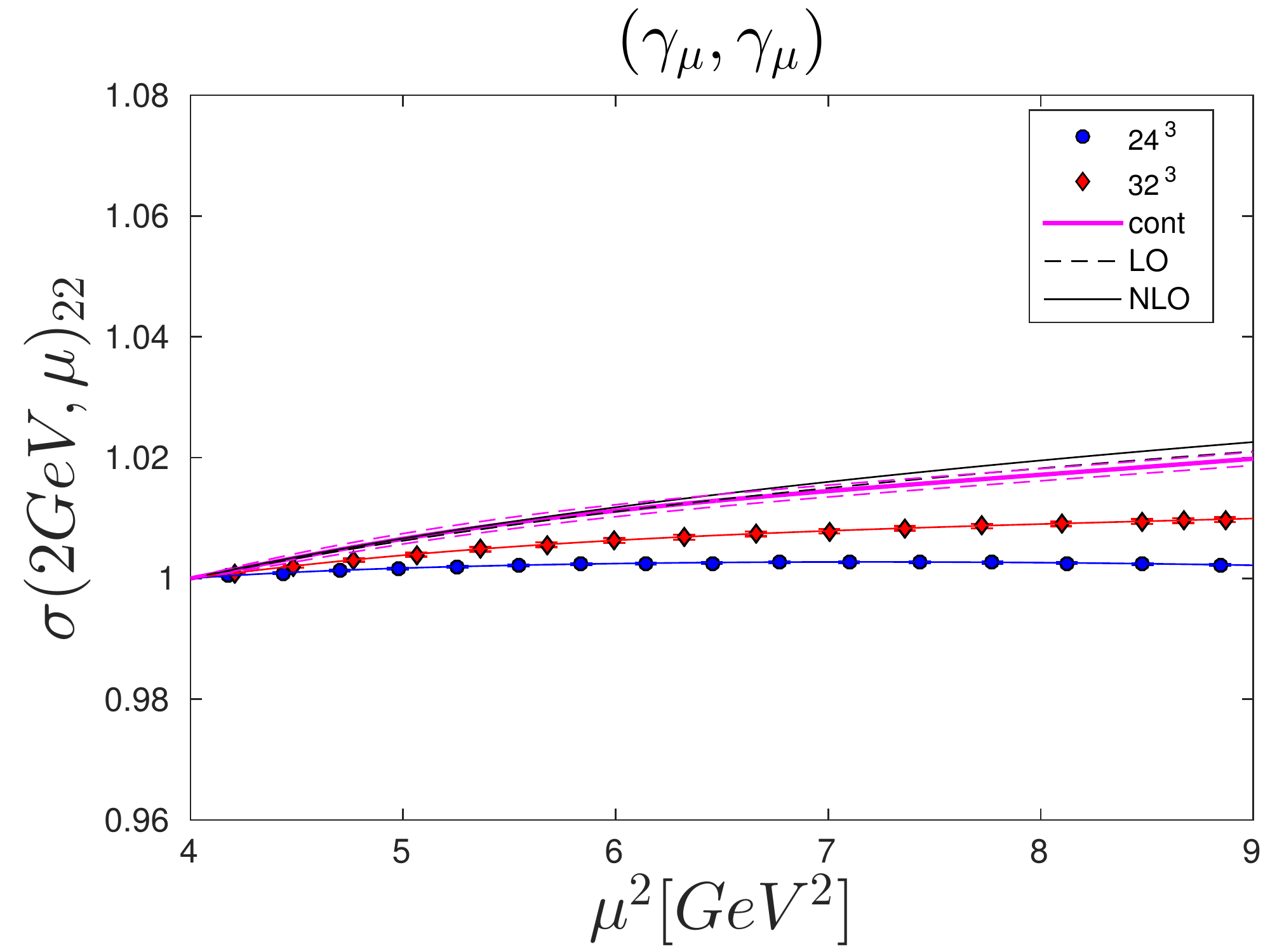} &
\includegraphics[type=pdf,ext=.pdf,read=.pdf,width=8cm]{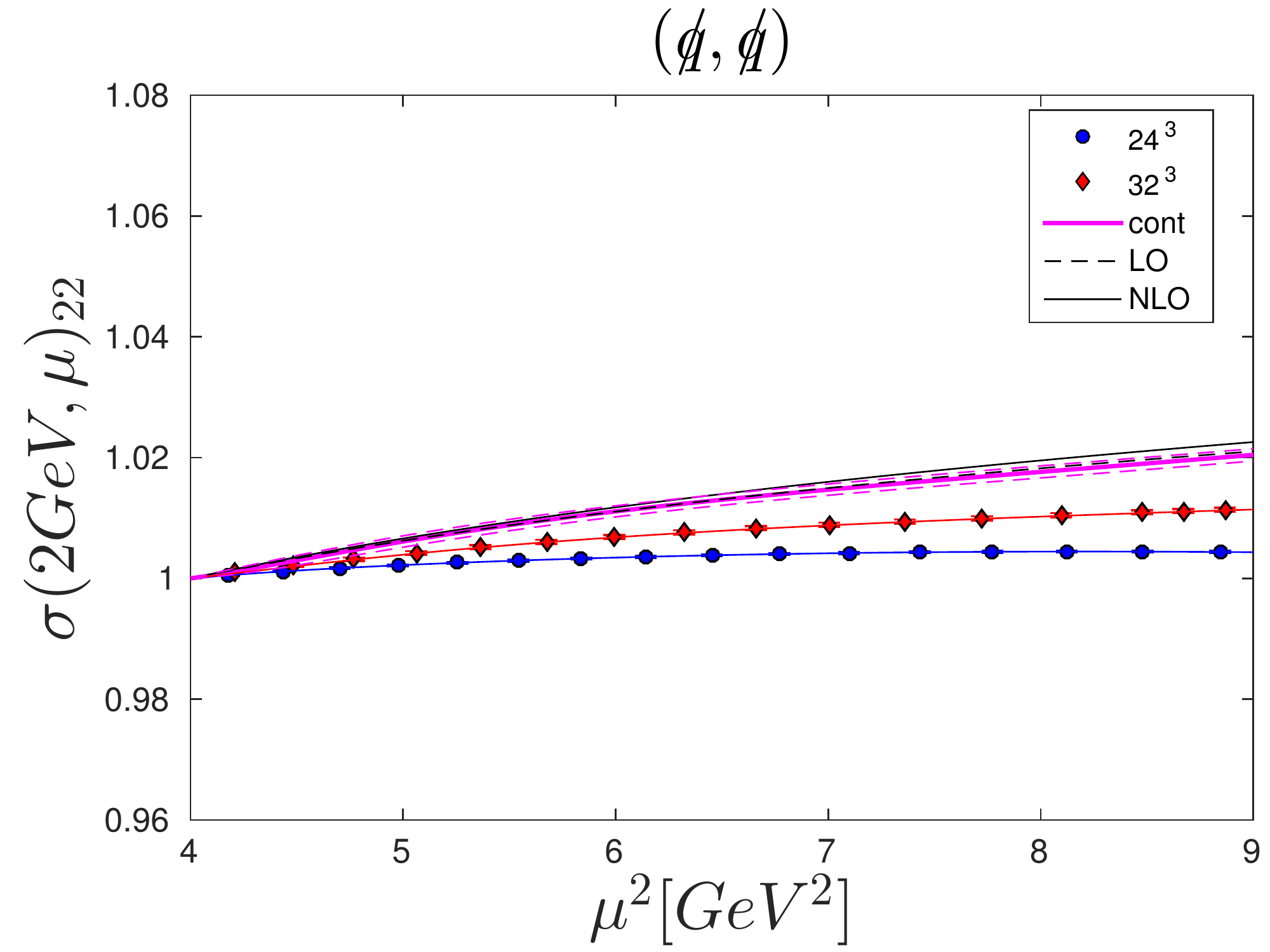} \\ 
\includegraphics[type=pdf,ext=.pdf,read=.pdf,width=8cm]{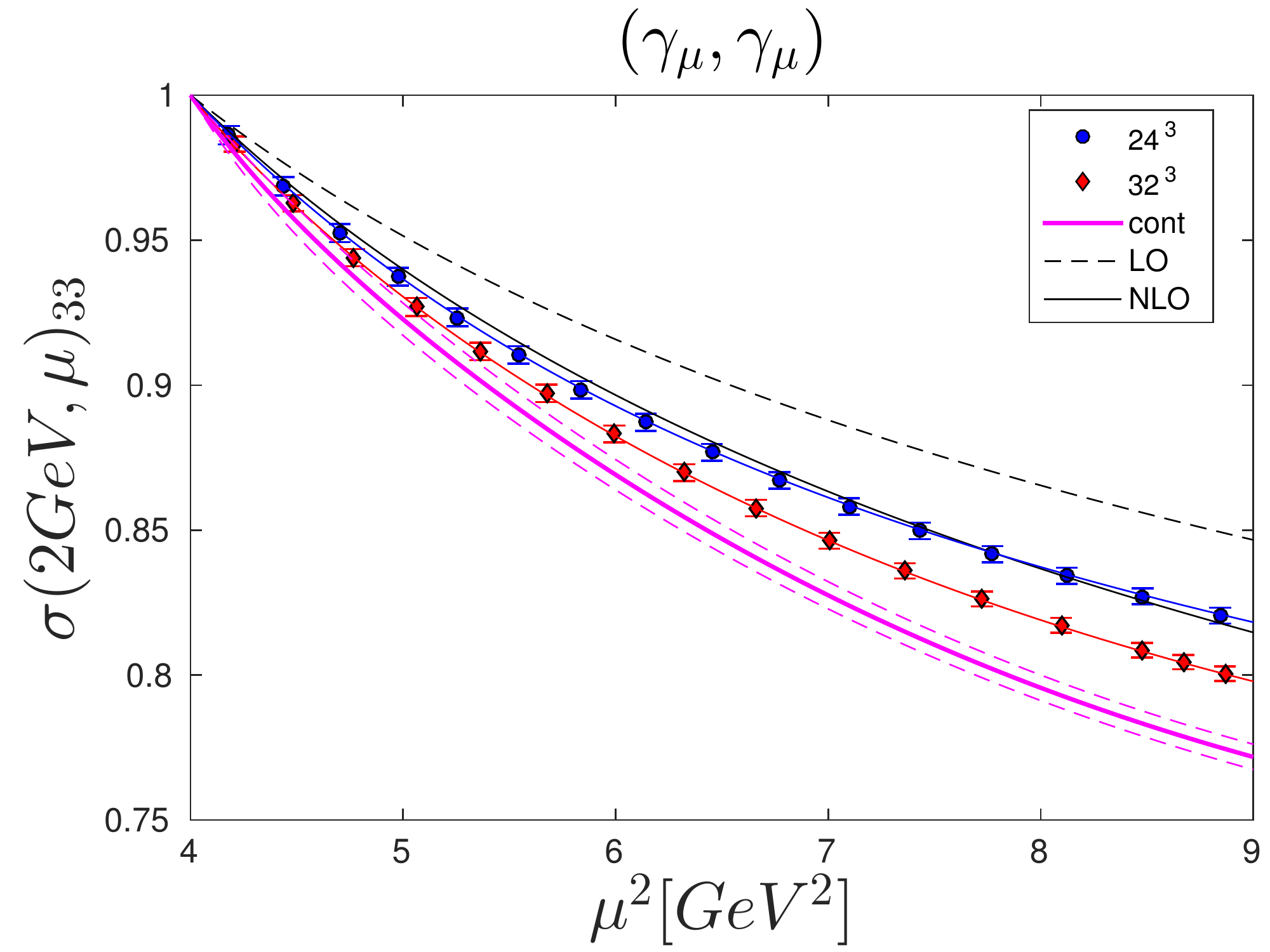} &
\includegraphics[type=pdf,ext=.pdf,read=.pdf,width=8cm]{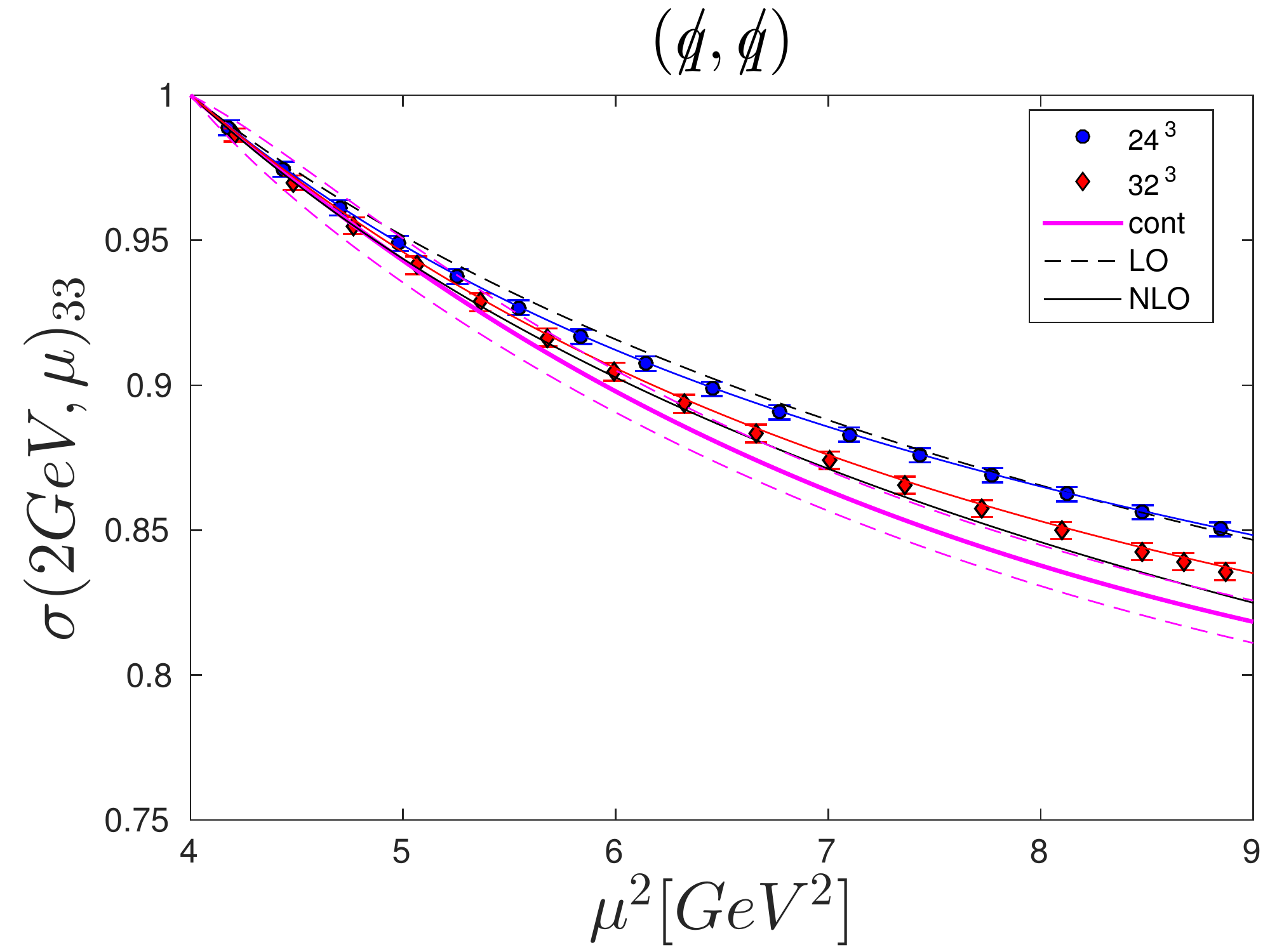}  
\end{tabular} 
\caption{Same as the previous plot for the scale evolution of the diagonal $(8,8)$ mixing matrix element
$\sigma_{22}$ and $\sigma_{33}$.}
\label{fig:ssf_22_2to3}
\end{center}
\end{figure}


\begin{figure}[htb]
\begin{center}
\begin{tabular}{cc}
\includegraphics[type=pdf,ext=.pdf,read=.pdf,width=8cm]{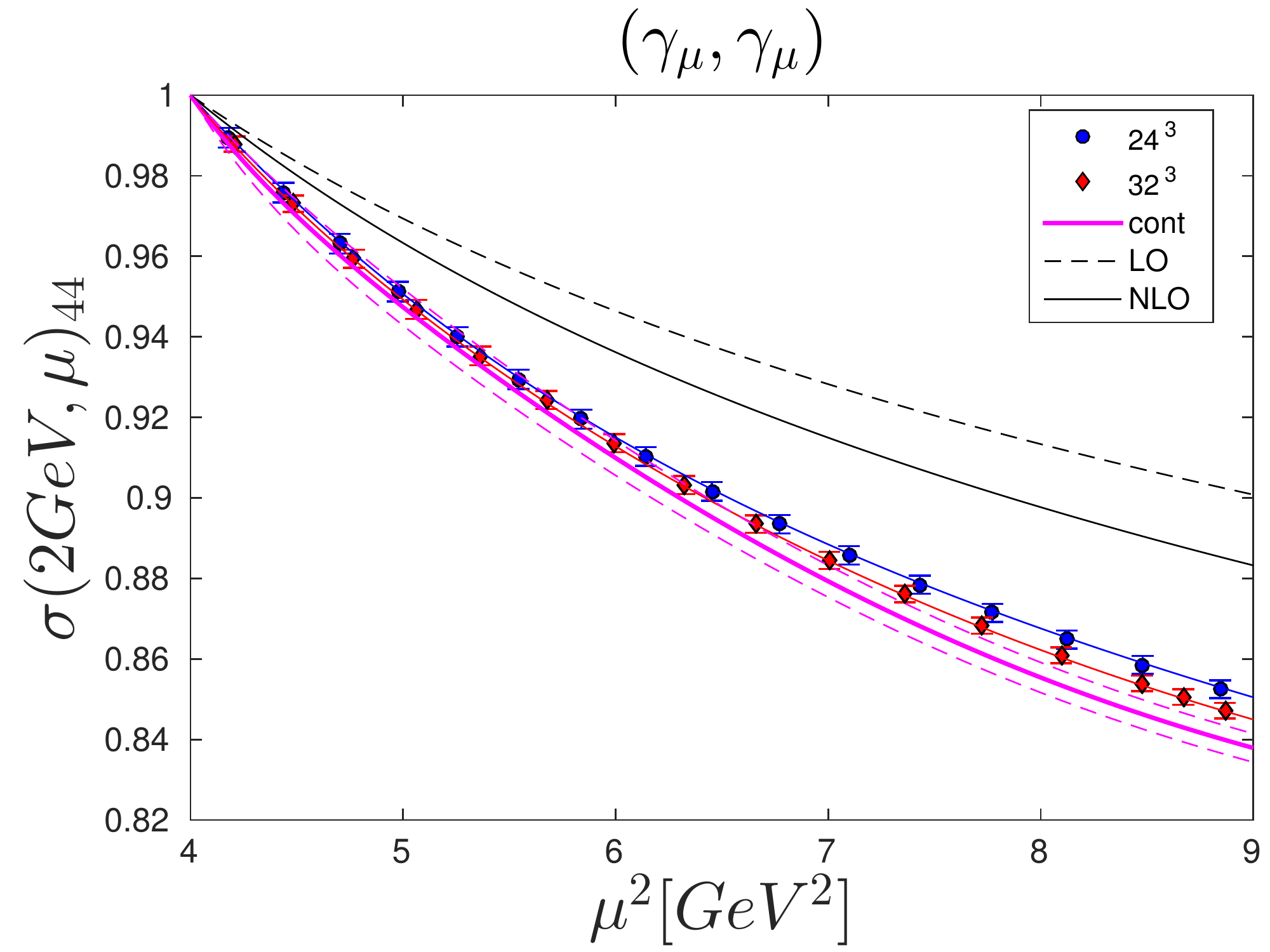} &
\includegraphics[type=pdf,ext=.pdf,read=.pdf,width=8cm]{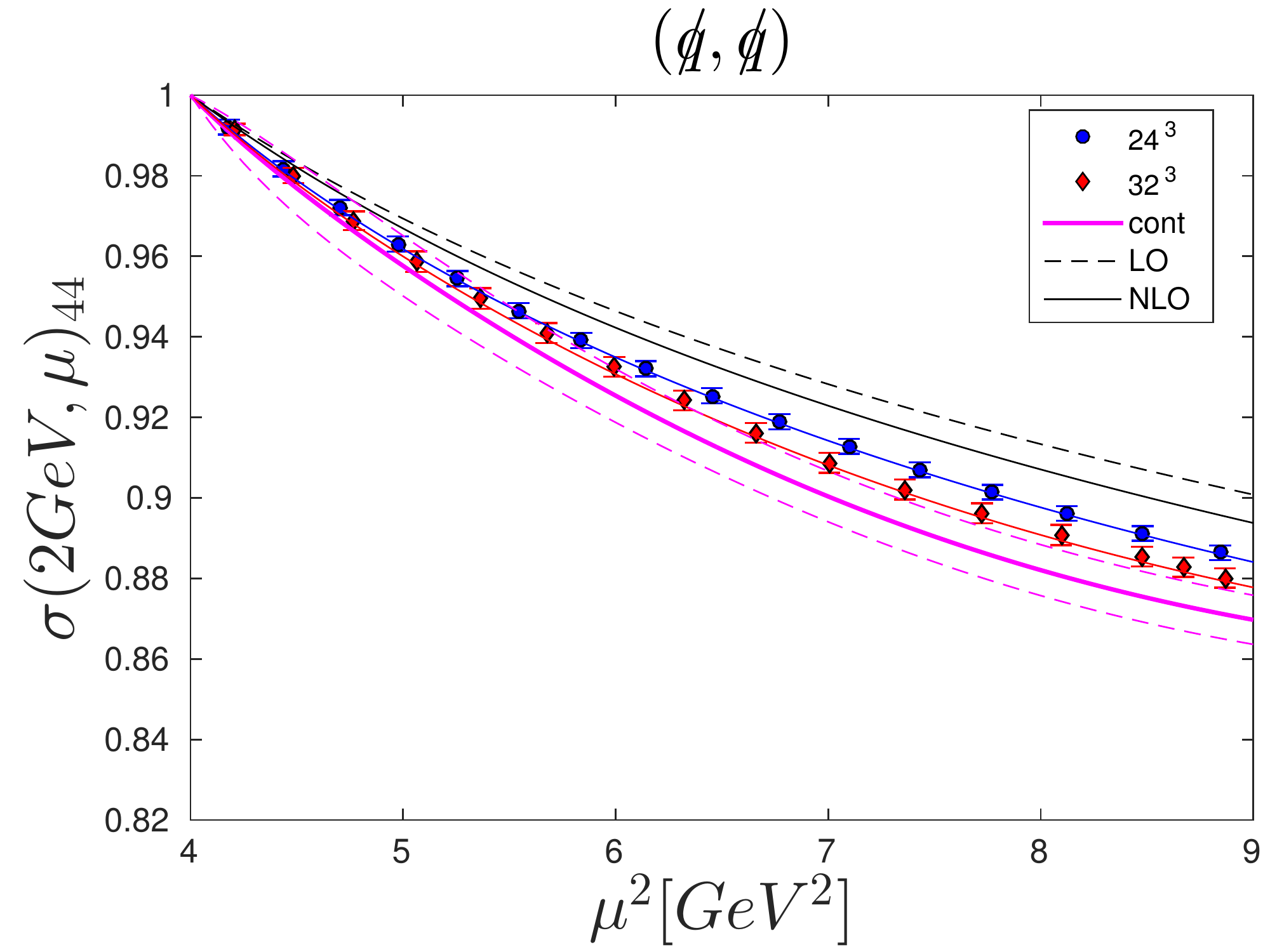} \\
\includegraphics[type=pdf,ext=.pdf,read=.pdf,width=8cm]{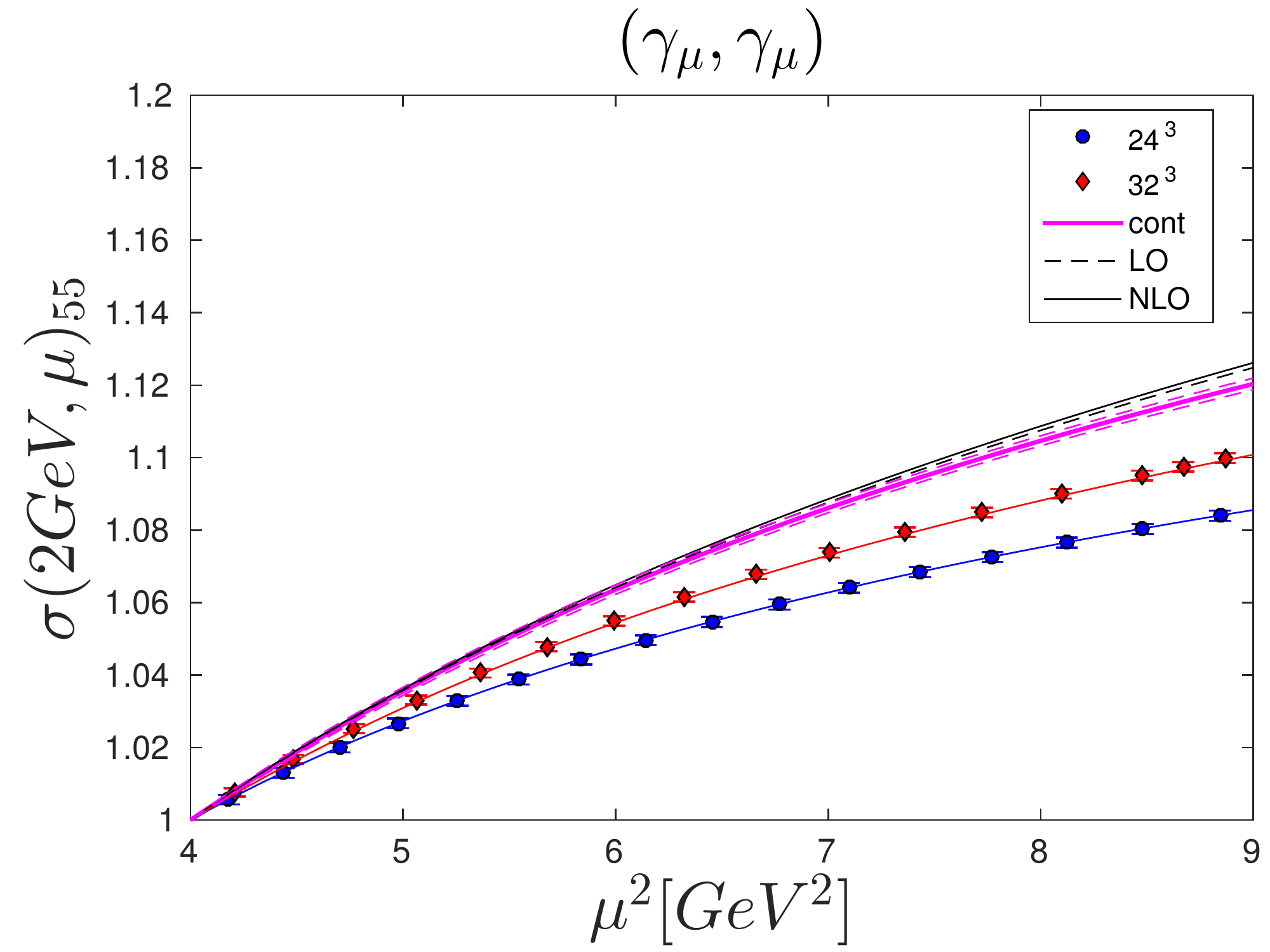} &
\includegraphics[type=pdf,ext=.pdf,read=.pdf,width=8cm]{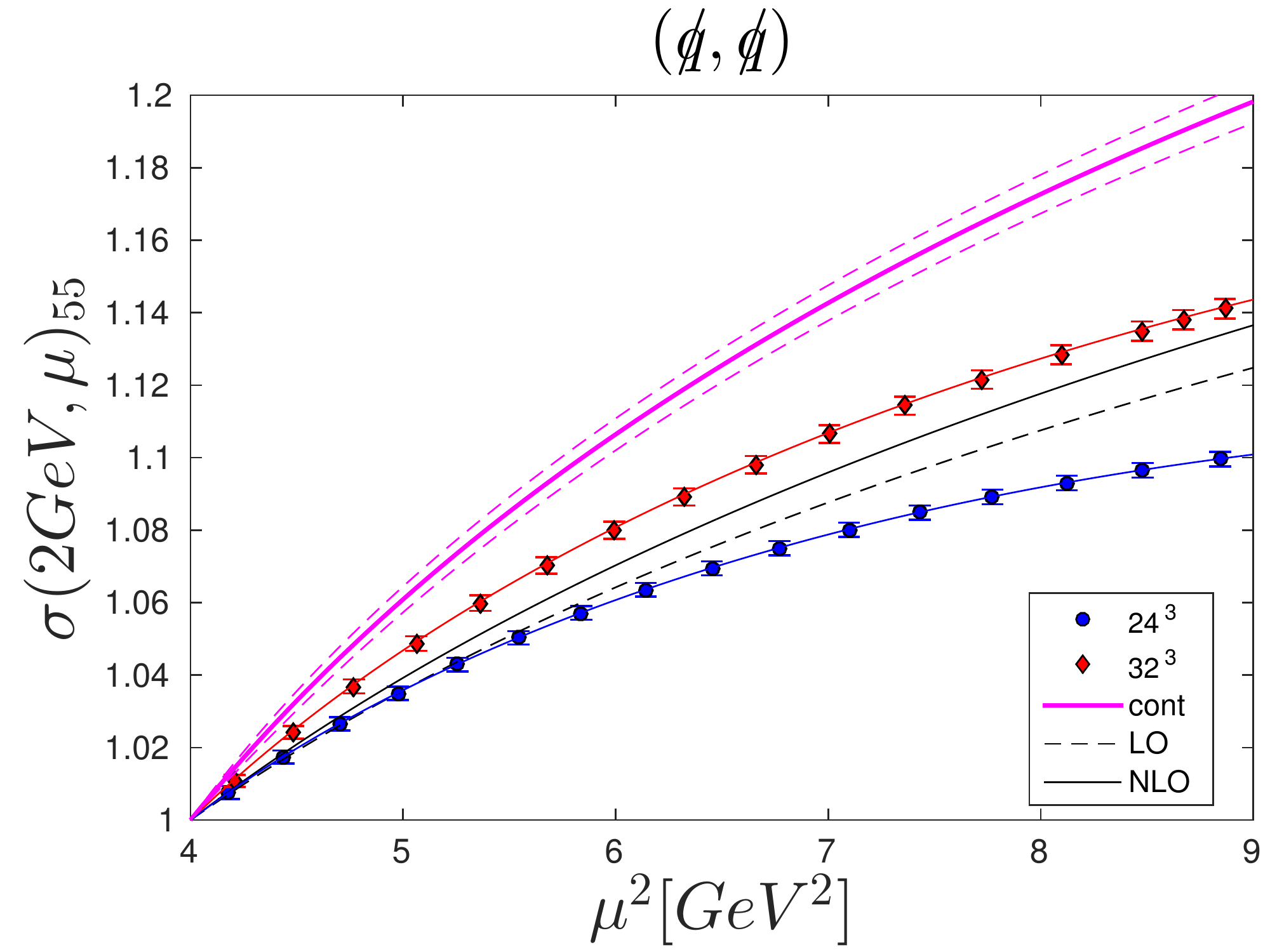} 
\end{tabular} 
\caption{Same as the previous plot for the scale evolution of the  diagonal $(6,\bar 6)$
mixing matrix element
$\sigma_{44}$ and $\sigma_{55}$.}
\label{fig:ssf_44_2to3}
\end{center}
\end{figure}

\begin{figure}[htb]
\begin{center}
\begin{tabular}{cc}
\includegraphics[type=pdf,ext=.pdf,read=.pdf,width=8cm]{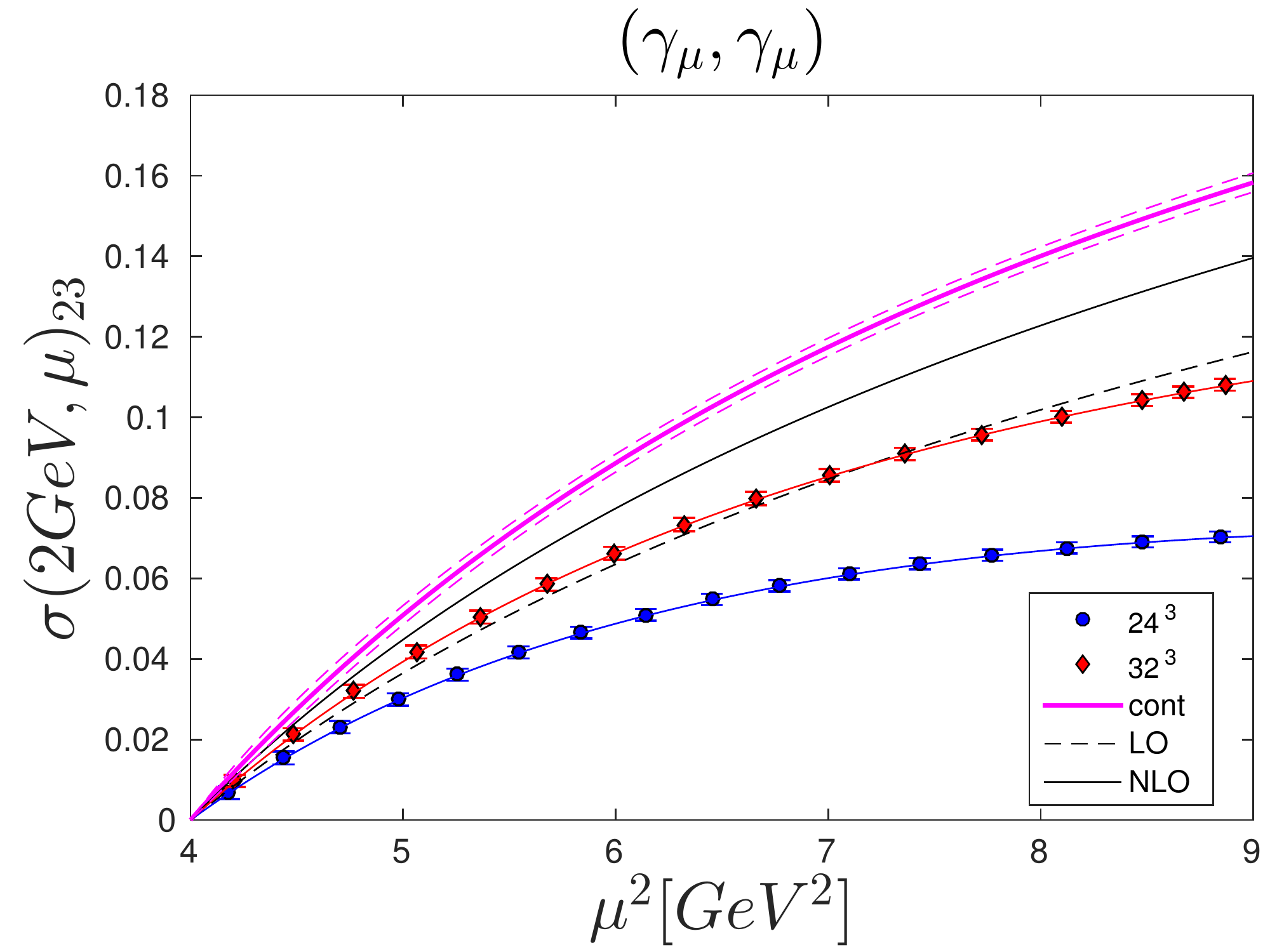} &
\includegraphics[type=pdf,ext=.pdf,read=.pdf,width=8cm]{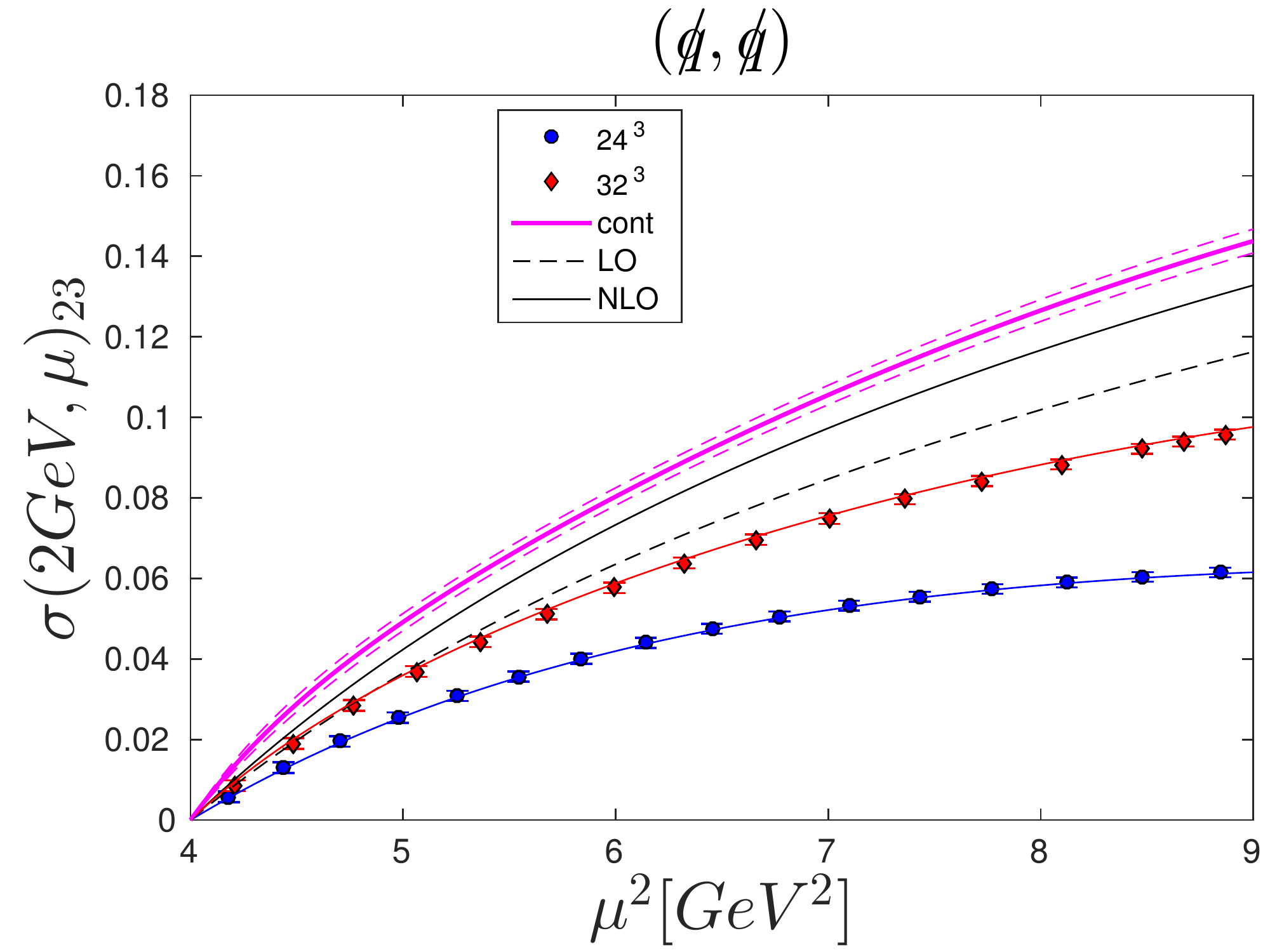} \\ 
\includegraphics[type=pdf,ext=.pdf,read=.pdf,width=8cm]{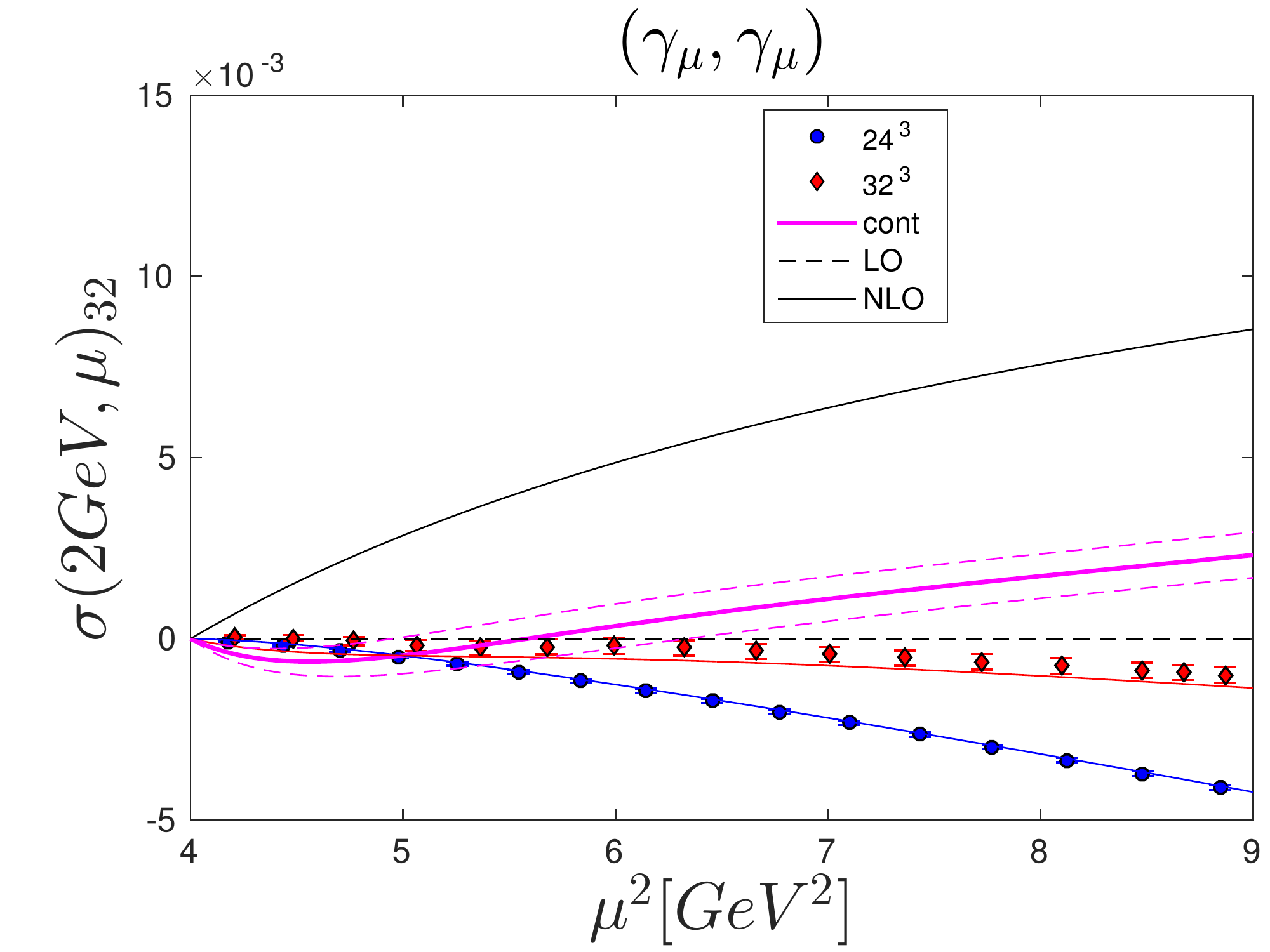} &
\includegraphics[type=pdf,ext=.pdf,read=.pdf,width=8cm]{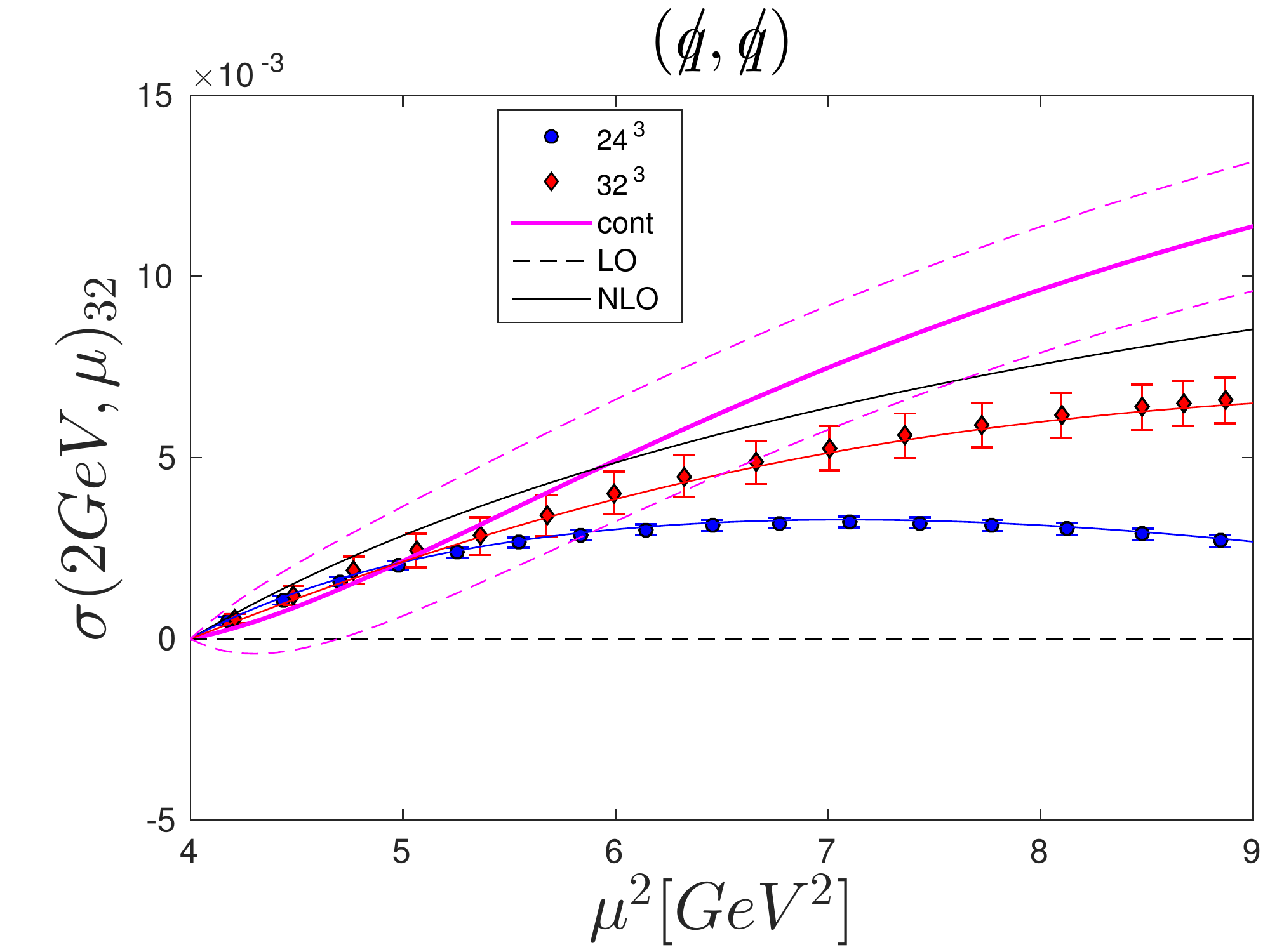}  
\end{tabular} 
\caption{Same as the previous plot for the scale evolution of the non-diagonal $(8,8)$ mixing matrix element
$\sigma_{32}$ and $\sigma_{33}$.}
\label{fig:ssf_32_2to3}
\end{center}
\end{figure}

We divide the non-perturbative running by the perturbative expectation, 
ie we compute
\be
\sigma(\mu_1,\mu_2) U(\mu_1,\mu_2)^{-1}
\ee
where  $\mu_1=\mu$ varies between $2$ and $3\,\GeV$, while $\mu_2=3\,\GeV$ is fixed.
$U$ is the same running computed either at leading order or at next-to-leading 
in perturbation theory. 
The results are shown in Figs.~\ref{fig:ssfoverpt_11}.
\begin{figure}[htb]
\begin{center}
\begin{tabular}{cc}
\includegraphics[type=pdf,ext=.pdf,read=.pdf,width=8cm]{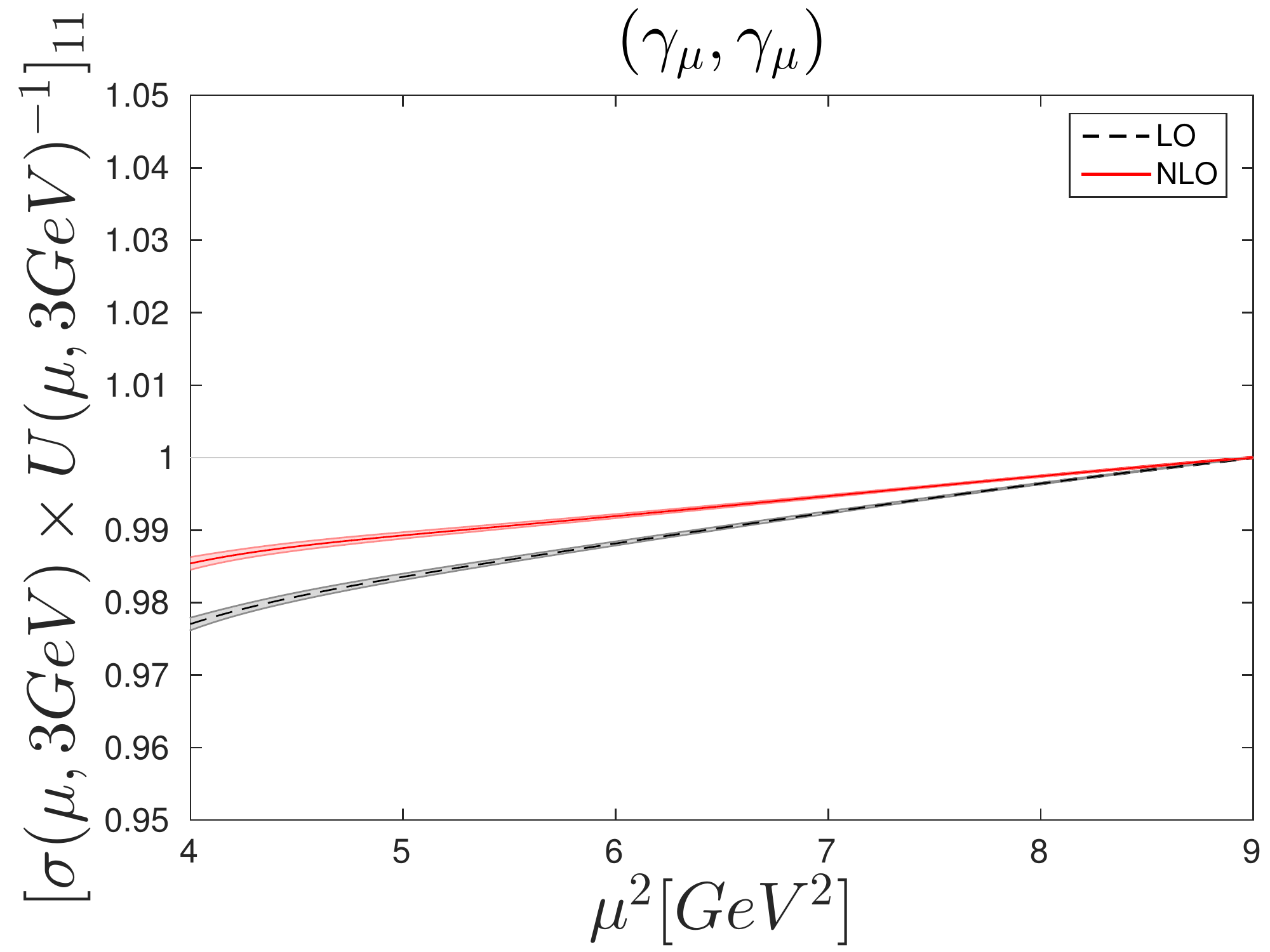} &
\includegraphics[type=pdf,ext=.pdf,read=.pdf,width=8cm]{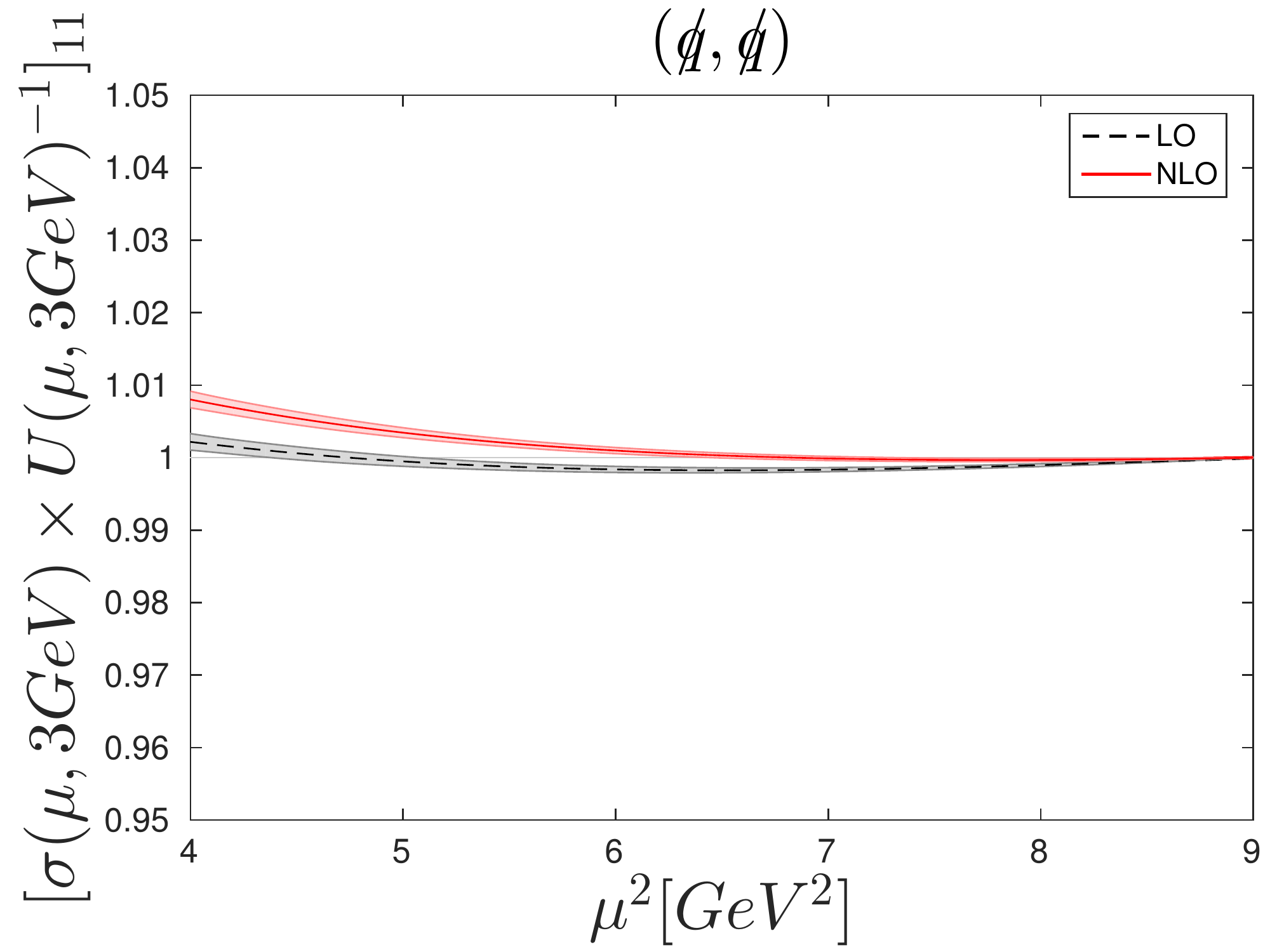}  
\end{tabular} 
\caption{Ratio $\sigma(\mu, 3\GeV)\times  U(\mu, 3\GeV)^{-1}$ for the $(27,1)$ operator 
and for the various schemes;
left: $(\gamma_\mu,\gamma_\mu)$, right: $(\s{q},\s{q})$.}
\label{fig:ssfoverpt_11}
\end{center}
\end{figure}

\begin{figure}[htb]
\begin{center}
\begin{tabular}{cc}
\includegraphics[type=pdf,ext=.pdf,read=.pdf,width=8cm]{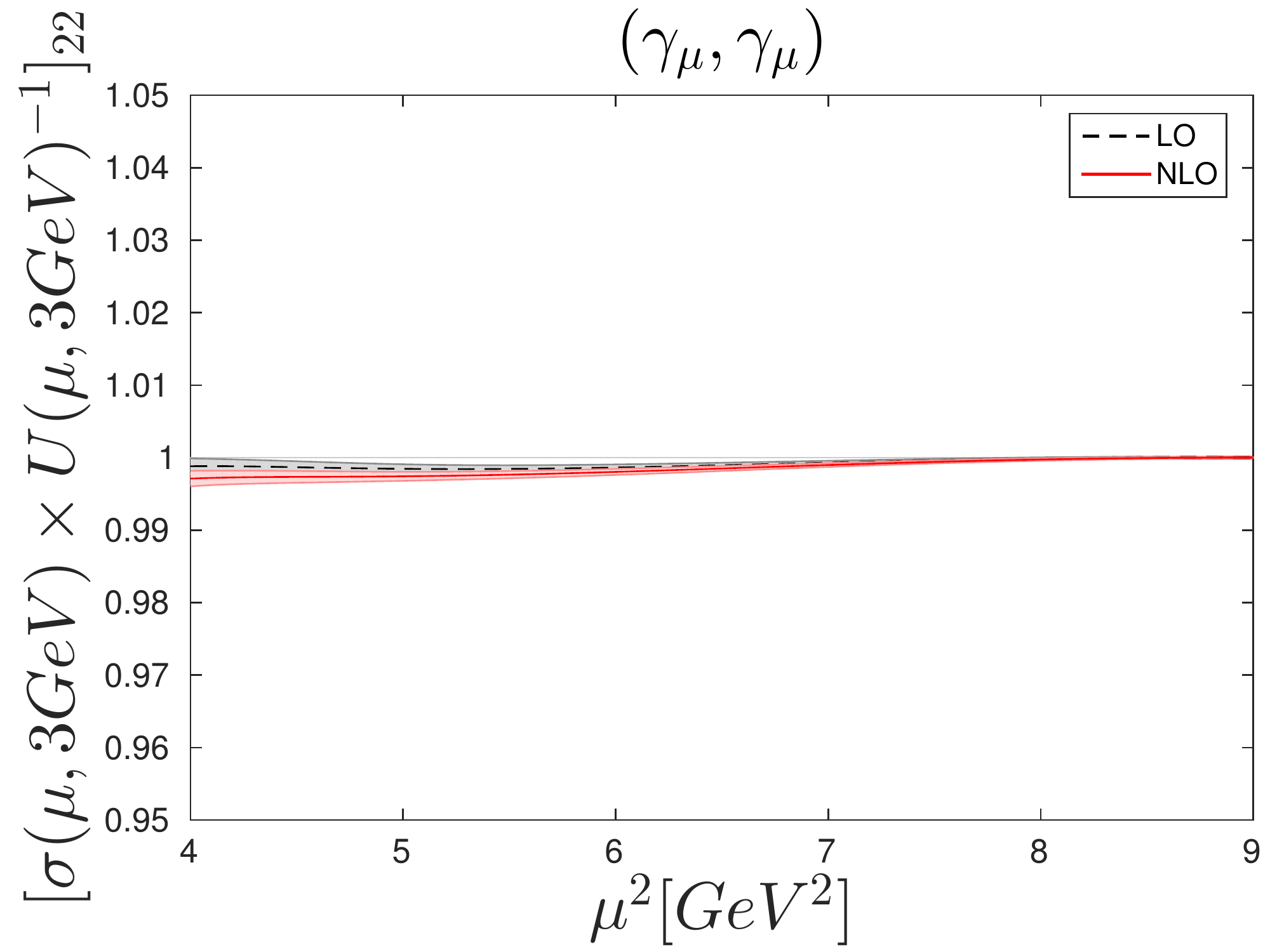} &
\includegraphics[type=pdf,ext=.pdf,read=.pdf,width=8cm]{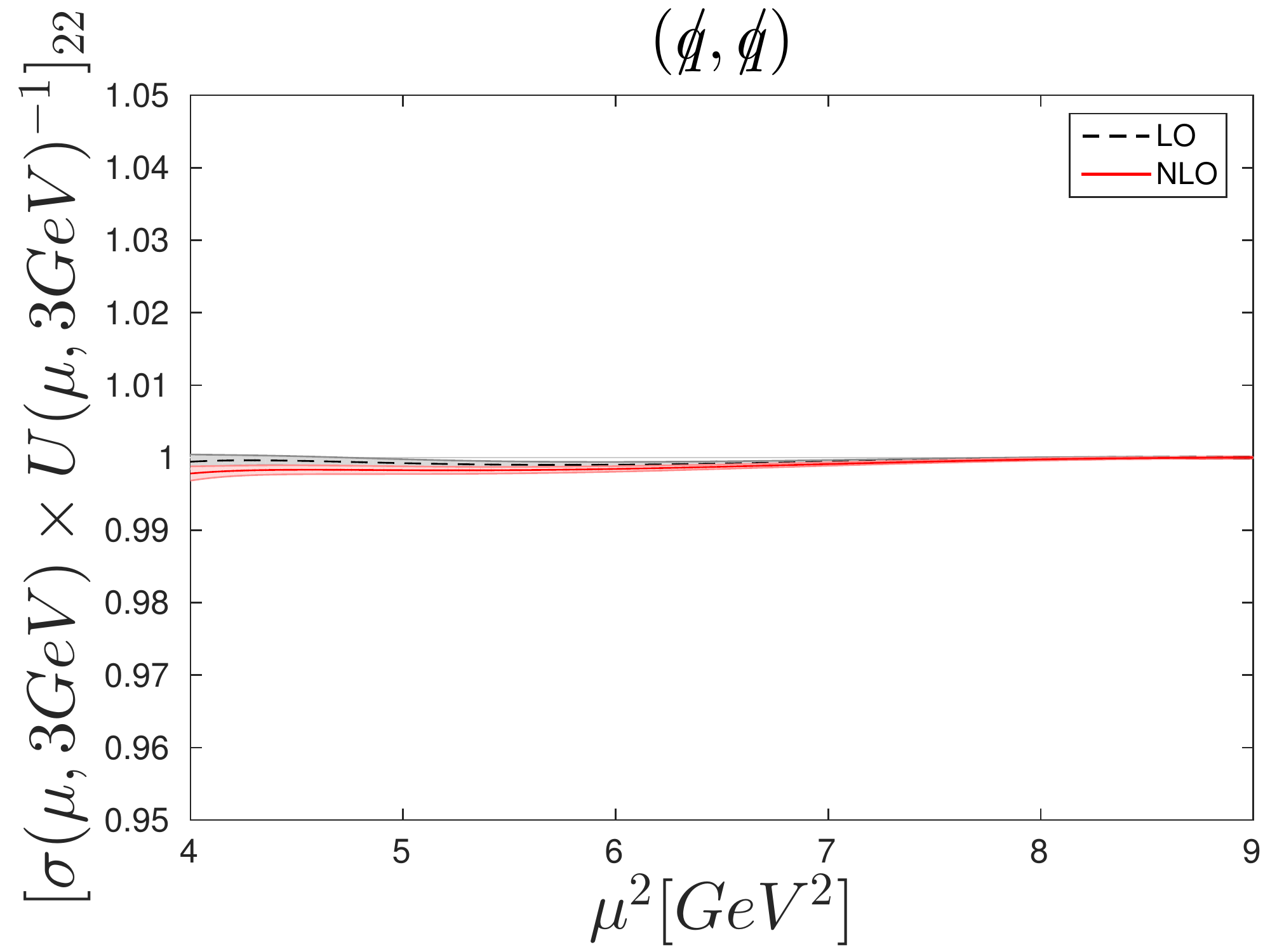} \\ 
\includegraphics[type=pdf,ext=.pdf,read=.pdf,width=8cm]{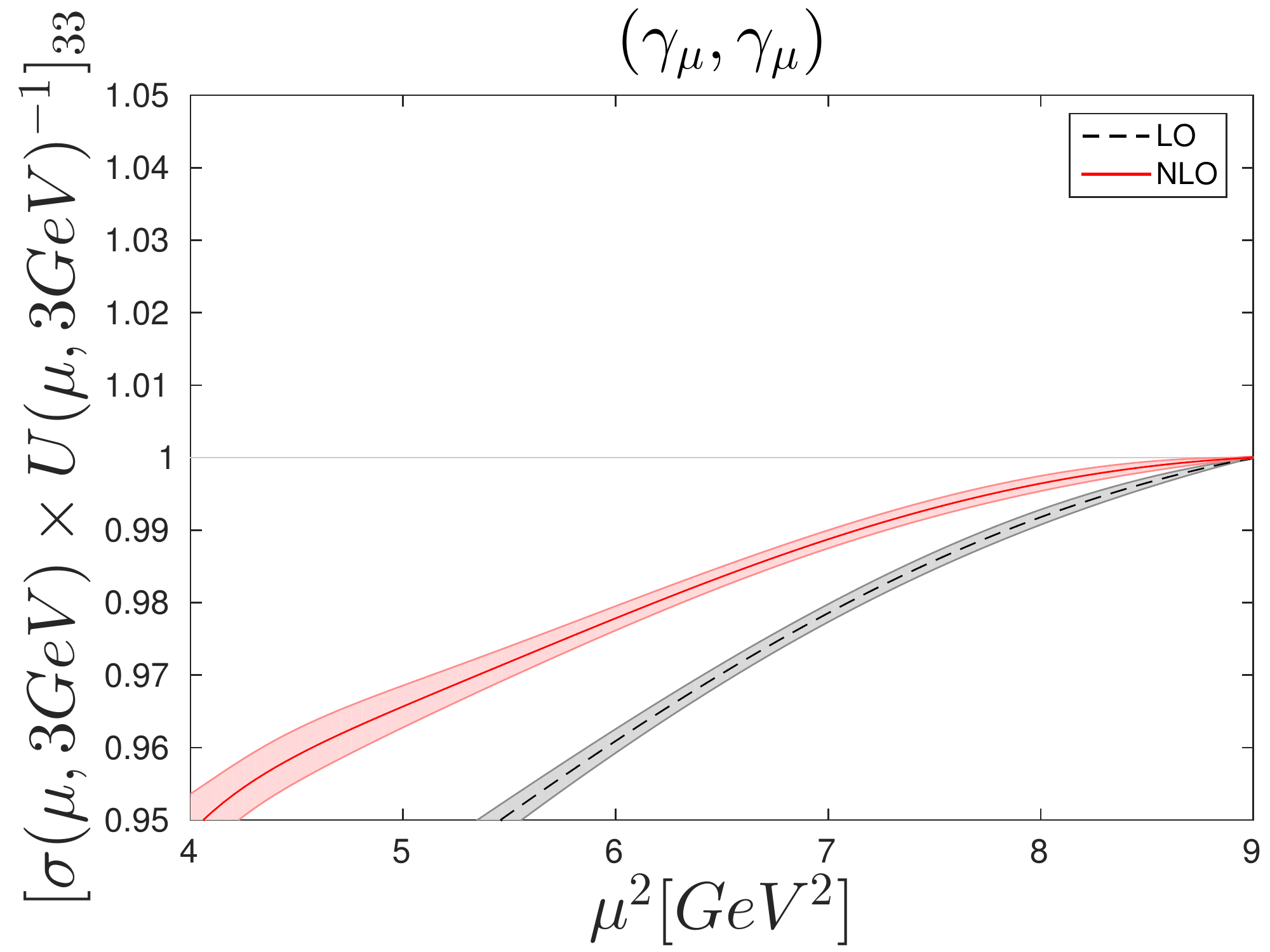} & 
\includegraphics[type=pdf,ext=.pdf,read=.pdf,width=8cm]{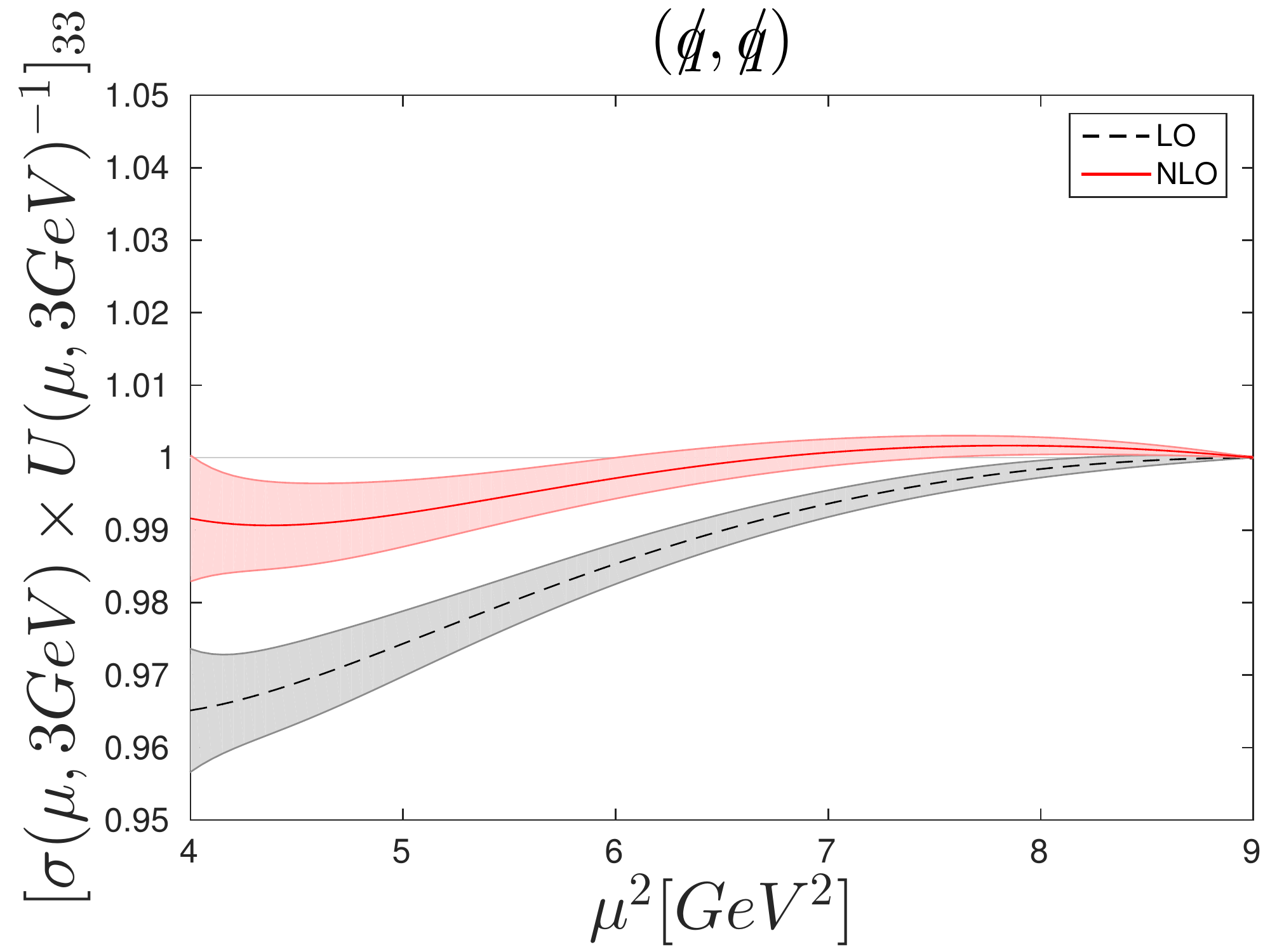}  
\end{tabular} 
\caption{Same as the previous plot for the diagonal $(8,8)$ mixing matrix element.}
\label{fig:ssfoverpt_22}
\end{center}
\end{figure}

\begin{figure}[htb]
\begin{center}
\begin{tabular}{cc}
\includegraphics[type=pdf,ext=.pdf,read=.pdf,width=8cm]{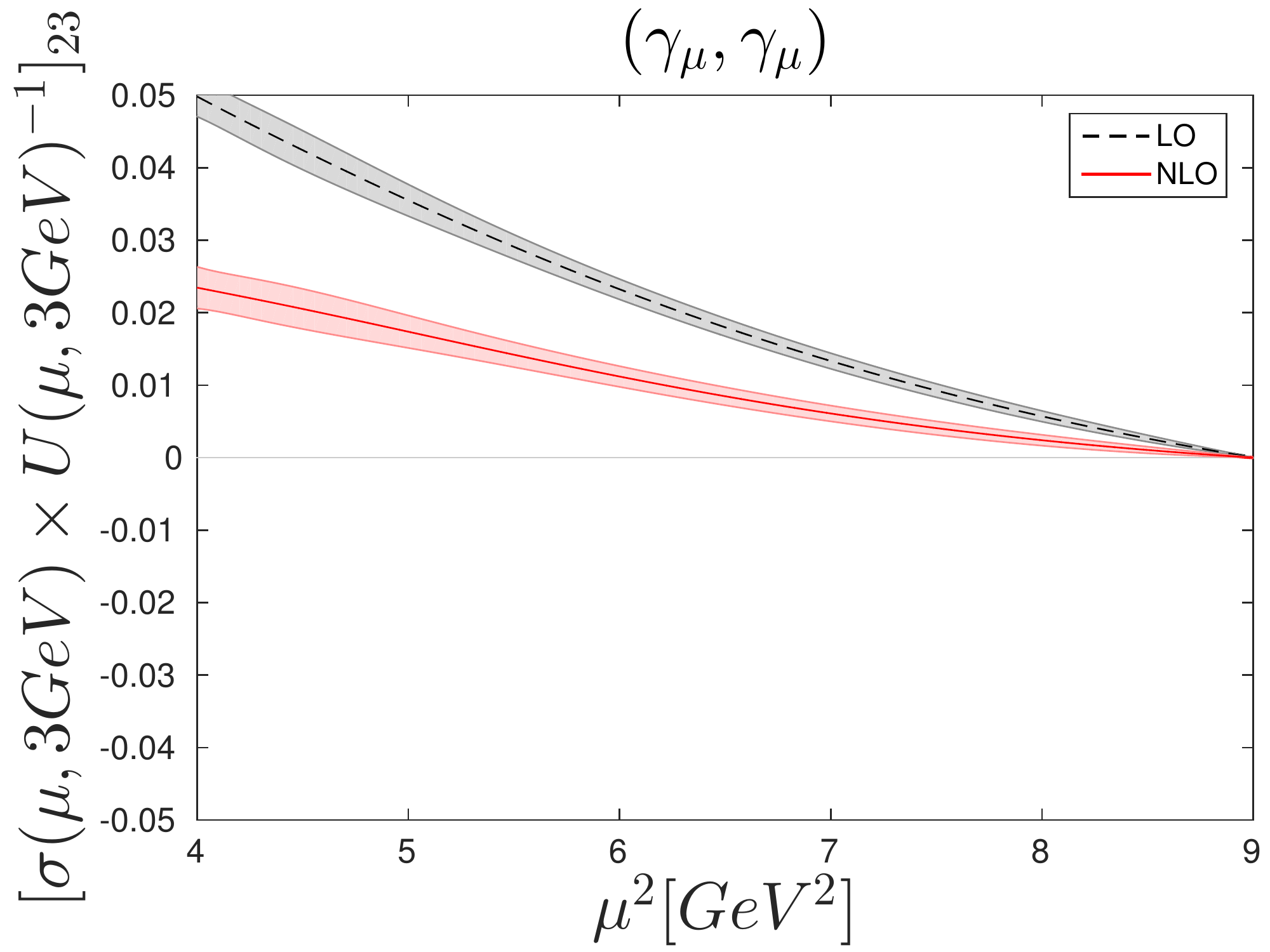} &
\includegraphics[type=pdf,ext=.pdf,read=.pdf,width=8cm]{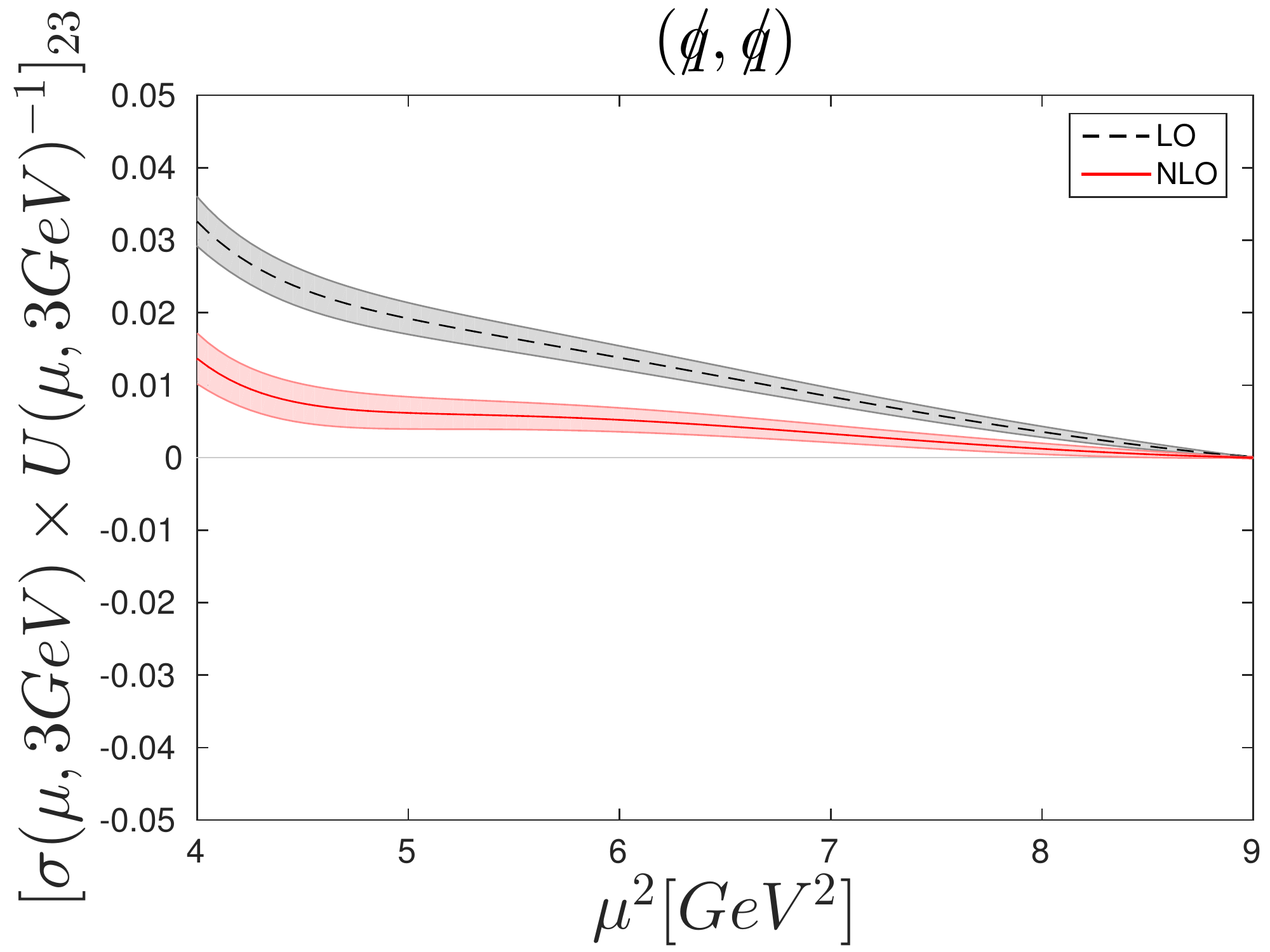} \\
\includegraphics[type=pdf,ext=.pdf,read=.pdf,width=8cm]{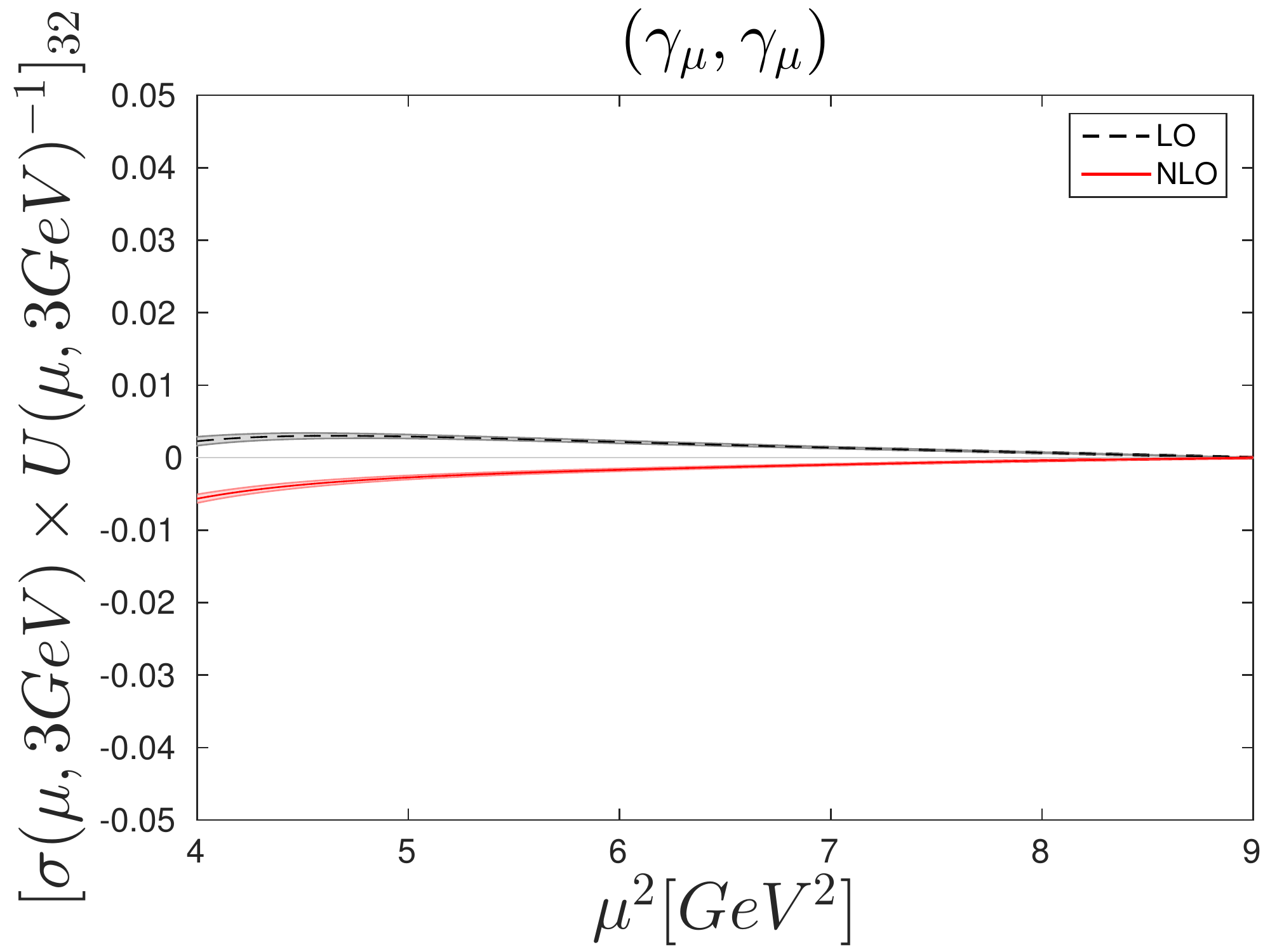} & 
\includegraphics[type=pdf,ext=.pdf,read=.pdf,width=8cm]{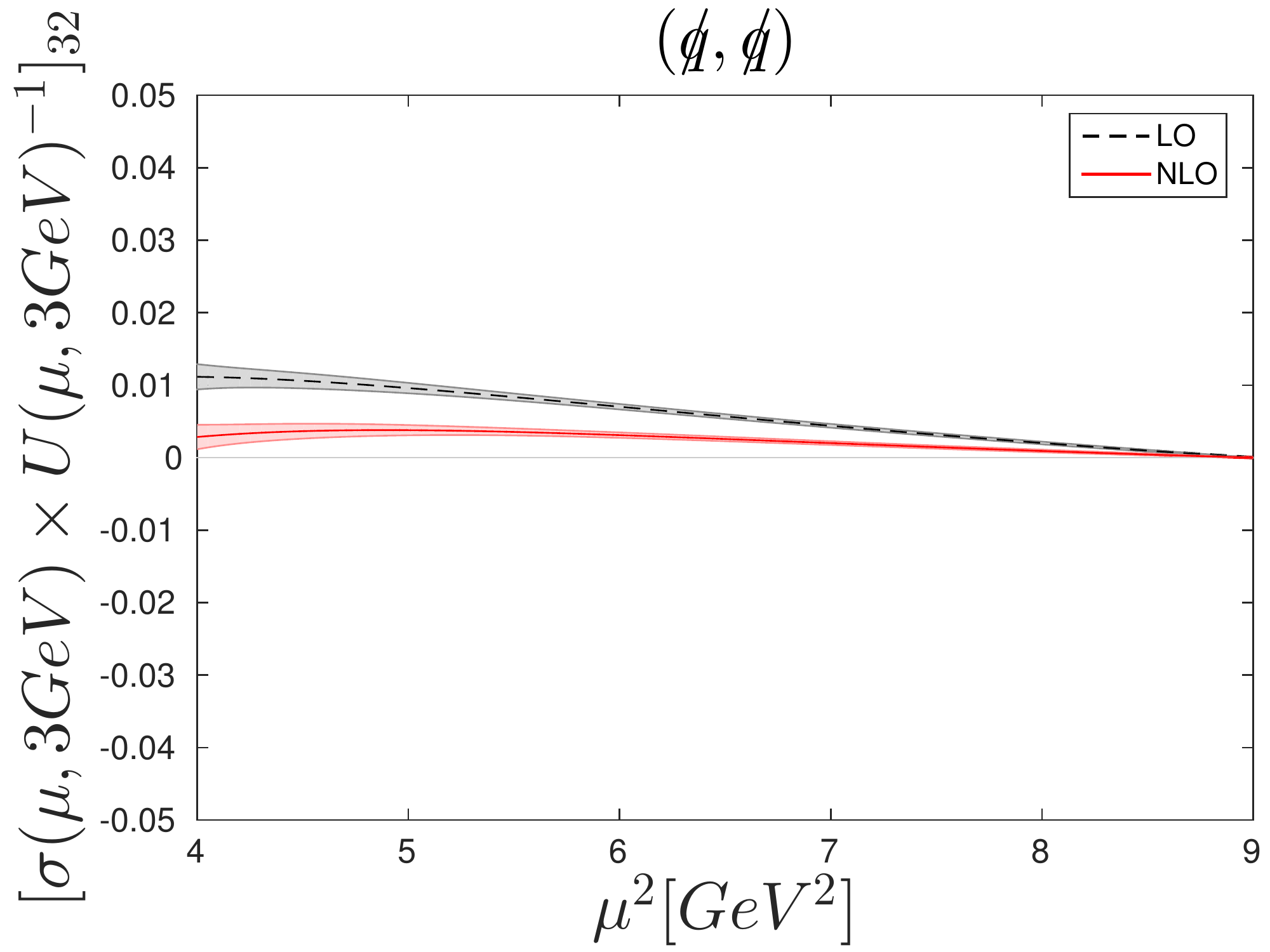}  
\end{tabular} 
\caption{Same as the previous plot for the off-diagonal $(8,8)$ mixing matrix element. }
\label{fig:ssfoverpt_23}
\end{center}
\end{figure}

\begin{figure}[htb]
\begin{center}
\begin{tabular}{cc}
\includegraphics[type=pdf,ext=.pdf,read=.pdf,width=8cm]{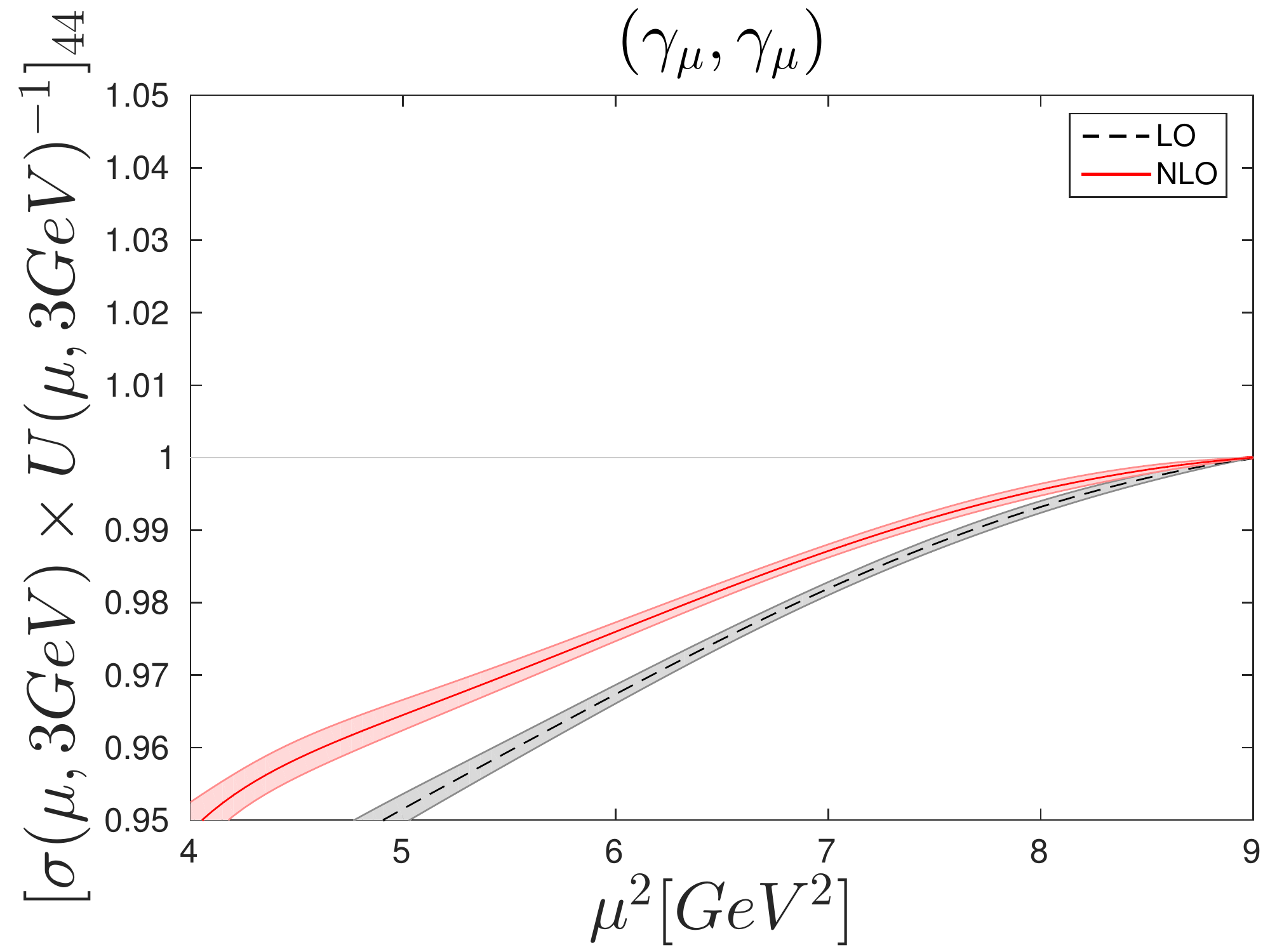} &
\includegraphics[type=pdf,ext=.pdf,read=.pdf,width=8cm]{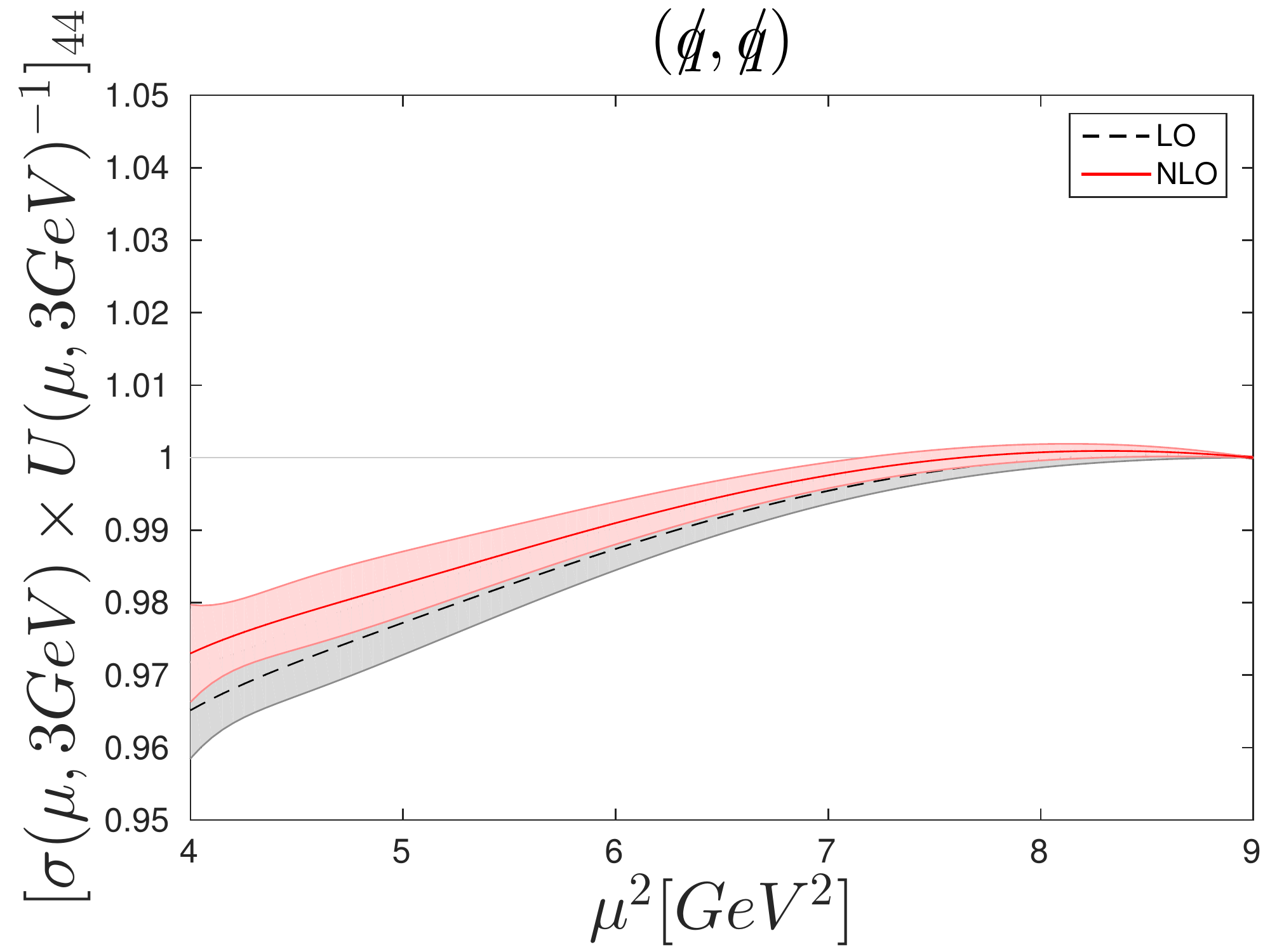} \\
\includegraphics[type=pdf,ext=.pdf,read=.pdf,width=8cm]{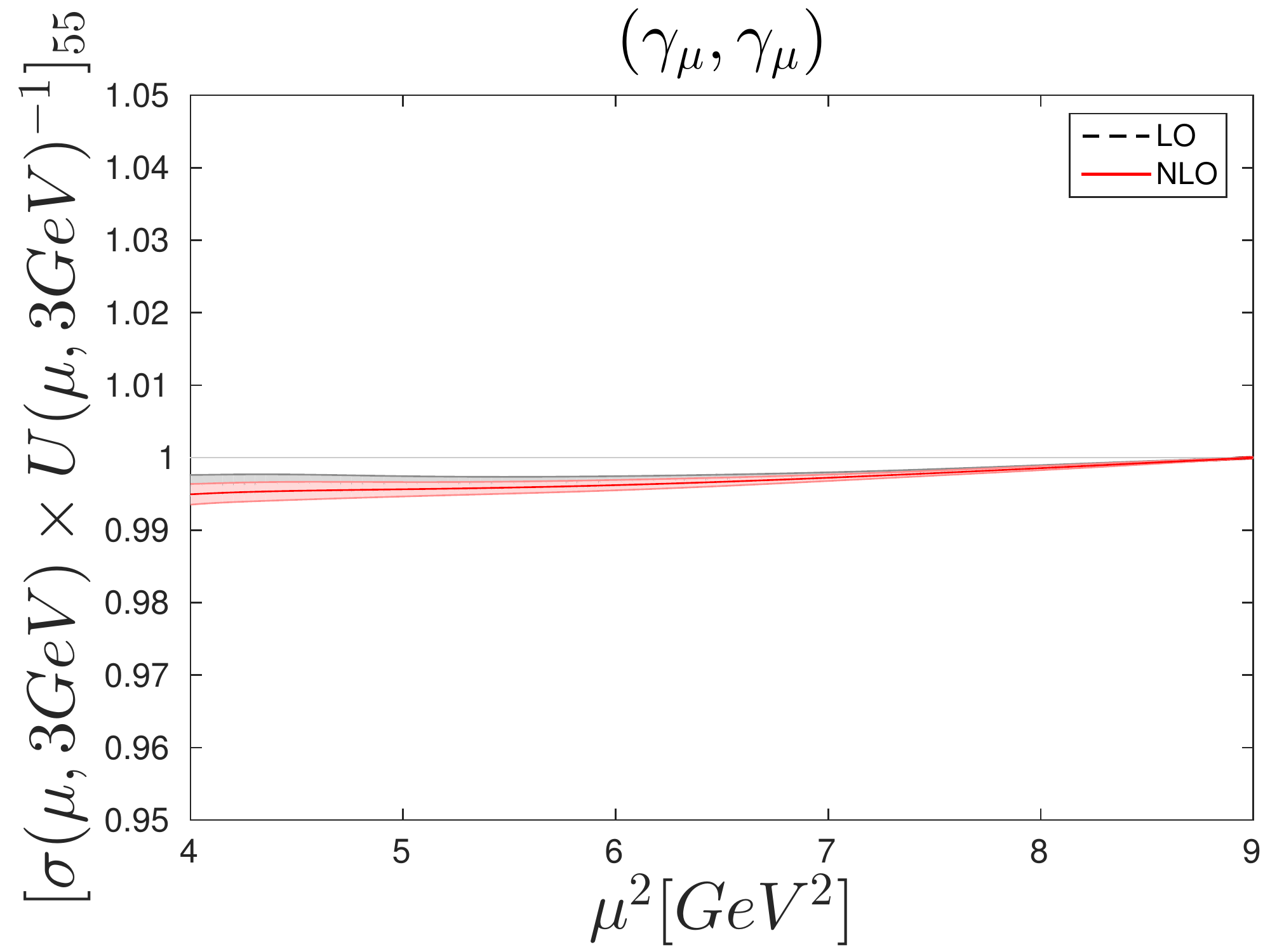} &
\includegraphics[type=pdf,ext=.pdf,read=.pdf,width=8cm]{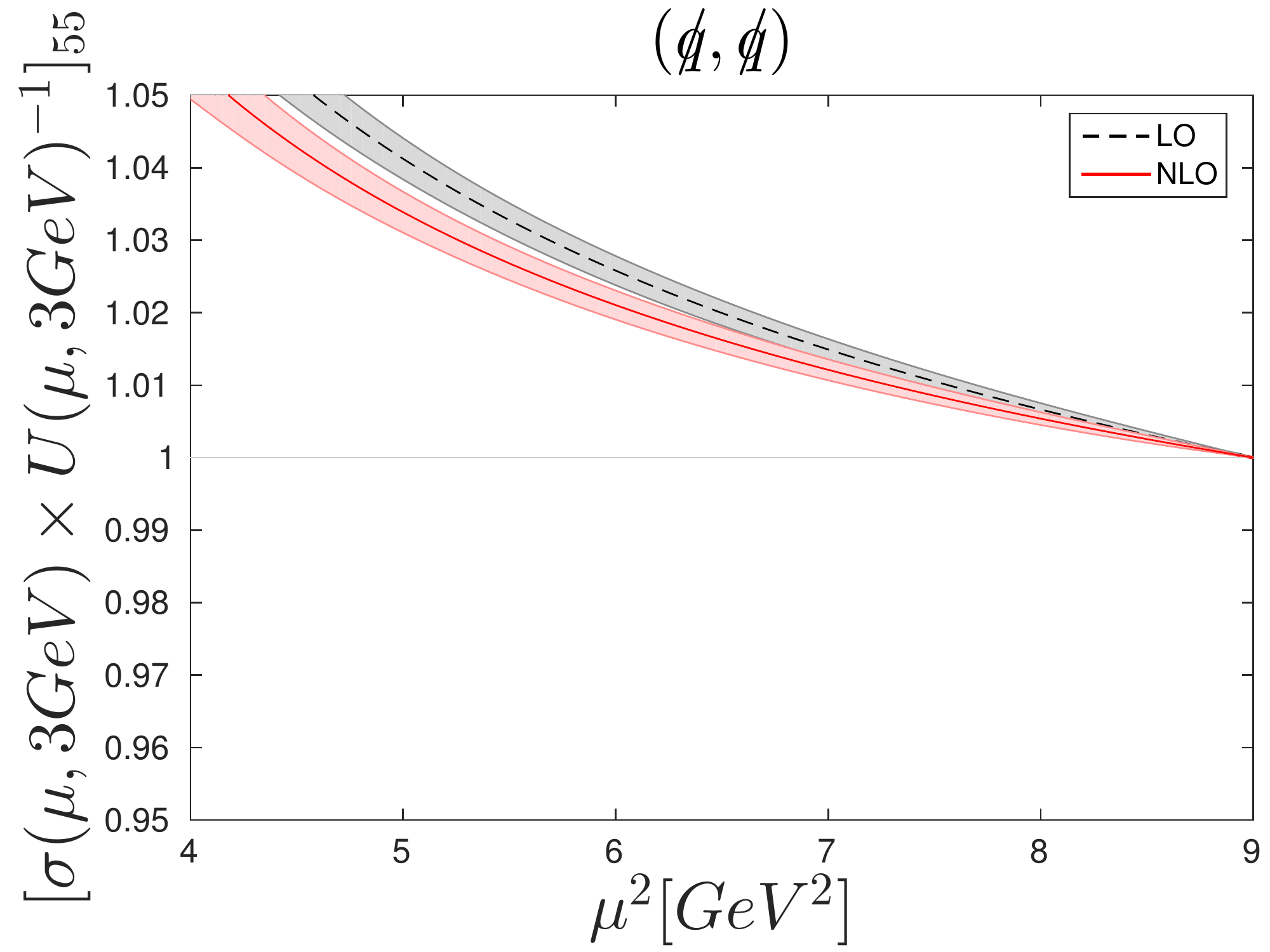} \\
\end{tabular} 
\caption{Same as the previous plot for the diagonal $(6,\bar 6)$ mixing matrix element.}
\label{fig:ssfoverpt_44}
\end{center}
\end{figure}

We observe the running can be relatively important, see for example $\sigma_{33}$ and $\sigma_{44}$,
this is expected from their anomalous dimension~\cite{Ciuchini:1997bw,Buras:2000if,Papinutto:2016xpq} .
Starting from $\mu=3$ GeV and running down to 2 GeV, the non-perturbative scale evolution is qualitatively
well-described by the Next-to-Leading perturbative prediction.
In the worse cases we observe a deviation of around 5 $\%$ at 2 GeV.
In a future we will include a finer lattice spacing to have a better handle on the discretisation effects.

\clearpage

\subsection{Fierz relations}
\label{appendix:fierz}

In Eq.(\ref{eq:renbasis_even}), we have only considered four-quark operators with
a colour-unmixed structure 
\be
(\bar s_a \Gamma d_a)(\bar s_b \Gamma d_b)
\equiv
(\bar s \Gamma d)(\bar s \Gamma d)_{\rm unm}
\ee
However the color partners
\be
(\bar s_a \Gamma d_b)(\bar s_b \Gamma d_a)
\equiv
(\bar s \Gamma d)(\bar s \Gamma d)_{\rm mix}
\ee
can be recovered by a Fierz transformation,
\be
Q^{\text{mix}}_i = F_{ij} Q^{\text{unm}}_j \;,
\ee
that we give explicitely here.
For the dirac structure, we introduce the standard notation
\be
\begin{aligned}
SS &= (\bar s d)(\bar s d) \; ,\\
VV &= (\bar s \gamma_\mu d)(\bar s \gamma_\mu d) \;,\\
TT &= \sum_{\nu > \mu} (\bar s \gamma_\mu \gamma_\nu d)(\bar s \gamma_\mu \gamma_\nu d) \;,\\
AA &= (\bar s \gamma_\mu \gamma_5 d)(\bar s \gamma_\mu \gamma_5 d) \;,\\
PP &= (\bar s \gamma_5 d)(\bar s \gamma_5 d)\;. 
\end{aligned}
\ee
For Euclidean $\gamma$ matrices the Fierz transformation
in the NPR basis reads
\be
\left(
\begin{array}{c}
  VV+AA \\
  VV-AA \\
  SS-PP \\
  SS+PP \\
  TT    \\
\end{array}
\right)_{(\rm mix)} =
\begin{pmatrix}
  1 &  0  &  0 &  0 &  0 \\
  0 &  0  & -2 &  0 &  0 \\
  0 &-1/2 &  0 &  0 &  0 \\
  0 &  0  &  0 &-1/2& 1/2 \\
  0 &  0  &  0 & 3/2& 1/2 \\
\end{pmatrix}
\times
\left(
\begin{array}{c}
  VV+AA \\
  VV-AA \\
  SS-PP \\
  SS+PP \\
  TT    \\
\end{array}
\right)_{(\rm unm)}
\ee

Results in the literature are often given in the SUSY
basis~\cite{Gabbiani:1996hi,Allton:1998sm,Ciuchini:1998ix},
which was also our choice in~\cite{Garron:2016mva}, 
\be
\label{eq:susybasis}
 \begin{aligned}
O_2&=\;(\overline s_a (1-\gamma_5) d_a)\,(\overline s_b (1-\gamma_5) d_b)\\
O_3&=\;(\overline s_a  (1-\gamma_5) d_b)\,(\overline s_b   (1-\gamma_5) d_a)\\
O_4&=\;(\overline s_a  (1-\gamma_5) d_a)\,(\overline s_b   (1+\gamma_5) d_b)\\
O_5&=\;(\overline s_a  (1-\gamma_5) d_b)\,(\overline s_b   (1+\gamma_5) d_a)\,,
 \end{aligned}
\ee
in addition to $O_1=Q_1$. In practice we only consider the parity even part of
these operators.
The relation between the NPR and the SUSY basis is then given by 
$O^+ = T Q^+$ where
\be
T= 
\begin{pmatrix}
  1 &  0   &  0 &  0  &  0    \\
  0 &  0   &  0 &  1  &  0    \\
  0 &  0   &  0 & -1/2&  1/2  \\
  0 &  0   &  1 &  0  &  0    \\
  0 & -1/2 &  0 &  0  &  0    \\
\end{pmatrix}
\;.
\ee

\bibliography{biblio}{}
\bibliographystyle{h-elsevier}

\end{document}